\documentclass[preprint]{aastex62}
\received{..., 2022}
\revised{...., 2022}
\accepted{\today}
\submitjournal{ApJ}

\shorttitle{Understanding variability of solar Balmer lines}

\shortauthors{Criscuoli et al.}

\begin{document}
\title{Understanding Sun-as-a-star variability of solar Balmer lines}

\correspondingauthor{Criscuoli}
\email{scriscuo@nso.edu}

\author[0000-0002-4525-9038]{Criscuoli, S.}
\affil{National Solar Observatory\\
3665 Discovery Dr. \\
Boulder, CO 80303, USA}
\author[0000-0002-9847-3388]{Marchenko, S.}
\affil{Science Systems and Applications, Inc.,\\ Lanham, MD, USA}
\affil{Goddard Space Flight Center, Greenbelt, MD, USA}
\author[0000-0003-3252-1882]{DeLand, M.}
\affil{Science Systems and Applications, Inc.,\\ Lanham, MD, USA}
\affil{Goddard Space Flight Center, Greenbelt, MD, USA}
\author[0000-0002-9308-3639]{Choudhary, D.}
\affil{Department of Physics and Astronomy, California State University,\\ Northridge (CSUN), CA 91330-8268, USA}
\author{Kopp, G.}
\affil{Univ. of Colorado / LASP\\
3665 Discovery Dr. \\
Boulder, CO 80303, USA}

\begin{abstract}

Precise, high-cadence, long-term records of stellar spectral variability at different temporal scales lead to better understanding of a wide variety of phenomena including  stellar atmospheres and dynamos, convective motions, and rotational periods. Here, we investigate the variability of solar Balmer lines (H-$\alpha$, -$\beta$, -$\gamma$, -$\delta$) observed by space-borne radiometers (OSIRIS, SCIAMACHY, OMI, and GOME-2), combining these precise, long-term observations with high-resolution data from  the ground-based NSO/ISS spectrograph. We relate the detected variability to the appearance of magnetic features on the solar disk. We find that on solar-rotational timescales (about 1 month), the Balmer line activity indices (defined as line-core to line-wing ratios) closely follow variations in the total solar irradiance (which is predominantly photospheric), thus frequently (specifically, during passages of sunspot groups) deviating from behavior of activity indices that track chromospheric activity levels. On longer timescales, the correlation with chromospheric indices increases, with periods of low- or even anti-correlation found at intermediate timescales. Comparison of these observations with estimates from semi-empirical
irradiance reconstructions helps quantify the contributions of different magnetic and quiet features. We conclude that both the lower sensitivity to network and in part the higher sensitivity to filaments and prominences, may result in complex, time-dependent relationships between Balmer and other chromospheric indices observed for the Sun and solar-like stars. The fact that core and wings contribute in similar manner to the variability, and current knowledge of Balmer-lines formation in stellar atmospheres, support the notion that Balmer lines core-to-wing ratios indices behave more like photospheric rather than chromospheric indices.

\end{abstract}

\section{Introduction}
\label{sec:intro}

The solar spectrum changes over different temporal scales, modulated by variations of the solar magnetic field \citep{domingo2009, ermolli2013}. It is very well known that these variations affect the physical and chemical properties of the Earth's atmosphere and of the heliosphere. Understanding the variability of the solar spectrum is therefore fundamental for a variety of applications which include space weather \citep[e.g.][]{lilensten2008,schwadron2010}, Earth's climate \citep[e.g.][]{gray2010, matthes2017}, ecosystems \citep[e.g.][and references therein]{frouin2018}, and renewable energies \citep[e.g.][]{myers2017}. 
%Understanding stellar spectral variability is paramount  for understanding and modeling stellar atmospheres and their variability and is fundamental for the detectability of exoplanets and for assessing their habitability. 
Similarly, the spectral energy distributions of stars affect the physical and chemical properties of planets' atmospheres, thus playing a major role in determining their habitability  \citep[e.g.][]{bordi2015,linsky2019,galluzzo2021}. Moreover, stellar spectral variability
may affect detection of exoplanets via the radial-velocity measurements \citep{lanza2011, oshagh2018}, the characterization of exoplanets' atmospheres obtained through transmission spectroscopy \citep[e.g.][]{kowa2019,yan2020, rackham2022}, and estimates of the abundance of molecules used as bio-markers 
\citep[e.g.][]{rackham2019}. 

Chromospheric indices, such as those derived from the Ca~II H\&K, \ion{Mg}{2} h\&k, and H$\alpha$ lines (a ratio of an average line-core flux to average line-wings flux), are often employed in many of the applications described above, as they are known to track the magnetic activity of the Sun and other stars very well  \citep[e.g.][]{egeland2017, linsky2017, lovric2017,degrijs2021}. In particular, in the case of the Sun, chromospheric indices are fundamental to reconstructing the solar irradiance (the radiant solar energy impinging at the top of the Earth atmosphere) at times when irradiance measurements were not available \citep[e.g.][] {coddington2016, yeo2017a, penza2022}, thus allowing production of the long and uniform time series necessary to estimate the effects of solar variability on the Earth's climate \citep[e.g.][]{,lean2017}. Chromospheric indices are also fundamental to estimating magnetic-activity variability over several decades, at times when full-disk magnetograms were not available \citep{chatz2022}, and thus constrain solar dynamo models \citep{pevtsov2016}. On stars other than the Sun, the areal contribution from magnetic regions cannot be directly observed, and one has to rely on secondary indicators, which are usually based on spectral signatures \citep[e.g.][]{fang2016, andretta2017,thompson2020}. Indices are also used as proxies of spectral variability when direct measurements of the latter are not available \citep[e.g.][]{youngblood2017}. 
Such estimates, when carried out over several  stellar rotations, provide unique information about stellar magnetic cycles and dynamos  \citep[e.g.][]{fabbian2017,faurobert2019}.

However, different chromospheric indices track the evolution of the magnetic field in different ways, and activity indices may present complex correlations among each other \citep[e.g.][]{dudok2009,salabert2017,tapping2017}. These differences provide observational constraints to understanding and modeling the physical mechanisms underlying the formation of chromospheric diagnostics \citep[e.g.][]{tilipman2021}. Of particular interest for modeling stellar chromospheres is the complex, non-linear relation observed between indices derived from the \ion{Ca}{2} (393 nm)  and Balmer lines, in particular H$\alpha$, found for F-G-K stars \citep[e.g.][]{strassmeier1990,cincunegui2007, gomesdasilva2014, flores2018, meunier2022}. These studies  showed that the variability of Balmer and \ion{Ca}{2} indices can be correlated, anti-correlated, or uncorrelated, with no clear trend depending on the fundamental stellar parameters.  More recently, \cite{meunier2022} showed that the correlation between the two indices is a function of time and of the level of activity, with only a small fraction (3\%) of low-activity stars presenting a clear anti-correlation. \citet{gomes2022} used a large sample of 152 FGK stars and studied the correlation between \ion{Ca}{2} and H$\alpha$ line indices using variable bandwidths, finding that the correlation is maximized while employing narrow, 0.06 nm, H$\alpha$ bandwidths. 

Understanding the variability of Balmer lines on different temporal scales is important because these lines are employed as diagnostics for a variety of phenomena, such as modeling stellar flares \citep{capparelli2017, kowa2017, kowa2022}, detecting and characterizing stellar coronal mass ejections \citep{vida2019,namekata2021}, and estimating stellar parameters \citep[e.g.][]{barklem2002,bergemann2016, amarsi2018}. Moreover, Balmer lines are fundamental diagnostics of the outer layers of exoplanetary atmospheres \citep{huang2017}, so that understanding the variability on stellar-rotational timescales is fundamental to estimating and disentangling the contribution due to stellar contamination \citep[e.g.][]{cauley2018, chen2020}.

The Sun offers a unique opportunity to improve our understanding of stellar magnetic variability for two main reasons. The first is that long time series spanning several solar cycles are available for many  chromospheric and coronal indices \citep{ermolli2014}, whereas only a few stars have been monitored over a few stellar cycles, so that stellar magnetic variability is typically investigated through statistical studies conducted using large surveys. The second is that spatially resolved solar observations help disentangle the contributions to variability from the different magnetic features observed on the solar disk. 

The longest time series of Sun-as-a-star observations of several photospheric and chromospheric lines is that acquired at the McMath-Pierce Solar Telescope between 1974 and 2010 \citep{livingston2007}. Analyses of these data show that the \ion{Ca}{2}~K and H$\alpha$ cores emissions are clearly correlated over the solar-cycle timescale \citep{livingston2007,livingston2010}.  By analyzing the same dataset, \citet{meunier2009} showed that the degree of correlation between the two indices is a strong function of magnetic activity, strengthening at the maximum and weakening at the minimum of the cycle.  The authors  ascribed these trends to the different sensitivities of these two lines to the contributions of plages and filaments. More recently, \citet{maldonado2019} analyzed three years of observations acquired with HARPS-N during the descending phase of Solar Cycle 24 and found that indices derived from H-$\alpha$, -$\beta$, and -$\gamma$ lines are anti-correlated with the S-index derived from the Ca II K line.  \citet{marchenko2014} also noted that H$\beta$, H$\gamma$, and H$\delta$ observed with the Ozone Monitoring Instrument (OMI) showed negligible variations with activity, as opposed to other lines observed in the range 265-500 nm, which in general grow shallower with increases of  the \ion{Mg}{2} and \ion{Ca}{2} indices.  

 \citet{marchenko2021} (Paper~I, hereafter) investigated the variability of H$\beta$, H$\gamma$, and H$\delta$  by analyzing long-term records of the OMI daily solar observations and the daily spectral irradiance measurements from the TROPOspheric Ozone Monitoring Instrument (TROPOMI) mission. Their analysis showed that on the solar-rotational timescale, an index constructed by combining the core-to-wing ratio of the three lines  closely follows changes of the inverted total solar irradiance (TSI), deviating from trends observed in chromospheric indices, thus suggesting that Balmer lines more closely track the variability of photospheric, rather than chromospheric, properties. 
 
 The main goal of this paper is to extend and complement the analysis presented in Paper~I by investigating the variability of H$\alpha$ on both solar-rotational and solar-cycle timescales, as measured during the descending phase of Cycle 23 and the ascending/maximum phases of Cycle 24. To the best of our knowledge,  this is the first work analyzing the variability of the H$\alpha$ core-to-wing index over several solar rotations and under different photospheric and chromospheric conditions.  The McMath time series employed in previous studies lacks indeed the necessary temporal cadence, and short-term analysis of these ground-based, disk-integrated observations are affected by large daily variations produced by various observational and instrumental effects, as noticed by \citet{livingston2007}. \citet{maldonado2019} also reported that the solar-rotational variability patterns could not be sufficiently investigated due to time gaps and insufficient signal-to-noise in HARPS-N observations. Such drawbacks are mitigated in this study by analyzing space-based, high signal-to-noise radiometric observations. To the best of our knowledge, this is also the first long-term (approximately 10 years) analysis carried out on observations other than those acquired at the McMath. This is an important aspect as the only other measurements presented in the literature, those obtained with HARPS-N (covering 3 years), produce discrepant results. 
 
 Comparison of these observations with estimates from the semi-empirical irradiance reconstructions allows quantifying the contributions of different quiet and magnetic features (i.e., network, plage, sunspots) and investigating the variability of the upper Balmer-line indices, for which observations on the solar-cycle timescale are not available yet. Because irradiance reconstructions are known to depend on the set of employed atmospheric models \citep{ermolli2013}, we also performed a detailed comparison of reconstructions obtained using two sets of such models and different sunspot umbra and penumbral models widely used for the modeling of both solar- and stellar-irradiance variability. The presented comparison provides important validation of models that are used for a variety of astrophysical purposes, which include producing long-term irradiance estimates to use as inputs in climatology studies \citep{jungclaus2017}, estimates of exoplanetary atmospheres \citep{rackham2022}, interpretation of spatially resolved solar observations \citep[e.g.][]{rutten2011,criscuoli2013b, schmit2015}, instrument characterizations \citep[e.g.][]{cohen2015, quintero2018}, and so on. 
 
The analyzed measurements and the model are described in Sections \ref{sec:measurements} and \ref{sec:model}, respectively.  Our results are described in Sec.~\ref{sec:results} and discussed in Sec.~\ref{sec:discussions}. Our conclusions are presented in Sec.~\ref{sec:conclusions}.

 \section{Measurements}
\label{sec:measurements}
Evaluation of solar Balmer-line behavior benefits from space-based observations because those are not affected by terrestrial atmospheric effects (e.g., clouds, seeing conditions). However, only selected instruments have the necessary wavelength coverage and spectral resolution to examine these features. We focus on measurements of the H$\alpha$ line at 656.3 nm, which we compare with the variability of upper Balmer lines obtained from OMI measurements (Paper~I). Here we adopt a similar approach and  construct a core-to-wing solar-activity index for various data sets by averaging solar fluxes at multiple wavelengths in the H$\alpha$ core and line wings.  This technique is broadly used in solar \citep[e.g.][]{maldonado2019} and stellar \citep[e.g.][]{meunier2022} studies  and exploits the advantage of efficient mitigation of various instrumental effects at the expense of reduced spectral resolution.

\subsection{GOME-2A and SCIAMACHY}
We first consider two instruments that measure SSI (spectral solar irradiance) with high spectral resolution on a daily basis to provide calibrations for their primary observations of backscattered atmospheric radiance. The GOME (Global Ozone Monitoring Experiment) instrument \citep{burrows1999} covers the spectral range 240-790 nm at 0.2-0.3 nm resolution. Multiple GOME instruments with overlapping data records have been operating since 1996. Here we analyze the solar irradiances collected by Metop-A/GOME (GOME-2A hereafter) since 2006 \citep{munro2016}. The SCIAMACHY (Scanning Imaging Absorption Spectrometer for Atmospheric Chartography) instrument \citep{skupin2005} covers the spectral range 240-2390 nm at moderate resolution (0.2 nm in UV to 1.5 nm in near-IR). SCIAMACHY operated from 2003 to 2012. GOME-2 and SCIAMACHY line indices are calculated (as averages) on 1 nm (H${\alpha}$ line core) and 2-3 nm (line wings) wavelength samples, thus falling close to 0.2-0.3 nm sampling of Balmer and \ion{Ca}{2} line cores and 1-2 nm averages in the line wings routinely applied to high-resolution solar data \citep[e.g.][]{maldonado2019}.  However, the spectral resolutions of these radiometers are much worse than those obtained with ground-based spectrographs employed in previous studies. Therefore, in order to investigate the effects of spectral resolution on the estimate of the variability of the H$\alpha$ index, we compared the synthetic index obtained with the models described in Sec.~\ref{sec:model} degraded to the spectral resolution of SCIAMACHY (R$\simeq10^3$) and of HARPS-N (R$\simeq10^5$), using the index definition described below and the one adopted in \citet{maldonado2019}, respectively. As shown in the Appendix (Fig.~\ref{corre_Harps_SCIA}, Sec.~\ref{sec:resolution}), the spectral resolution mostly affects the amplitude of the variations, but the two estimates are highly correlated, indicating that both estimates are equally suitable to evaluate how the H$\alpha$ index changes with respect to other indices, which is one of the main objectives of this study. Note that similar results, not shown, were found comparing the $H\alpha$ index variability obtained with simulated OMI-SCIAMACHY measurements with simulated ISS measurements (see Sec. \ref{sec:iss/solis}.)

The H$\alpha$ core (655.93-657.00 nm) and wing (652.30-654.24 nm and 659.98-662.96 nm) from SCIAMACHY irradiance data each show a consistent increase throughout the data record that is suggestive of uncorrected calibration drift, possibly due to stray light, since it is greater in the line-core time series. The index time series shows quasi-annual oscillations and multiple step changes as well as occasional short-term periodicity that does not correspond to rotational modulation regardless of the phase of solar cycle activity. GOME-2A irradiances (averages in the line wings, 652.37-654.34 nm and 660.06-662.96 nm,  and the line core,  655.90-656.95 nm) are affected by similar trends that likely arise from the yearly and long-term changes of the instrument spectral-response function \citep{munro2016}. We conclude that neither data set provides a useful indication of the H$\alpha$ behavior at temporal scales longer than a few solar rotations. Nevertheless, these space-based, practically uninterrupted, long-term, high S/N observations offer a unique opportunity to unambiguously resolve and follow the rotationally induced variability patterns that elude detection in the available ground-based, high-resolution data.  For this purpose we construct a composite record by producing the individual daily GOME-2A and SCIAMACHY line indices, detrending them with a 61-day running-mean filter, smoothing each detrended record with a 3-day running mean, then averaging the detrended and smoothed data.

\subsection{OSIRIS}
There are also long-term data sets of the radiance available from instruments that view the Earth’s limb, which provide vertically resolved radiance spectra throughout the dayside of each orbit. One source of such measurements is the OSIRIS (Optical Spectrograph and Infrared Imaging System) instrument \citep{Llewellyn2004, McLinden2012}, which was launched in 2001 on the Odin satellite. OSIRIS measures radiances between 280-800 nm with  approximately 0.8 nm spectral resolution, and scans the atmosphere vertically from 10 km up to at least 70 km (sometimes higher). OSIRIS radiance data are available in a L1B product through specific arrangement with the instrument team at the University of Saskatchewan. For this analysis, we average individual radiance spectra over two broad altitude bands (15-35 km and 35-55 km) to evaluate potential altitude dependence in these results. Creation of a core-to-wing ratio for a solar-activity index uses the same approach applied to direct solar measurements. Initial review of H$\alpha$ solar-activity index values derived from OSIRIS measurements over individual orbits showed a small but persistent variation during each day. Because of this behavior, we have chosen to use only dates with a complete set of OSIRIS orbits (typically 15 in a single day) for our analysis. This choice significantly reduces the number of individual dates available, because OSIRIS shared spacecraft operations with the SMR (Submillimeter Radiometer) astronomy instrument on Odin during the early part of its mission (2002-2007), and has experienced spacecraft power issues that limit measurements since 2012. We also see evidence of a small seasonal variation during each year, possibly due to the semi-annual shift in the primary OSIRIS measurement location from the northern to southern hemisphere. Therefore, we group the OSIRIS H$\alpha$ activity indices into annual averages, based on approximately 10-15 days roughly evenly spread over each calendar year.

\subsection{ISS/SOLIS}
\label{sec:iss/solis}
 In order to investigate the long-term (decadal) H$\alpha$ variability, we supplemented the OSIRIS data with high-resolution spectra from the Integrated Sunlight Spectrometer (ISS) at SOLIS. This ground-based spectrograph acquires daily disk-integrated observations of the Sun in ten spectral bands in the optical region. In this study, we analyze observations acquired in the H$\alpha$ spectral range (656.28$\pm$0.825) at a spectral resolution of approximately 0.0025 nm. Details about SOLIS/ISS operations and data reduction can be found in \citet{bertello2011} and \citet{pevtsov2014}. From daily spectra, we computed the H$\alpha$ core-to-wing ratio. The core intensity was estimated by a polynomial fit of the spectral region around the intensity minimum of the line, and the continuum intensity was estimated as the average intensity at 655.935 nm. This region was carefully selected to avoid contamination from other solar and telluric lines abundant at this wavelength range. The ISS acquired data in the H$\alpha$ range between 2007 and 2017. However, here we consider data acquired between 2008 and 2014.6, since the earlier observations were affected by calibration issues that produced large fluctuations of the estimated index, and data acquired afterwards lead to a large discontinuity related to relocation of the instrument from Kitt Peak to Tucson. We also found that ISS observations are affected by daily variations induced by variable observing conditions. These problems %, coupled to the insufficiently high S/N of the individual ISS line-index measurements, 
 prevent us from studying the rotational variability patterns, thus echoing the conclusions of Maldonado et al. (2019), who were able to see the rotational signal in the frequency spectra, but did not resolve the rotational modulations in the index series.  We further tested the ISS indices by detrending them with 61-day running means and correlating with the same-day detrended TSI measurements during the epoch of prominent TSI changes induced by the passage of large sunspot groups from July 2011 to August 2012. This resulted in negligible TSI-ISS correlation, $r=0.03$ (n=237 observations), while the same-epoch correlation between the GOME-SCIAMACHY indices and TSI gave highly significant $r=-0.58$ (n=392), thus closely matching $r\le -0.6$ from Paper I. Hence we employ the ISS data to study the variability of the H$\alpha$ line exclusively on the decadal temporal scales.
 
\subsection{Supplementary data sets}
The rotational scale variability of the H$\alpha$ index is compared in this study with the variability of the "Balmer index" derived from OMI measurements of the upper Balmer lines H$\beta$, H$\gamma$, and H$\delta$. As described in Paper~I, this index is defined as the average of the core-to-wing ratio indices of the three lines, computed after detrending the observed spectra with a 61-day running mean.  The OMI data clearly resolved and followed the solar-rotational variability patterns, but stopped short of unraveling the long-term trends due to the gradual increase of instrumental noise. Additionally, OMI does not cover the H$\alpha$ line, the subject of this study.
%Unfortunately, OMI measurements turned out to be unsuitable to investigate long term variability of the upper Balmer lines, OMI and TROPOMI measurements being affected by long term, instrument-induced variations similar to those affecting GOME and SCHIAMACHY described above.  
%Preliminary analysis of the OMI data from 2007-2013 pointed to different (compared to the Mg II – like lines) behavior of the Balmer lines on the solar-cycle timescales \citep{marchenko2014}. More refined approach (Paper I) considered the OMI data acquired over full solar cycle, however stopping short of unraveling the long-term trends due to the gradual increase of the instrument noise. On the other hand, the regular TROPOMI observations commenced in 2018, thus not providing the necessary long-term baseline (Paper I). Moreover, both OMI and TROPOMI do not cover the H$\alpha$ spectral range.
We were also not able to find long-term, Sun-as-a-star observations of other Balmer lines from ground-based observations at times overlapping with OMI and the Precision Solar Photometric Telescope \citep[PSPT,][]{rast1999} observations employed for the irradiance reconstruction model (see Sec.~\ref{sec:model}). For this reason, the analysis over the decadal temporal scale is limited to measurements of the H$\alpha$ only. 
The variability of Balmer lines was compared with other activity indices. In particular, we employed: the daily TSI measurements from the Total Irradiance Monitor \citep[TIM,][]{{kopp2005}, {kopp2021b}} aboard the SORCE mission \citep{rottman2005}\footnote{available at:https://lasp.colorado.edu/lisird/}; the \ion{Ca}{2} K Plage Index provided by the San Fernando Observatory \citep{walton1999};  the Dark Photometric Index (DPI) derived from UASF-SOON observations \citep{coddington2016}; the solar filaments and prominence areas, derived from Meudon spectroheliograms\footnote{available http://voparis-helio.obspm.fr/hfc-gui/} \citep{fuller2005}; the latter were also derived from from the LSO/KSO international H$\alpha$ prominence catalogue\footnote{https://www.astro.sk/~choc/open/lso\_kso\_h\_alpha\_promimence\_catalogue/lso\_kso\_h\_alpha\_promimence\_catalogue.html} \citep{rybak2011}; the \ion{Mg}{2} index derived from OMI measurements \footnote{available at:https://lasp.colorado.edu/lisird/} for the solar-rotational timescales; and the Bremen \ion{Mg}{2} index composite \citep{snow2014} \footnote{available at https://www.iup.uni-bremen.de/gome/gomemgii.html} to investigate the variability at the decadal timescale.

\section{Model description}
\label{sec:model}
%[Fontenla 99, page 491, discussion of formation height of Ha wrt other chrompsheric lines. States scarce sensitivity to network and plage of Ha. This is corroborated by IBIS observations by Cauzzi et l. 2009 (especially) and Molnar et al. 2019]

The cores of Balmer lines probe the upper layers of the solar photosphere and low-to-middle chromosphere \citep[e.g.][]{avrett2008}. Therefore, the finding in Paper~I that their variability more closely resembles that of the TSI rather than of chromospheric indices, and, similarly, the anti-correlation with the \ion{Ca}{2} index found by \citet{maldonado2019}, are somewhat unexpected.  In order to understand the processes driving the observed variability, we compare  observational results with variability obtained from an irradiance reconstruction model based on a semi-empirical approach.
In such approaches \citep[e.g.][]{penza2006,ermolli2011,fontenla2011,haberreiter2014,criscuoli2018}, the irradiance is reconstructed by combining the daily variations of area coverage of quiet and magnetic regions with synthetic spectra representing the radiative emission of the various features. Synthetic spectra are typically derived by using one-dimensional static atmospheric models, although models based on the use of 3D radiative magneto-hydrodynamic simulations of the solar atmosphere have been recently developed \citep{yeo2017b, criscuoli2020a}.  We refer the interested reader to the recent reviews on solar irradiance modeling by \citet{petrie2021} and \citet{kopp2021}.

In order to estimate the area coverage of magnetic and quiet features, we employed full-disk observations acquired between 2005 and 2015 with the PSPT in the red (607$\pm$0.45 nm) and \ion{Ca}{2} K filters (393.4$\pm$0.27 nm). Specifically, eight different magnetic and quiet areas (dark center cell, quiet Sun,  quiet network, active network, plage, hot plage, sunspot umbra, and sunspot penumbra) and their positions on the disk were identified using the Solar Radiation Physical Modeling \citep[SRPM][]{harder2005, fontenla2011}, which makes use of $\mu$-dependent (where $\mu$ is the cosine of the heliocentric angle) intensity thresholds applied on red images for identification of sunspot umbrae and penumbrae and of \ion{Ca}{2} K images for identification of the remaining features. The spectral synthesis was performed in Non Local Thermodynamic Equilibrium (NLTE) at 21 heliocentric positions with the Rybicki and Hummer code \citep[RH, ][]{Uitenbroek2001, kowa2017,kowa2022}.  This radiative transfer code allows synthesis in LTE and NLTE in different types of geometries and thus is one of the most widely used in solar physics. Recently, the sets of opacity in RH have been updated for solar-irradiance reconstruction applications \citep[e.g.][]{criscuoli2018, criscuoli2019, berrilli2020, criscuoli2020b}. Balmer lines were synthesized at variable spectral resolution, denser in the line cores (approximately 0.1~pm) and coarser in the wings (approximately 1~pm), using the 20-level hydrogen model atom described in \citet{kowa2017}. 

\subsection{Two sets of atmosphere models}
\begin{figure*}
   \centering

  \includegraphics[width=5.3cm,trim={3.6cm 0 0 0}]{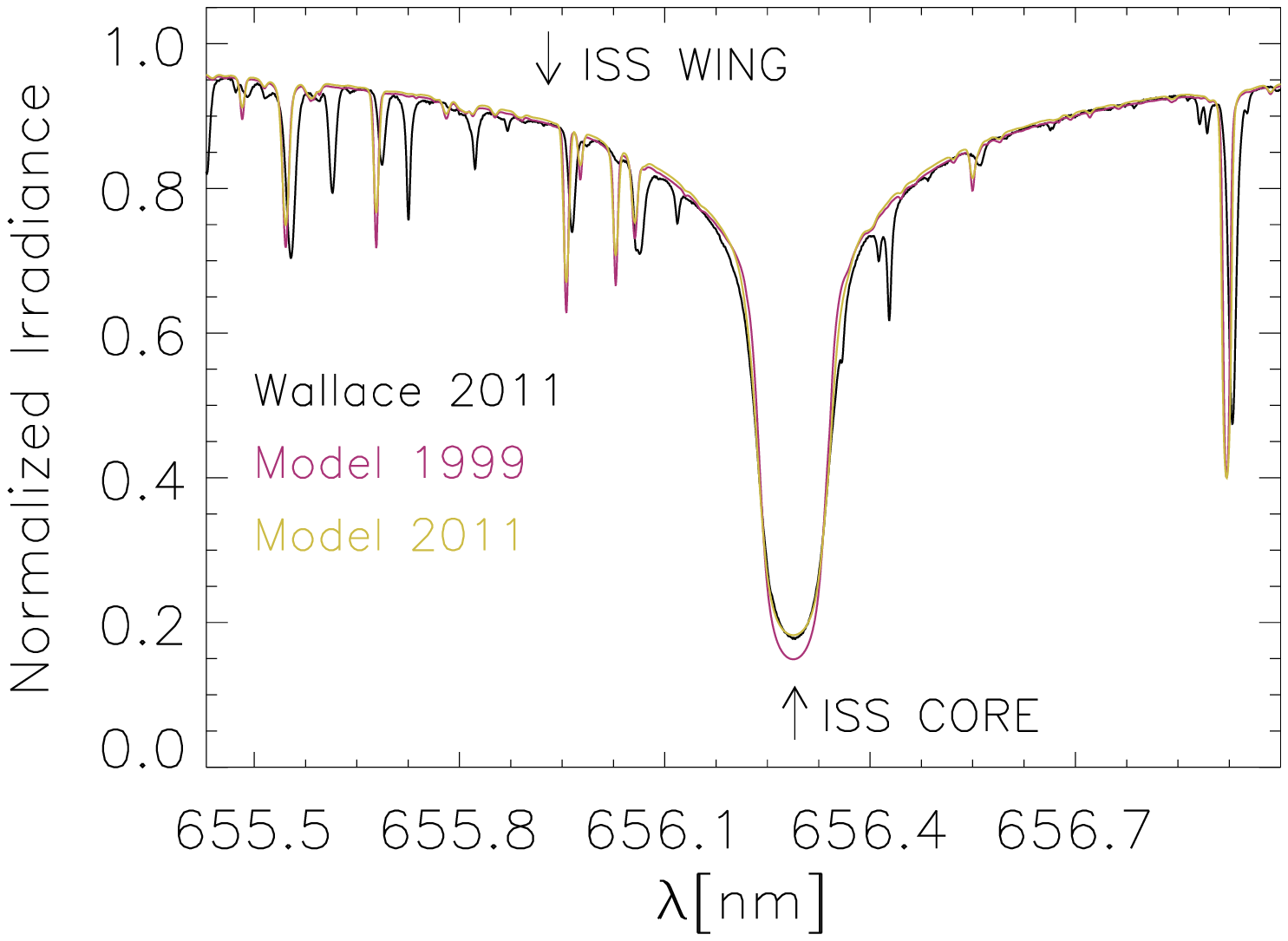}
  \includegraphics[width=5.8cm,trim={2cm 0 0 0}]{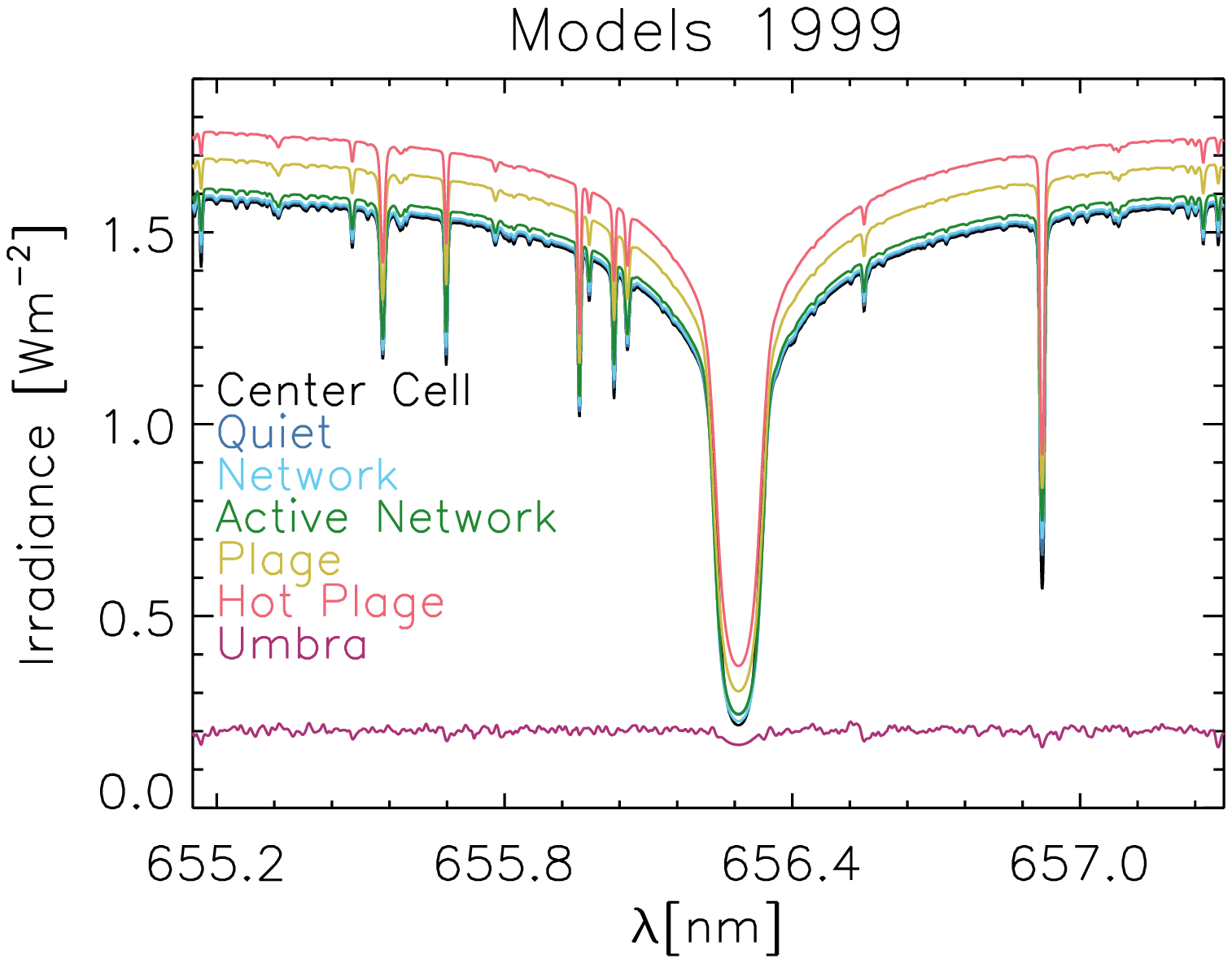}
  \includegraphics[width=5.8cm, trim={2cm 0 0 0}]{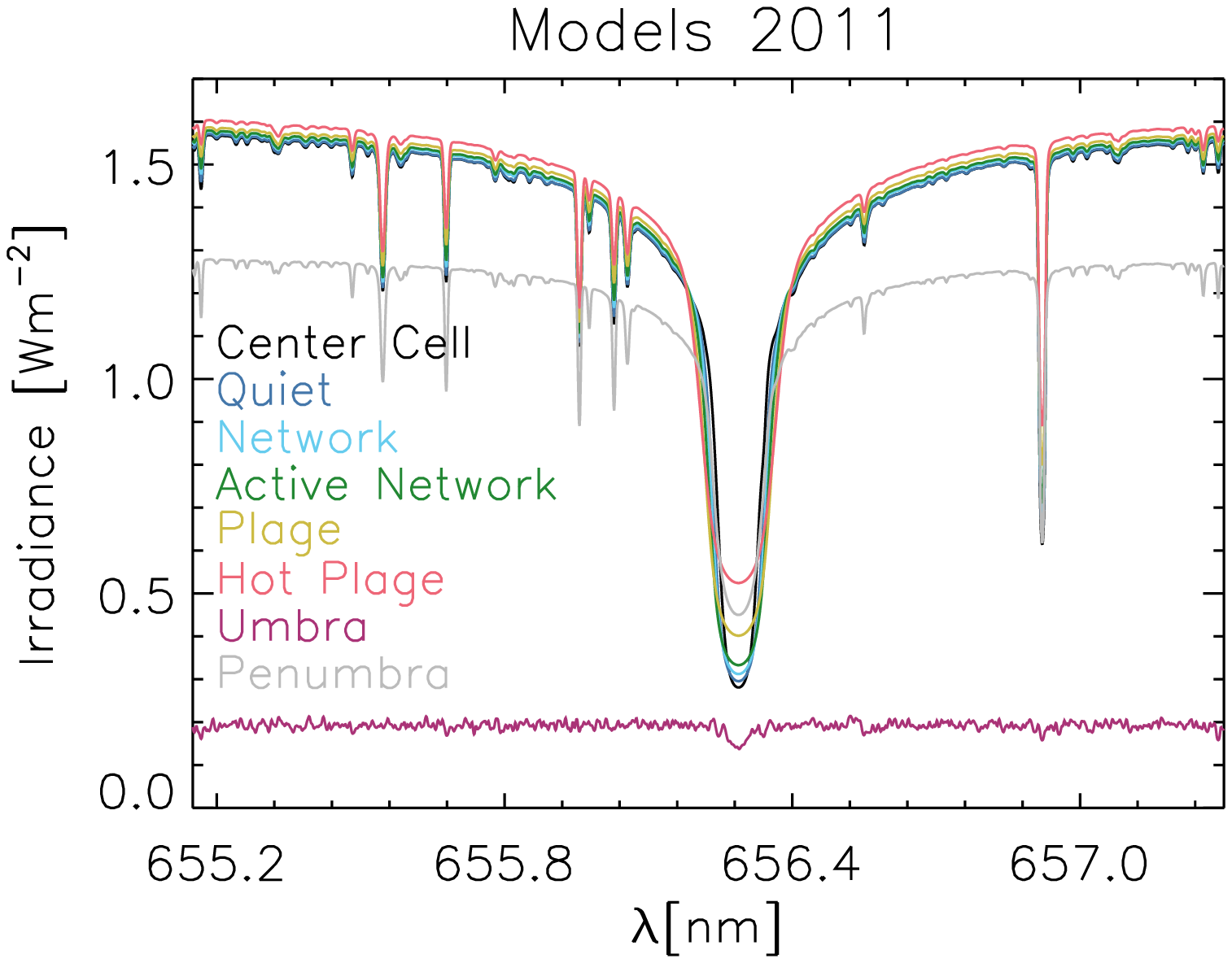}

   \includegraphics[width=5.3cm,trim={3.6cm 0 0 0}]{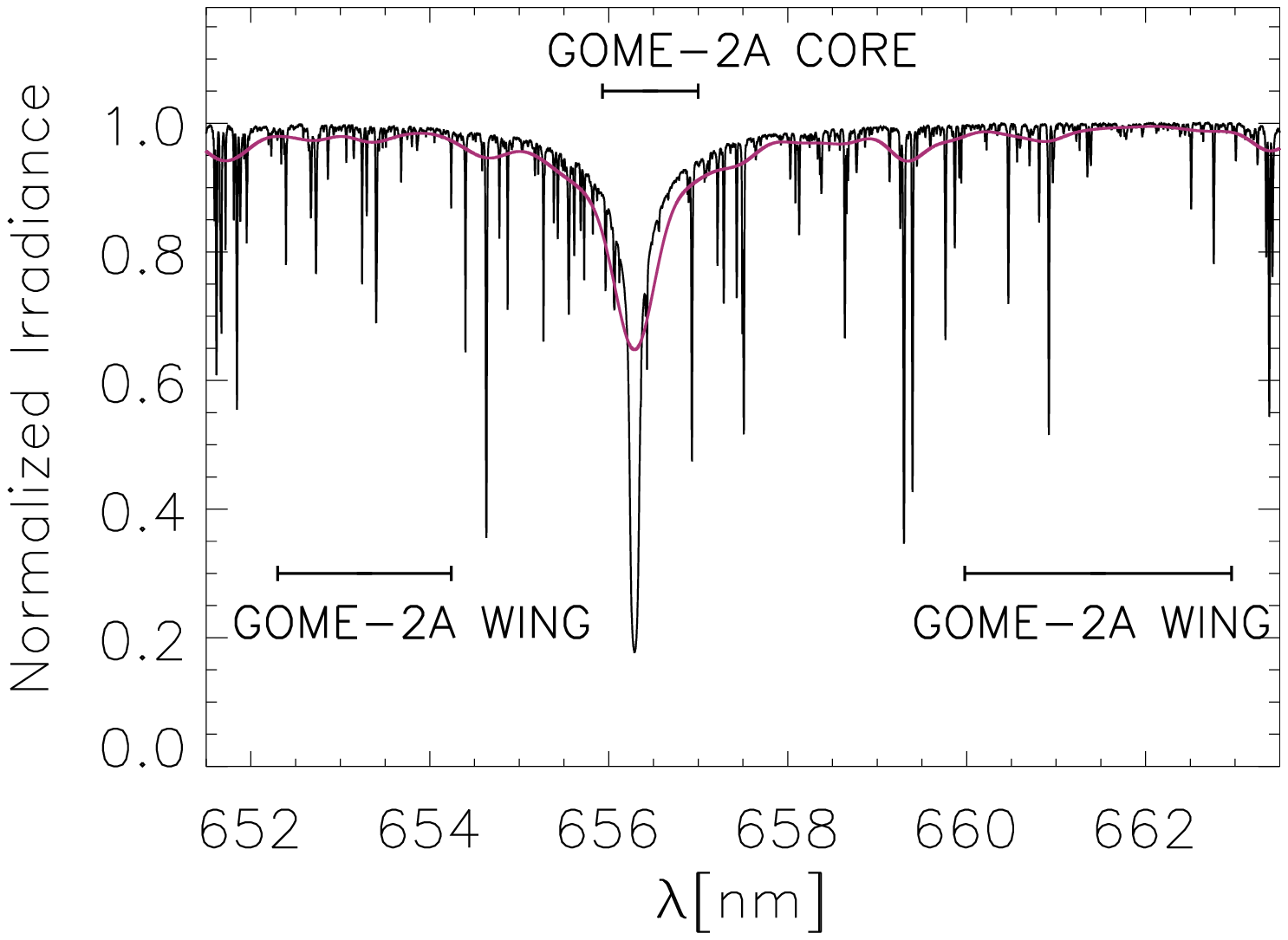}
  \includegraphics[width=6cm,trim={0.7cm 0 0 0}]{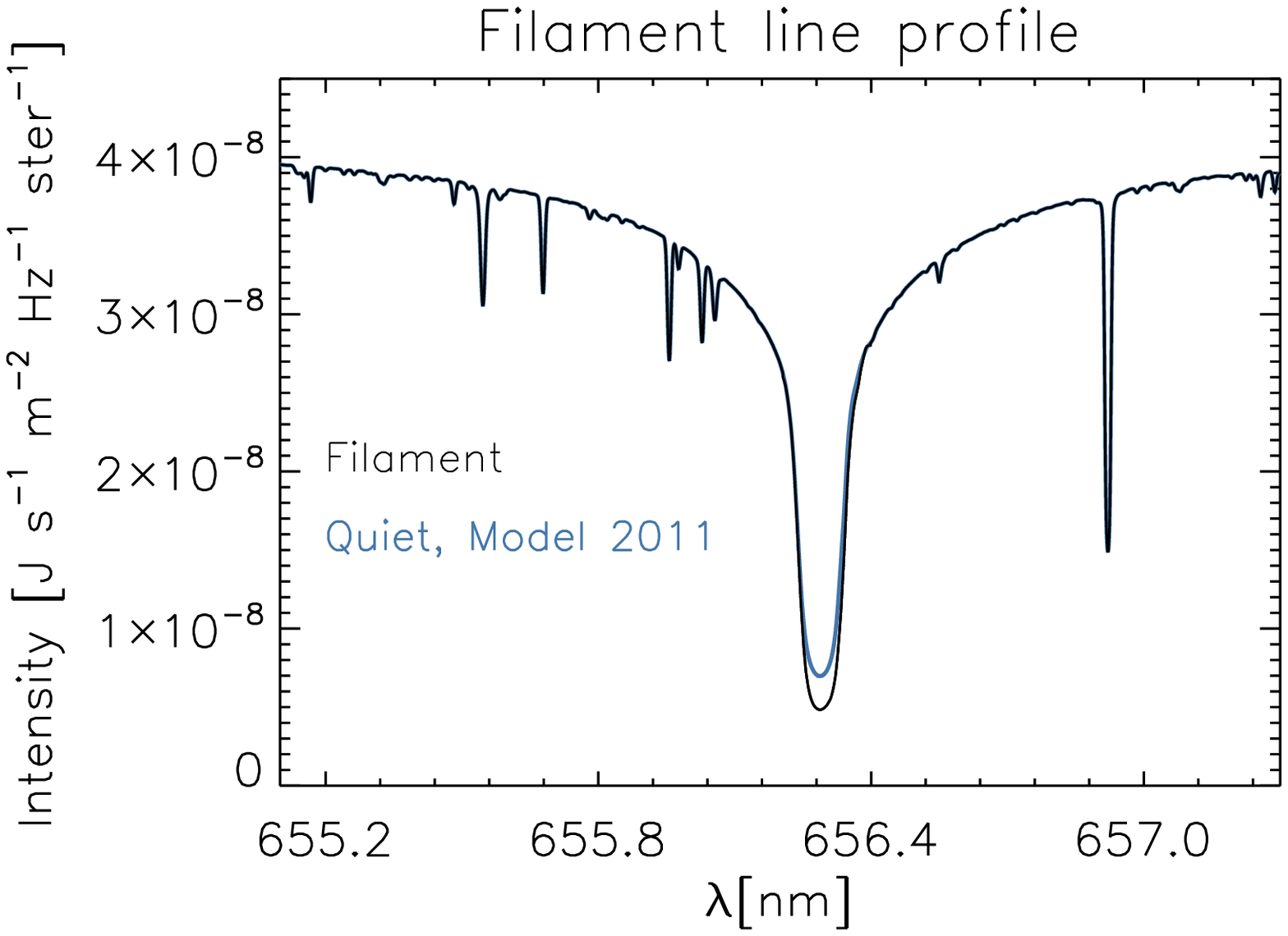}
  \includegraphics[width=6cm, trim={0.7cm 0 0 0}]{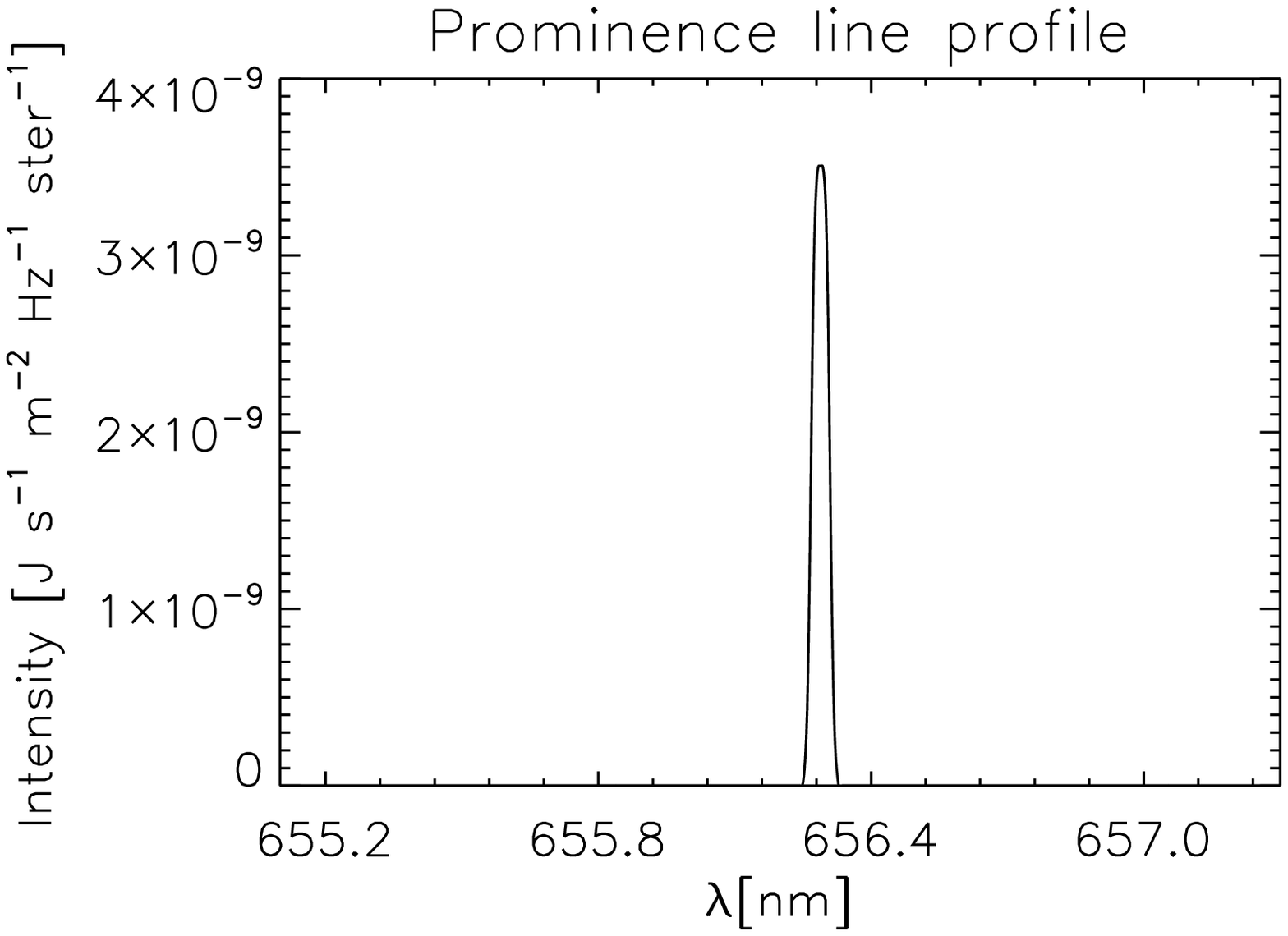}
   \caption{Top Left: Comparison of H$\alpha$ profiles (normalized to nearby continuum) obtained using the quiet model from the set of FAL1999 and FAL2011 and the observed reference flux \citep{wallace2011}.  Top center and right: H$\alpha$ irradiance line profiles obtained using the two sets of FAL atmospheric models. Bottom Left: Wallace reference spectrum at the original resolution (black) and degraded to the spectral resolution of GOME-2A (red). The spectral ranges used to define the core and wings are also shown. Bottom center and right: filament and prominence line profiles employed in the models described in Sec.~\ref{sec:filaments}. The bottom center panel shows for comparison also the disk center quiet sun profile from Models 2011.
              \label{Fig_comp_synt_atlas}}%
    \end{figure*}

 Because the use of specific sets of atmospheric models is a source of uncertainties in irradiance reconstructions \citep[see, for instance,][]{ermolli2010, uitenbroek2011, schmit2015}, we compared synthetic Balmer-line variability obtained with two sets of semi-empirical atmospheric models, i.e., the set by \citet{fontenla1999} (FAL1999, hereafter) and the set by \citet{fontenla2011} (FAL2011, hereafter), both of which have been previously employed for solar irradiance reconstructions \citep[e.g.,][]{ermolli2013, egorova2018, criscuoli2018}. 
 Syntheses of the H$\alpha$ line obtained with the two sets of models are shown in Fig.~\ref{Fig_comp_synt_atlas}, together with a comparison of the quiet-Sun models syntheses with the observed reference spectrum by \citet{wallace2011}. Similar plots for the H-$\beta$, -$\gamma$, and -$\delta$ lines are shown in Figs.~\ref{models_vs_wallace} and ~\ref{models_beta_gamma_delta} (Appendix). 

The two sets of models present some similarities but also some clear differences: In FAL2011, the network models consistently produce shallower profiles than the quiet models, while in FAL1999, the network models present similar, if not slightly deeper, profiles than the quiet ones; overall, FAL2011 produces shallower profiles than FAL1999, with the exception of the sunspot models, which are similar for the two sets; and finally, the core widths of FAL2011 models increase with the amount of activity, while in FAL1999 the core width remains substantially invariant or slightly decreases. Unfortunately, observations of Balmer lines in different types of quiet and active structures that would allow validating the adopted atmospheric models are rather scarce. However, we note that the FAL1999 quiet-Sun model systematically produces profiles that are slightly deeper than the observed reference spectrum, while the quiet-Sun model of the FAL2011 set reproduces the observed profiles very well. The FAL2011 also qualitatively reproduces the increase of the H$\alpha$ core width with activity reported by \citet{cauzzi2009}, \citet{molnar2019}, and \citet{cauley2017}. On the other hand, \citet{molnar2019}  find that the plage's $H\alpha$ core intensity is only slightly elevated compared to the 'quiet' Sun (their Fig. 2), in contrast to the approximately 30\% shallower FAL2011 profiles, but more in line with FAL1999. Similarly, the contrast of the plage profile in H$\alpha$ reported in \citet{cauley2017} has values similar to those of the plage model of FAL1999 in both the line core and far wings. Finally, \citet{hao2020} observed no significant variation of the H$\alpha$ line-core width and intensity in small bright points with respect to the nearby quiet regions, which are, even in this case, characteristics better reproduced by the FAL1999. 
Therefore, considering the controversial estimates from the recent observations, here we compare outputs from both sets of models.

Note that the set of FAL1999 atmospheres does not include a penumbral model, so, for the FAL1999 reconstructions, we employed the corresponding model from the FAL2011 set. In the following, the reconstructions obtained using the FAL1999 and the FAL2011 are referred to as Model 1999 and Model 2011, respectively.

%%%%%%%%%%%%%%%%%%%%%%%%%%%%%%%%%%%%%%%%%%%%%%%%%%%%
%%%%%%%%%%%%%%%%%%%%%%%%%%%%%%%%%%%%%%%%%%%%%%%%%%%%

The daily high-resolution model spectra were convolved with Gaussian functions of appropriately chosen full-width at half-maxima, thus conforming to the relatively low spectral resolution of each instrument, and the indices were derived as described in Sec.~\ref{sec:measurements}. Table \ref{tab_core_wings_models} shows the irradiance values in bands of the H$\alpha$-line core and wings (also shown in Fig.~\ref{Fig_comp_synt_atlas}), as defined in SCIAMACHY and GOME-2 observations, obtained from the different atmospheric models.

%%%%%%%%%%%%%%%%%%%%%%%%%%%%%%%%%%%%%%%%%%%%%%%%%%%%%%%%%%%%%%%%%%%%%%%
    \begin{table}[h]

       \begin{tabular}{p{0.22\linewidth}cccc}
            \hline
            \noalign{\smallskip}
         Model    &  \multicolumn{3}{c}{IRRADIANCE [$Wm^{-2}$]}  \\
         \hline
              & Left Wing  &   Right Wing  &  Core  \\
                      &[652.3-654.24] & [659.98-662.96] & [655.93-657.00] &\\
           \noalign{\smallskip}
            \hline
            \noalign{\smallskip}
            
            &  \multicolumn{3}{c}{FAL1999}  \\

          Center Cell.  &  1.614    & 1.598  &    1.323   \\
          Quiet  & 1.621     & 1.605  &   1.326    \\
          Network  &  1.626    & 1.610  &    1.330   \\
          Act. Network& 1.644     & 1.627  &   1.346     \\
          Plage  &  1.721    & 1.701  &    1.424   \\
          Hot Plage  & 1.788     & 1.766  &   1.499     \\
          Umbra  &  0.202    & 0.205  &    0.199   \\
                  \noalign{\smallskip}
            \hline
&\multicolumn{3}{c}{FAL2011} \\
          Center Cell.  &  1.591    & 1.575  &    1.315   \\
          Quiet  & 1.596     & 1.581  &   1.309    \\
          Network  &  1.602    & 1.587  &    1.313   \\
          Act. Network& 1.604      &  1.588  &  1.314       \\
          Plage  & 1.611      & 1.594   &  1.324     \\
          Hot Plage  &  1.627    & 1.609   &  1.350       \\
          Umbra  &   0.1905    &  0.197  &   0.190     \\
          Penumbra  &  1.283     &  1.270  &  1.130      \\
          \noalign{\smallskip}
            \hline
       \end{tabular}
   % $$  
        
     \caption {Irradiance values in the core and wings of the H$\alpha$ line obtained for the different atmosphere models degraded to the spectral resolution of SCIAMACHY and GOME-2.} 
       
    \label{tab_core_wings_models}       
  \end{table}

\subsection{Statistical testing: observations vs. models}

Comparing the observed and modelled indices, we apply various statistical criteria. All metric values are reported in Sec.~\ref{sec:App_metrics} of the Appendix.
The Pearson correlation coefficients between the models and measurements and their significance were computed using IDL (Interactive Data Language){\footnote {IDL® is registered trademark of L3Harris Technologies, Inc.}} routines (namely \textit{correlate.pro} and \textit{pcorre.pro}). Results are reported in Table~\ref{table_metrics} for the solar-rotational timescale and in Table~\ref{tab:metrics_decadal} for solar-cycle timescales. Note that in all cases we found a probability close to zero that the  correlations are generated by chance.  

We also employed the Mean Absolute Percentage Error (MAPE) defined as:
\begin{equation}
MAPE=\frac{100}{N}\sum_{n}^N \frac{|I_{n}^{model}-I_{n}^{obs.}|}{I_{n}^{obs.}}
\end{equation}

Results obtained over the solar-rotational timescales (i.e., on the 61-day detrended time series) are shown in Table~\ref{table_metrics}. Another metric often employed to evaluate the observation-model agreements  is the standard deviation computed over the detrended data. Standard deviation values computed over the period 2005-2015 are reported in Table~\ref{table_stddev}.  

\subsection{Model sensitivity to the sunspot representation}
As explained above, the two sets of models employed in this study present different quiet and network/plage models, while the sunspot models are rather similar. In this respect, we note that \citet{loukitcheva2017} showed that the sunspot model from the \citet{fontenla2009} set (which is similar to the sunspot models used in our study) overestimates the sunspot brightness temperature measured with ALMA in the micrometer spectral range. In order to test the sensitivity of our results to uncertainties in the sunspot models, we reconstructed the Balmer-line variability making use of two other sunspot models published in the literature, namely a Kurucz model at 4500~K\footnote{specifically, we used a model with effective temperature of 4500~K, null velocity, and $\log g$ = 4.5.} and the model by \citet{avrett2015}, while keeping the rest of the model parameters intact. 
The Kurucz model at 4500~K was chosen because it better reproduces the observed average effective temperature of sunspots, as discussed in \citet{tagirov2019}. The Kurucz model at 4500 K is also employed in SATIRE reconstructions to successfully reproduce both TSI and SSI variability \citep[e.g.][]{krivova2003}.
Kurucz atmospheres have been also employed to model the contribution of starspots to stellar magnetic variability in Balmer lines \citep{rackham2019}.  The sunspot model by \citet{avrett2015} was utilized because, 
according to \citet{loukitcheva2017}, it underestimates the observed ALMA brightness temperature at spectral regions forming in the high chromosphere, so that the \citet{fontenla1999} and \citet{avrett2015} sunspot models likely represent  "upper" and "lower" boundaries of the temperature stratification of a "real" sunspot. The synthetic spectra obtained for the different sunspot models are shown in Fig.~\ref{sunspot_models_flux} of the Appendix. 
Finally, to test the sensitivity of our results to the choice of the penumbral model, we recomputed the reconstructions substituting the sunspot penumbral model with synthetic spectra obtained with the model published in \citet{ding1989}. In all cases, the base set of models considered is the FAL1999, for which we expect the contribution of sunspots to be larger than for the FAL2011.
Figure~\ref{Fig_Ha_v2_v3_v4} in the Appendix compares the variability of the measured indices on the solar-rotational timescale with those obtained with the different atmospheric models described above.  Not surprisingly, larger variations are found at the rotations strongly dominated by sunspots, as, for instance, the one occurring at 2005.06 (first peak in left panels of  Fig.~\ref{Fig_Ha_v2_v3_v4}), for which the reconstruction obtained using the \citet{avrett2015} model seems to better reproduce the observations. Overall, the plots show that the reconstructions are only marginally sensitive to the choice of the employed sunspot and penumbra model.  Indeed, we found that both the MAPE and the standard deviations of the new reconstructions are largely the same as the one found for the original Model 1999, thus indicating that the discrepancies between the modelled and measured variability should be mostly ascribed to uncertainties in the modeling of plage and network, and that uncertainties in the sunspot atmosphere model play a secondary role.

\section{Results}
\label{sec:results}
  The variability of the observed and modelled Balmer indices was investigated on solar-rotational timescales (weeks/months) as well as on a decadal (solar cycle) timescale. To focus on the rotational modulation, both observed and synthetic data were detrended with a 61-day running mean.

 %%%%%%%%%%%%%%%%%%%%%%%%%%%%%%%%%%%%%%%%%%%%%%%%%%%%
\subsection{Variability on Solar-Rotational Timescales}
\label{sec:results1}
The analysis of the H$\alpha$ index derived from the SCIAMACHY and GOME-2 composite confirms results obtained in Paper~I that the variability of Balmer lines closely follows the inverted TSI.  

 %%%%%%%%%%%%%%%%%%%%%%%%%%%%%%%%%%%%%%%%%%%%%%%%%%%%
   \begin{figure*}
   \centering
  \includegraphics[width=14cm]{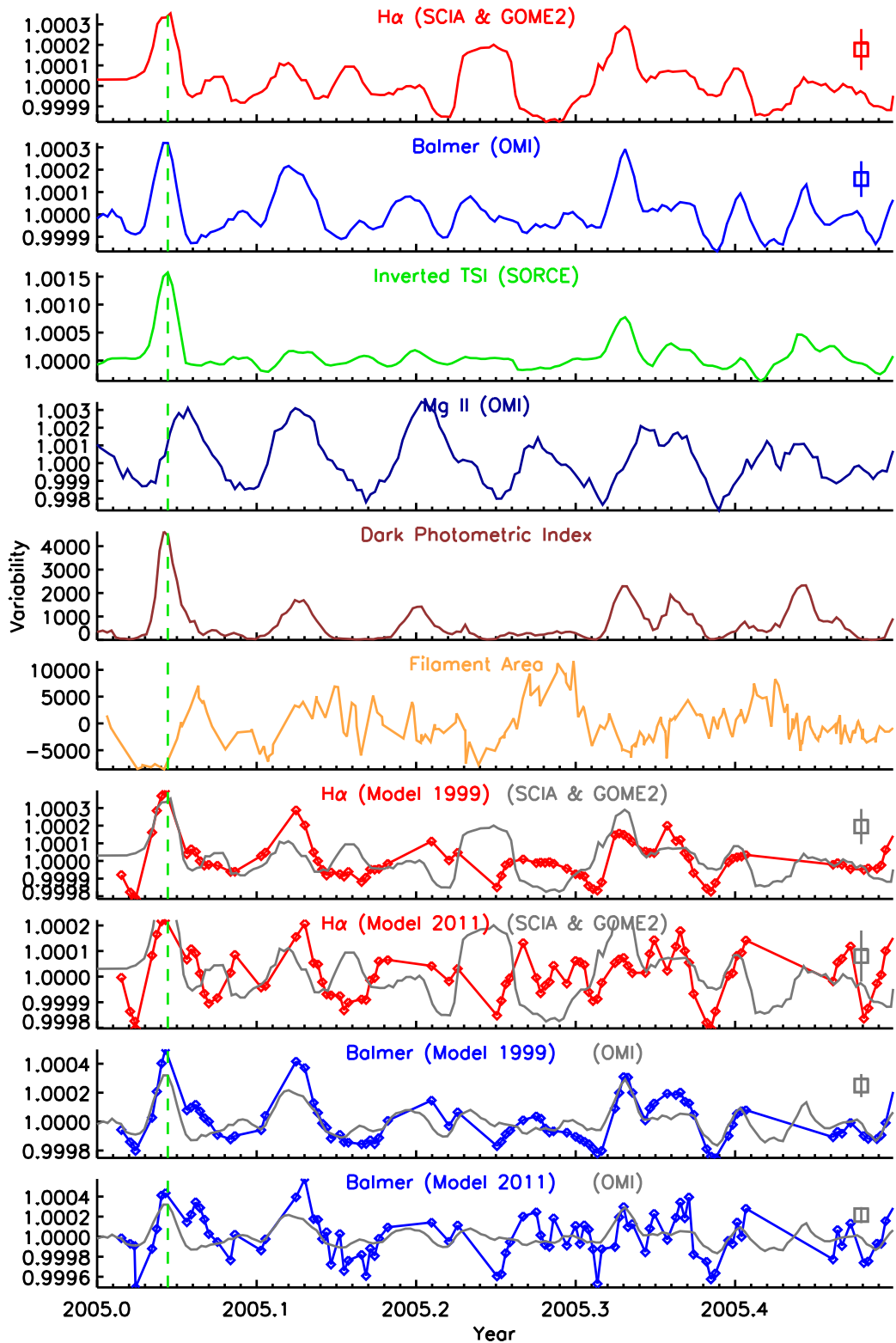}

   \caption{Comparison of variability of the H$\alpha$ index and Balmer index  with the variability of other activity indices during the descending phase of Cycle 23. All data were detrended with a 61-day running mean. The top two panels show the indices obtained from measurements, which are compared to the models in the four bottom panels. The vertical, green, dashed lines highlight a rotation in which models show very good agreement with the observations. ±1-sigma error bars are shown for OMI, SCIAMACHY, and GOME2 data.}            \label{Fig_rotscale_2005}%
    \end{figure*}
 %%%%%%%%%%%%%%%%%%%%%%%%%%%%%%%%%%%%%%%%%%%%%%%%%%%%
 
 %%%%%%%%%%%%%%%%%%%%%%%%%%%%%%%%%%%%%%%%%%%%%%%%%%%%
   \begin{figure*}
   \centering
  \includegraphics[width=14cm]{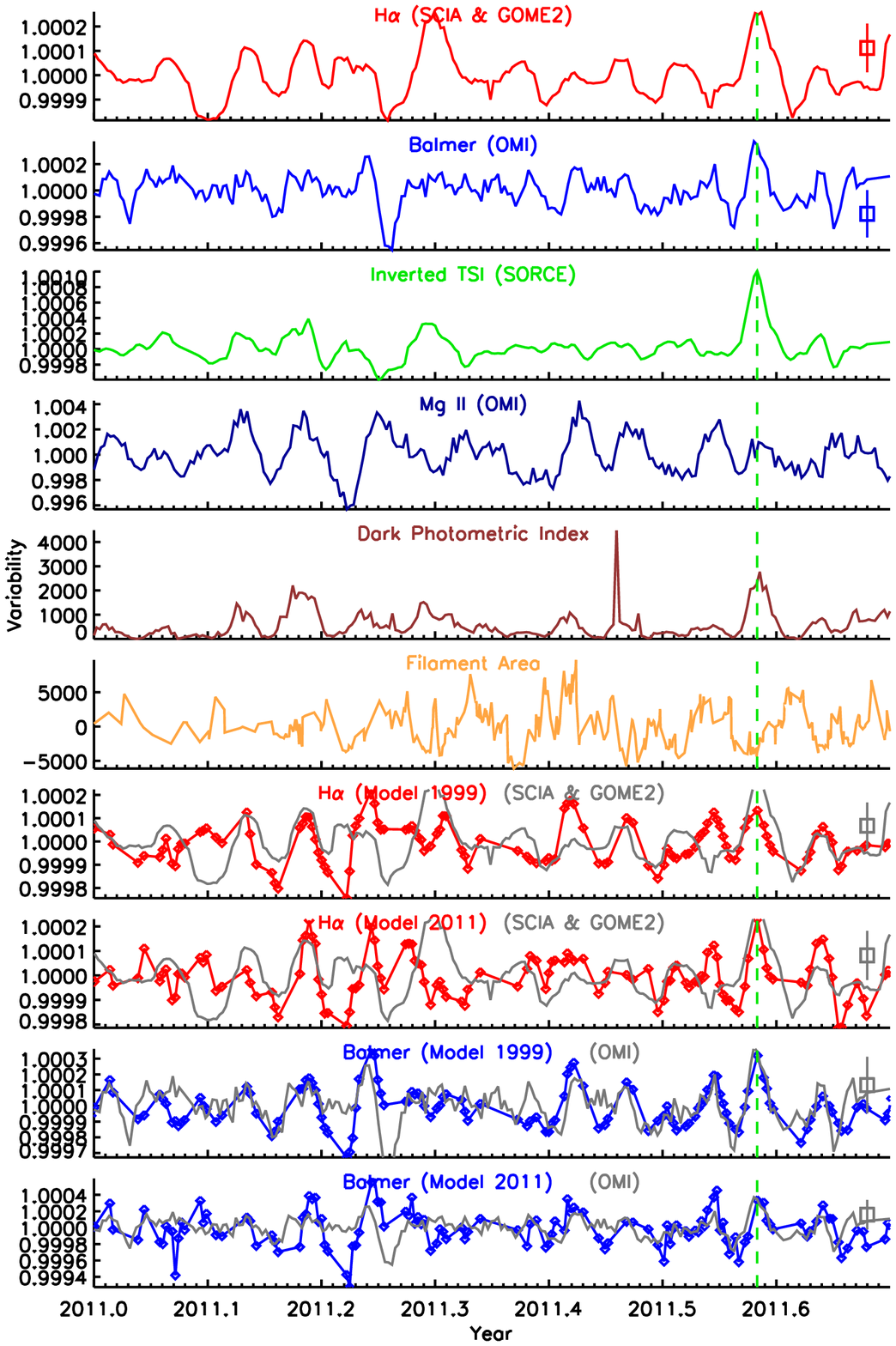}

   \caption{As in Fig.~\ref{Fig_rotscale_2005} but for the ascending phase of Cycle 24.}            \label{Fig_rotscale_2011}%
    \end{figure*}
 %%%%%%%%%%%%%%%%%%%%%%%%%%%%%%%%%%%%%%%%%%%%%%%%%%%%

Figures~\ref{Fig_rotscale_2005} and \ref{Fig_rotscale_2011} compare the variability of the H$\alpha$ index  (derived from SCIAMACHY and GOME-2) and of the Balmer index with the variability of the \ion{Mg}{2} index, the inverted TSI, the filament area, and the DPI over a few solar rotations observed in 2005 and 2011, respectively. The 2005  epoch is characterized by the presence of relatively large sunspot groups and pronounced rotational modulations caused by the passage of plage/network fields. Note, however, that maxima in the former are frequently displaced both in amplitude and time from the periodic modulation of the latter, thus accentuating the predominantly sunspot-driven variability of the line indices.
We note that during the first three rotations of these 2005 data (Fig.~\ref{Fig_rotscale_2005}), the amplitude of the peaks of the \ion{Mg}{2} index, which is a chromospheric index,  are fairly constant while the indices derived from the Balmer lines show decreasing peak amplitudes, in agreement with the inverted TSI and the DPI, which are photospheric indices. A stronger correlation of the Balmer and H$\alpha$ indices with the DPI and the inverted TSI is also suggested by visual inspection of the 2011 observations (Fig.~\ref{Fig_rotscale_2011}). 
%%%%%%%%%%%%%%%%%%%%%%%%%%%%%%%%%%%%%%%%%%%%%%%%%%%%  
\begin{table}[h]

\begin{center}

\begin{tabular}{ |l|c|c|c| } 
 \hline
     & \ion{Mg}{2} & inv. TSI & DPI\\
     \hline
 Balmer & 0.25 & 0.55 & 0.57 \\ 
 Model 1999 - Balmer & 0.62 & 0.34 & 0.66 \\ 
 Model 2011 - Balmer & 0.56 & 0.19 & 0.52 \\ 
 H$\alpha$ & 0.2 & 0.52 & 0.58 \\ 
 Model 1999 - H$\alpha$ & 0.57 & 0.29 & 0.60 \\ 
 %Model 2011 - H$\alpha$ & 0.45 & 0.14 & 0.41 \\
 Model 2011 - H$\alpha$ & 0.45 & 0.27 & 0.49 \\
 \hline
\end{tabular}
\caption{Rotational timescales: Pearson correlation coefficients between the Balmer and H$\alpha$ indices and other activity indices computed in the period 2005-2015. Indices have been detrended with a 61-day average. \label{table1}}

\end{center}
\end{table}
%%%%%%%%%%%%%%%%%%%%%%%%%%%%%%%%%%%%%%%%%%%%%%%%%%%%  
The observed Balmer indices (both Balmer and H$\alpha$: Table~\ref{table1}) show higher correlations with the two photospheric indices (inverted TSI and DPI) than with the \ion{Mg}{2} index, thus confirming the conclusions inferred from visual inspection of Figs.~\ref{Fig_rotscale_2005} and \ref{Fig_rotscale_2011}. The DPI describes the sunspot contribution to the TSI variability and is directly related to the sunspot area coverage. The relatively high correlation with this index in particular corroborates the conclusion drawn in \citet{marchenko2021} that on rotational timescales, the variability of Balmer lines is predominantly modulated by sunspots.

\begin{figure}
   \centering
  \includegraphics[width=10cm]{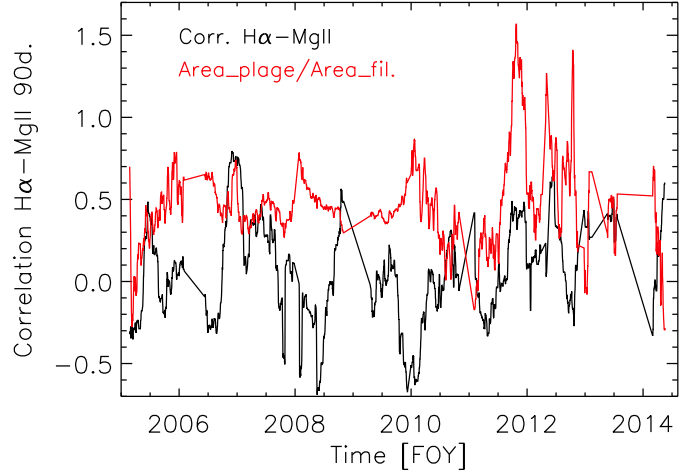}
   \caption{Black line: Correlation coefficient between the H$\alpha$ index derived from SCIAMACHY and GOME-2 and the \ion{Mg}{2} index derived from OMI radiometric observations computed over 90 days.  Red line: ratio between plage and filament area, scaled to fit the figure.}              \label{Fig_filaments_corHaMgII}%
    \end{figure}

These results indicate that the relatively low correlation between the observed Balmer and H$\alpha$ indices and the chromospheric indices may be related to the higher sensitivity of the Balmer lines to sunspots than to plage and networks. However, this does not necessarily rule out the contribution from filaments that might reduce the correlation; this is especially the case for H$\alpha$. In order to investigate this possibility, we compared the trends of the ratio of the area of plage and the area of filaments with the correlation coefficients between the \ion{Mg}{2} index and the H$\alpha$ index. All data were detrended as explained at the beginning of Sec.~\ref{sec:results}. The comparison is shown in Fig.~\ref{Fig_filaments_corHaMgII}. If, as claimed in previous studies \citep[e.g.][]{meunier2009, maldonado2019}, the appearance of filaments over the solar disk decreased the correlation between the H$\alpha$ and the \ion{Mg}{2} index, the two curves in the figure would show some degree of correlation.  However, the plot clearly shows several instances in which the correlation between the two indices decreases while the area ratio between plage and filaments increases, and vice versa, which results in a null correlation, $r\simeq0.076$ (nevertheless significant at p$<$0.01 due to the large sample size, n=2529 sample).  The plot also shows some dependence on the level of activity, the correlation being 0.2 (p$<$0.01 and n=732) and 0.25 (p$<$0.01 and n=923 sample) during the descending solar-cycle phase (2005-2007.5) and the ascending/maximum phase (2010.5-2015), respectively, and the two quantities showing an anticorrelation of -0.51 (p$<$0.01 and n=874 sample) during the period of minimum (2007.5-2010.5), which indicates that at times of higher magnetic activity, on the solar-rotational timescale filaments might have a small contribution in reducing the correlation between the Balmer and other chromospheric indices. This result is in part explained by the fact that, although variations of filament properties (i.e. number, length) correlate well with other activity indices,  they also usually present rather flat maxima, with peaks occurring often during, or right before, the descending phase of the solar cycle \citep{mazumder2018, mazumder2021}. The contribution of filaments to the H$\alpha$ variability is further discussed in Sec.~\ref{sec:filaments} and \ref{sec:models_contributions}.
   %%%%%%%%%%%%%%%%%%%%%%%%%%%%%%%%%%%%%%%%%%%%%%%%%%%%
%%%%%%%%%%%%%%%%%%%%%%%%%%%%%%%%%%%%%%%%%%%%%%%%%%%%  

Comparison between the modelled and observed variability (four bottom panels in Figs.~\ref{Fig_rotscale_2005} and \ref{Fig_rotscale_2011}) provides further insight into the sources of rotational modulations. In particular, because the modelled variability of Balmer indices does not take into account the filaments' contribution, we would expect the models to deviate from measurements the most at times when the filament area increases. 
Visual inspection of measured and modelled trends reveals several instances in which this is indeed the case, with the most noticeable one occurring around 2005.28. The filament area shows a broad maximum that closely corresponds to a minimum in the measured H$\alpha$ index, as one would expect from the line’s high sensitivity to filaments. The measured Balmer indices remain less affected, since their relatively lower filament sensitivity is counter-balanced by the rising plage input: note that the plage maximum coincides with the broad \ion{Mg}{2} peak. During this period, the models produce slightly increased, albeit noisy, index values. Something similar also happens around 2005.06 when the increase of the filament area produces a dimming in the measured Balmer indices not reproduced by the models. Similarly,  the filament maximum around 2011.62 (Fig.~\ref{Fig_rotscale_2011}) is not accompanied by any \ion{Mg}{2} enhancement. As anticipated, the measured H$\alpha$ shows a corresponding dip while both the measured Balmer indices and the models remain fairly flat. %Th
Note, however, that the relative difference between the models and the measurements does not show any correlation with the filament area (r=0.052, p$<$0.01, n=2529), thus suggesting that overall the discrepancies between the models and the observations are to be imputed to different effects. In this respect, it is important to note that there are several rotations in which measurements are well reproduced by the models, as for instance the two marked by vertical, green, dashed lines in Figs.~\ref{Fig_rotscale_2005} and \ref{Fig_rotscale_2011}. 

Overall, for DPI, the models produce correlation coefficients (Table~\ref{table1}) that are  comparable to the ones derived from measurements. %However,  the models correlate better with the \ion{Mg}{2} index than the inverted TSI. 
However, the models correlate better than measurements  with the \ion{Mg}{2} index than the inverted TSI. These results suggest that the contributions of plage and, in part, of the networks (both are known to modulate chromospheric indices) are most likely overestimated, and that therefore the synthesized plage and network line profiles are too shallow. The models show systematically better performance in the Balmer index case that may point to either some problems with the H$\alpha$ (or the Balmer index) observations, or the heightened H$\alpha$ sensitivity to filaments that is not captured by the models. The modelled contribution of each feature to the Balmer-line variability is further discussed in Sec.~\ref{sec:discussions}.
%%%%%%%%%%%%%%%%%%%%%%%%%%%%%%%%%%%%%%%%%%%%%%%%

Figure~\ref{Fig_rotscale_2011} shows that the H$\alpha$ index closely follows the trends of the inverted TSI in the 2011.1-2011.3 time range. However, at about 2011.1, sunspots are practically absent and the plage component (the \ion{Mg}{2} index) shows a very modest increase while the H$\alpha$ index goes through a deep minimum, matching the inverted TSI. Accordingly, the models produce a modest increase of the signal, reacting to the small rise in Mg II. The filament area remains practically constant to within the expected measurement noise levels. Similarly, the significant increase of the H$\alpha$ index at about 2011.29 closely follows the inverted TSI, while the models produce lower values, following the minimum in the Mg II index. Note that around 2011.10 and 2011.30 the measured Balmer index remains almost constant, being well reproduced by the models. This discord in the H$\alpha$ and Balmer index behavior may signify some problems in the SCIAMACHY and GOME-2 data. However, one may point, yet again, to the close match between the H$\alpha$ and the inverted TSI variability patterns. This prompted the introduction of the exclusively TSI-driven model in \citet{marchenko2021}. 

 \subsubsection{Sensitivity of the observed and simulated line indices to the model components}
 \label{sec:multireg}
 To summarize and clarify these findings, we run a multiple linear regression analysis with the IDL  routine {\it {regress.pro}}.  We use all the 'active' model ingredients (see Section 3), namely the networks, active networks, plage,  hot plage, umbra, and penumbra, along with the filament areas, and regress this 7-component mix against various time series (the 1st column in Table~\ref{tab_multi_reg}).  All area-filling factors and fitted values are detrended (61-day running means, as described above) and re-scaled: 
 \begin{equation}
                                     X_i=(x_{i} - x_{av}) / \sigma(x) , i=1,n
\end{equation}
where $x_{av}$ and $\sigma(x)$ are the the average and standard deviation in the 61-day temporal window, respectively. 

Multiple-regression fits are performed on a gradually truncated list of variables, eliminating the least significant variable from the outcome of each run and running the Wald test on the outcome. The tests are performed accounting for the effect of the correlated fitting residuals, following \citet{lean2020}: for each assessed characteristic (the rows in Table~\ref{tab_multi_reg}), we calculate an auto-correlation function of the residuals (observed - fitted) and use the auto-correlation values at the 1-day lag. This effectively (by ~2-9 times, depending on the tested characteristic) reduces the degrees of freedom (dof).  We also calculate the significance (p=0.01) of the individual Pearson correlation coefficients  with the revised dof. Thus, we eliminate the statistically insignificant variables in the multi-regression fits. In Table~\ref{tab_multi_reg} we list the retained individual correlation coefficients and the resulting linear multi-regression correlation (the last column). We note that all the rejected variables carry statistically insignificant correlation coefficients, while practically all the retained variables lead to significant (cf. 3rd column in Table \ref{tab_multi_reg},  p=0.01) correlations.

We also fit the observed TSI and \ion{Mg}{2} values, thus evaluating the sensitivity of the testing routine. As demonstrated in numerous studies \citep[see][and references therein]{lean2020}, detrended TSI shows heightened sensitivity to the hot plage (or facular) and sunspot components, while the \ion{Mg}{2} index primarily depends on the two plage components and, to a lesser extent, on the active network and sunspots. 

Comparing the models (1999, 2011) to observations (Balmer, H$\alpha$ in Table~\ref{tab_multi_reg}), one may note the complete lack of sensitivity (Balmer) or relatively low sensitivity (H$\alpha$) to the plage component. %This is anticipated in the model case. 
The models show much higher sensitivity to the network and plage components, in particular for the Balmer case. We return to these findings in Sec.~\ref{sec:discussions}.

The lack of sensitivity of Model 2011 - H$\alpha$ and Model 2011 - Balmer to the sunspot area filling factor (umbra) may seem surprising. Note that the penumbra values were retained in the fits. Both the umbra and penumbra factors initially carry statistically significant correlation coefficients, with the former only slightly trailing the latter. However, since the two are highly correlated, both are not needed in the final, multi-component fit. The applied statistical testing approach iteratively eliminates the fitting components that do not provide a statistically significant contribution to the quality of multi-component fits. Hence, formally, the umbra factors were eliminated first because their effects are well represented by the penumbra components. Retaining them and eliminating the penumbra component leads to the same (statistical) quality of fits. This relative insensitivity of the 2011 models to the umbra/penumbra inter-change can be explained by considering that these particular area-filling factors are highly correlated, r=0.96. Moreover, the 2011 Model is over-reliant on the networks and plage, thus diminishing the influence of sunspots, whereas the H$\alpha$ and Balmer indices are heavily weighed by the sunspot darkening. We will return to this dichotomy in Sec.~\ref{sec:discussions}.

One analogous point is that practically all the activity factors (columns in Table~\ref{tab_multi_reg}) are spatially and temporally (active solar regions and networks) inter-related, thus statistically correlated: e.g., umbra/penumbra. Ideally, the applied factor-elimination scheme should be run on the ensemble of independent variables. In practice, however, we have to consider that, though inter-related, the factors carry different spectroscopic signatures (Figs.~\ref{Fig_comp_synt_atlas} and \ref{models_beta_gamma_delta}) that must be viewed as independent during the spectral synthesis. Thus, we run the multi-regression analysis on a full list of variables and account for all the components in the synthetic models.   

Finally, we note that the multiple linear correlation coefficients r$_{multi}$ (last column in Table~\ref{tab_multi_reg}) are higher for the TSI and the \ion{Mg}{2} index (for the reasons explained above) than for the measured Balmer and H$\alpha$ indices, further confirming that the Balmer indices have a non-obvious and complex dependence on the solar-surface and -chromospheric magnetism. The r$_{multi}$ coefficients for the modelled Balmer indices are (not unexpectedly) high because, with the exception of the filaments, the models are constructed using the magnetic-feature filling factors against which they are regressed.

%%%%%%%%%%%%%%%%%%%%%%%%%%%%%%%%%%%%%%%%%%%%%%%%%%%%%%%%%%%%%%%%%%%%%%%%55
     \begin{table*}[h]
    % $$
        % \begin{array}{p{0.22\linewidth}llllllllll}
        \begin{tabular}{p{0.22\linewidth}llllllllll}
            \hline
            \noalign{\smallskip}
                 & dof  &   r\_{cr} (p=0.01)  &  E & F & Fil. & H & P & Pen. & S & r$_{multi}$ \\
            \noalign{\smallskip}
            \hline
            \noalign{\smallskip}

          Balmer                           &  351    & 0.137  &    --    & --        & --           & --        & 0.413 & 0.580 & 0.561 & 0.585 \\
          Model 1999 - Balmer & 554     & 0.109  &   --     & 0.471 & --           & 0.857 & 0.913 & 0.694 & 0.669 & 0.955 \\
          Model 2011 - Balmer & 1135    & 0.076  & 0.528 & 0.881  & --          & 0.821 & 0.726 & 0.498 &           --        & 0.973 \\
          H$\alpha$                         & 1676    & 0.063 & -0.06 & -0.027 & 0.074 & 0.193 & 0.44 & 0.585 & 0.573 & 0.681 \\
          Model 1999 - H$\alpha$ & 425  & 0.125 & 0.102  & 0.544  & --           & 0.781 & 0.810 & 0.5617 & 0.617 & 0.860 \\
          Model 2011 - H$\alpha$ & 1178& 0.075 & 0.403  & 0.720  & --           & 0.687 & 0.594 & 0.427 &                           --       & 0.795  \\
          TSI                                    & 673  & 0.099 & -0.042 & -0.094 & --  & --        & 0.241 & 0.779 & 0.771 & 0.889 \\
          \ion{Mg}{2} index                      & 460  & 0.120 & -0.104 & 0.486   & --           & 0.788 & 0.710 & 0.437 & 0.421 & 0.823 \\
            \noalign{\smallskip}
            \hline
       %  \end{array}
       \end{tabular}
   % $$   
     \caption {Linear multi-regression correlation coefficients derived from the detrended, re-scaled data acquired in 2005-2015 (n=1859). dof denotes degrees of freedom; r\_{cr} denotes the correlation coefficient from the rejected variables; E denotes network, F – active network, H – plage, P – hot plage, Pen – penumbra, S – sunspot, with all these surface-filling factors derived from the PSPT images; Fil. denotes the filament-filling factor; r$_{multi}$ is the multiple linear correlation coefficient.}
         %\label{indices}
    \label{tab_multi_reg}       
  \end{table*}
  
%%%%%%%%%%%%%%%%%%%%%%%%%%%%%%%%%%%%%%%%%%%%%%%%%%%%%%%%%%%%%%%%%%%%%%%%%%%%%%5
\subsection{Variability on Decadal Timescales}
\label{sec:results2}

\begin{figure}
   \centering
   \includegraphics[width=9.5cm]{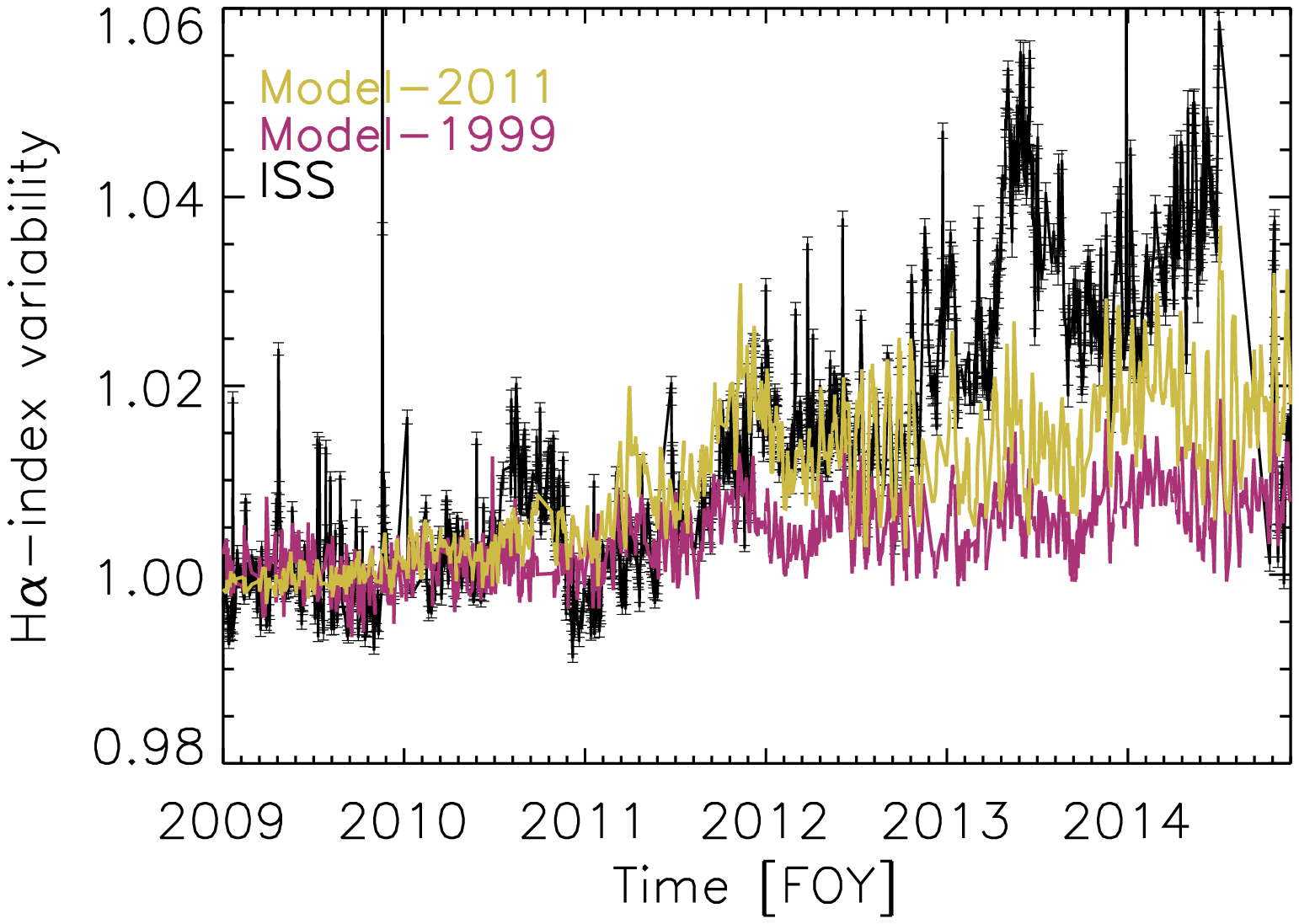}\\
     \includegraphics[width=10cm,trim={0.71cm 0  0.1cm 0.2cm},clip]{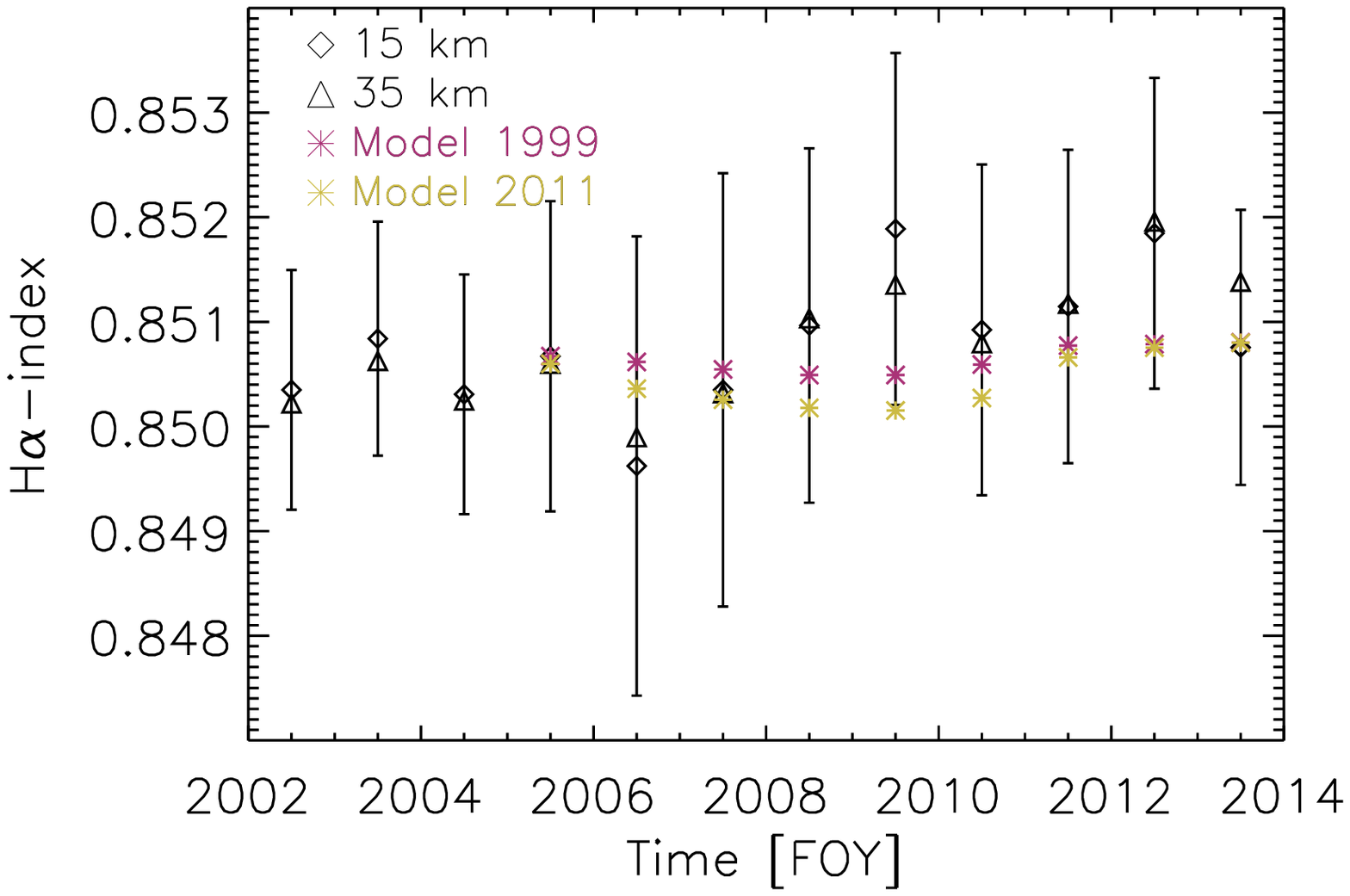}\\
  \includegraphics[width=9.5cm,trim={0.1cm 0  0.1cm 0.1cm},clip]{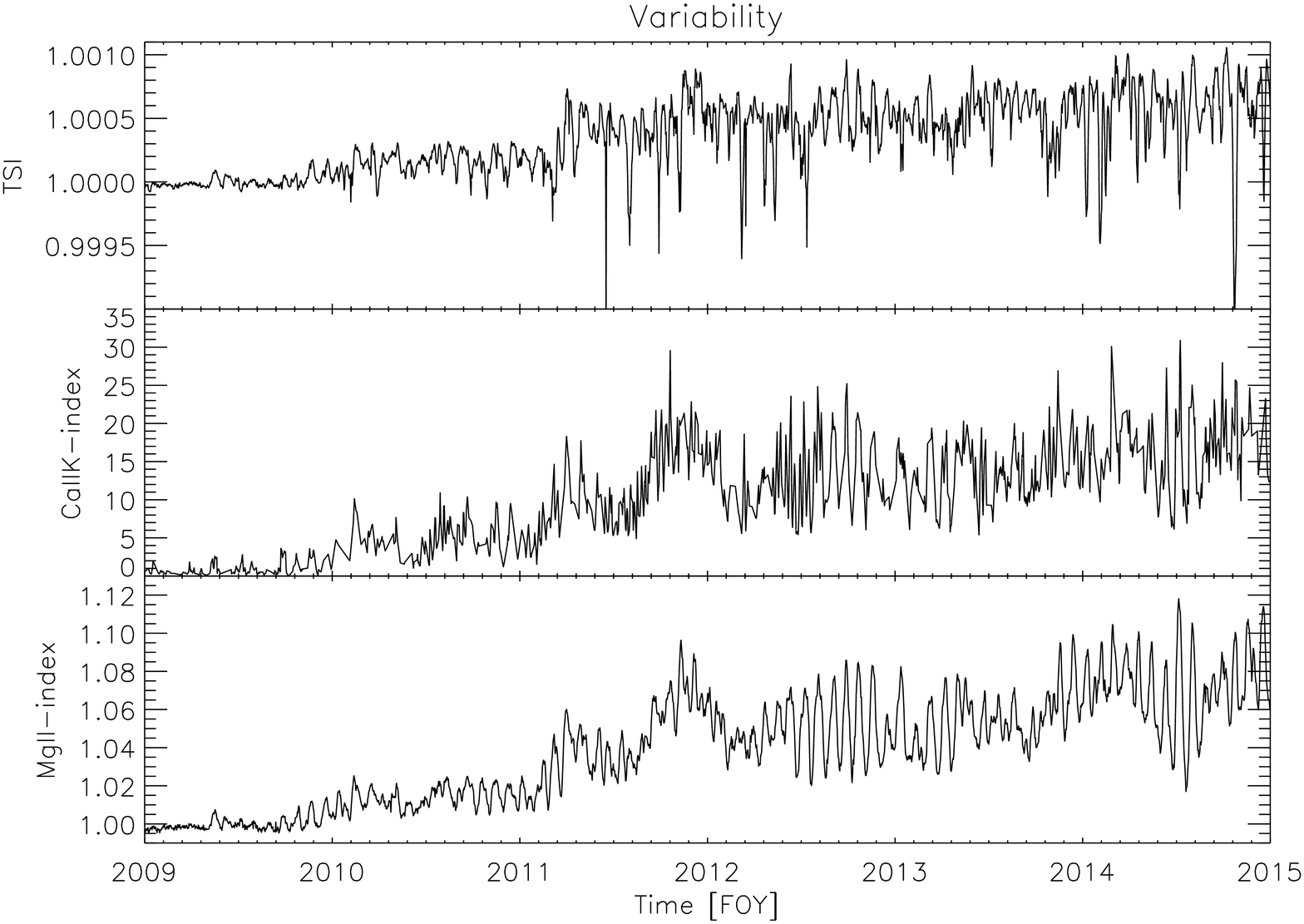}
   \caption{Variability of different activity indices with respect to year 2009. Top: variability of the  H$\alpha$ index derived from ISS measurements together with results obtained from the two semi-empirical models. Middle: variability of the  H$\alpha$ index derived from OSIRIS measurements. Bottom:  variability of the TSI (top), \ion{Ca}{2} K-SFO index (middle), and \ion{Mg}{2}-Bremen index (bottom). }              \label{long_term_indices}%
    \end{figure}

The variability of the $H\alpha$ index obtained from the analysis of SOLIS/ISS observations is shown in the top panel of Fig.~\ref{long_term_indices} along with the variability obtained from the two irradiance models. Here the variability is defined as the ratio between the $H\alpha$ index and the average index measured in 2009, a year in which magnetic activity was at minimum levels. The plot shows that, in agreement with results presented in \citet{meunier2009} and in \citet{livingston2010}, the $H\alpha$ index increases with solar activity.  The maximum variations measured by the ISS occur in 2013 and 2014 and are approximately $4\%$, which are comparable with variations obtained by \citet{meunier2009} between the beginning and the maximum of Cycle 22. This Cycle 24 rise, however, seems unrealistically large, considering that Cycle 24 was less intense than Cycle 22 analyzed by \citet{meunier2009}. Supposing that the $H\alpha$ index scales linearly with the sunspot number, and considering that the number of sunspots in Cycle 22 was roughly double the number of sunspots in Cycle 24, one would expect a variability of approximately $2\%$, which roughly agrees with the results obtained with the reconstruction based on the set of FAL2011 atmospheric models. Overall, this reconstruction seems to reproduce the data relatively well until the end of 2013, while the FAL1999 model produces less variability. It is also worth noting that other activity indices produce a rather 'flat' maximum between 2012 and the beginning of 2015, as illustrated, for instance, in the bottom panel of Fig.~\ref{long_term_indices}, thus corroborating our conclusion that the  increase of the index measured by the ISS in 2013 and 2014 most likely results from instrumental effects not corrected by the standard ISS calibration pipeline. 
A positive correlation of the H$\alpha$ index with the magnetic activity  was also found  by analysis of the OSIRIS observations, as shown in the middle panel of Fig.~\ref{long_term_indices}. In the figure, the 35-55 km time series has been shifted downwards with a single absolute offset to demonstrate its consistency with the 15-35 km time series. %The change from Cycle 23/24 minimum in 2006-2007 to Cycle 24 maximum in 2012 is +0.31\%. 
These data give a modest positive correlation (r = 0.28) with annual-average Bremen \ion{Mg}{2} index values calculated using concurrent dates if the year 2002 sample is excluded (due to differences in OSIRIS instrument operation during the first few months of its mission).
As a validation of this approach, we have also calculated OSIRIS solar-activity index values for the Ca II K absorption feature at 393.4 nm. These data give a solar-cycle amplitude of 2.3\% from 2007 to 2013 and are well-correlated with annual-average Bremen \ion{Mg}{2} index values (r = 0.76). However, the increase of the index measured between 2007 and 2009 suggests that the adopted procedure does not entirely remove the spurious trends  described in Sec.~\ref{sec:measurements}. For comparison, the plot shows the index obtained from the two models (with their spectra degraded to OSIRIS resolution and the employing same definition of index) shifted to match OSIRIS observations in 2005. The plot shows that, overall, the measurements and models agree within the uncertainties of the measurements.   

The Pearson correlation coefficients between the H$\alpha$ index derived from ISS measurements and other activity indices are reported in Table~\ref{table3}, along with the correlation coefficients derived from the models. Note that 2005-2012 OMI data show r=0.98 correlation between the Mg II and Ca II indices \citep{deland2013}. Results in the table confirm the result obtained by visual inspection of Fig.~\ref{long_term_indices} that, on the decadal temporal scale, the  H$\alpha$ index is highly correlated with the two chromospheric indices \ion{Mg}{2} and \ion{Ca}{2} and with the TSI, a photospheric index.  

\begin{figure}
   \centering
     \includegraphics[width=5.5cm,trim={0.7cm 0  1.3 0},clip]{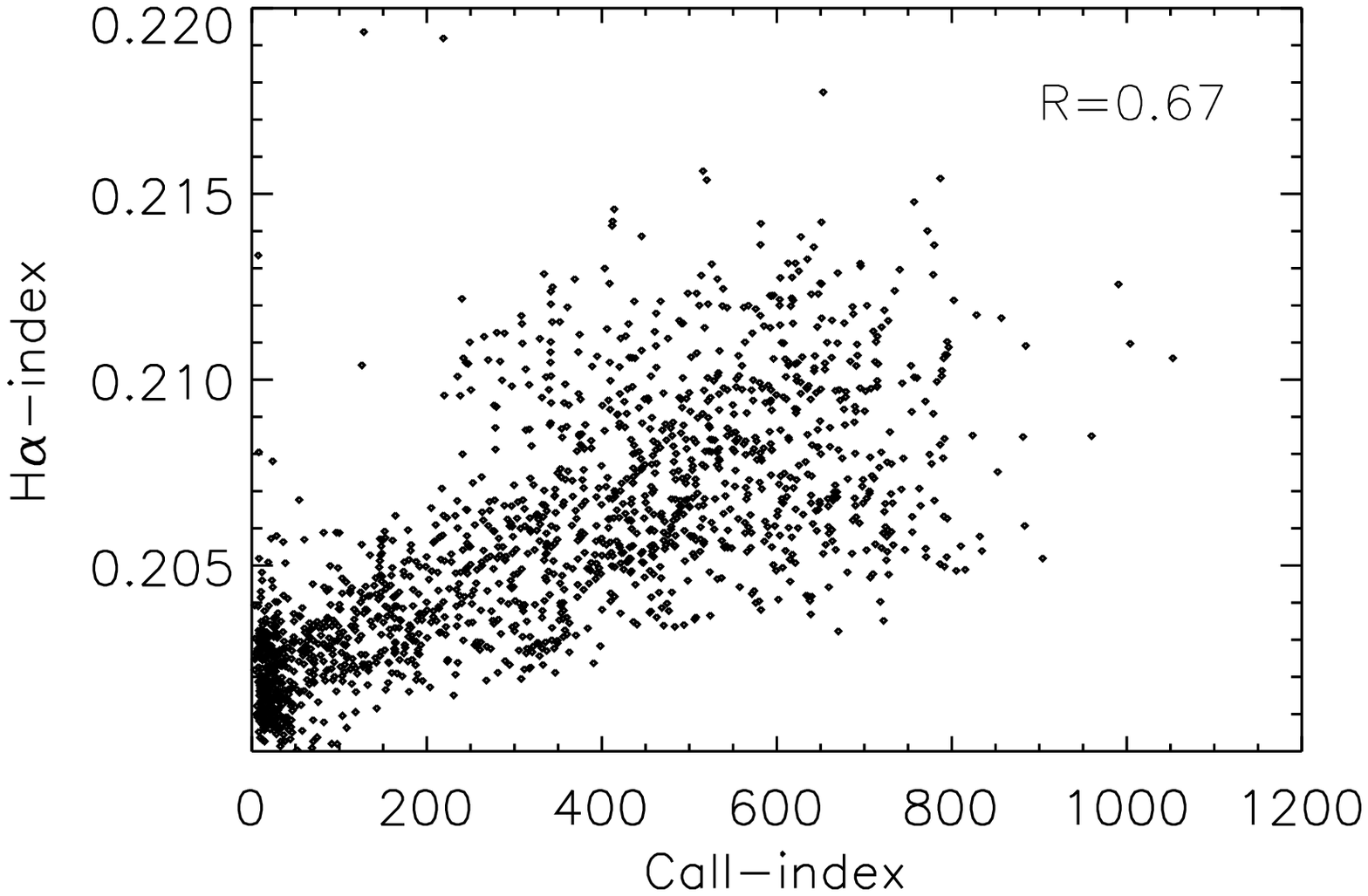}
  \includegraphics[width=5.5cm,trim={1.1cm 0  1.3 0},clip]{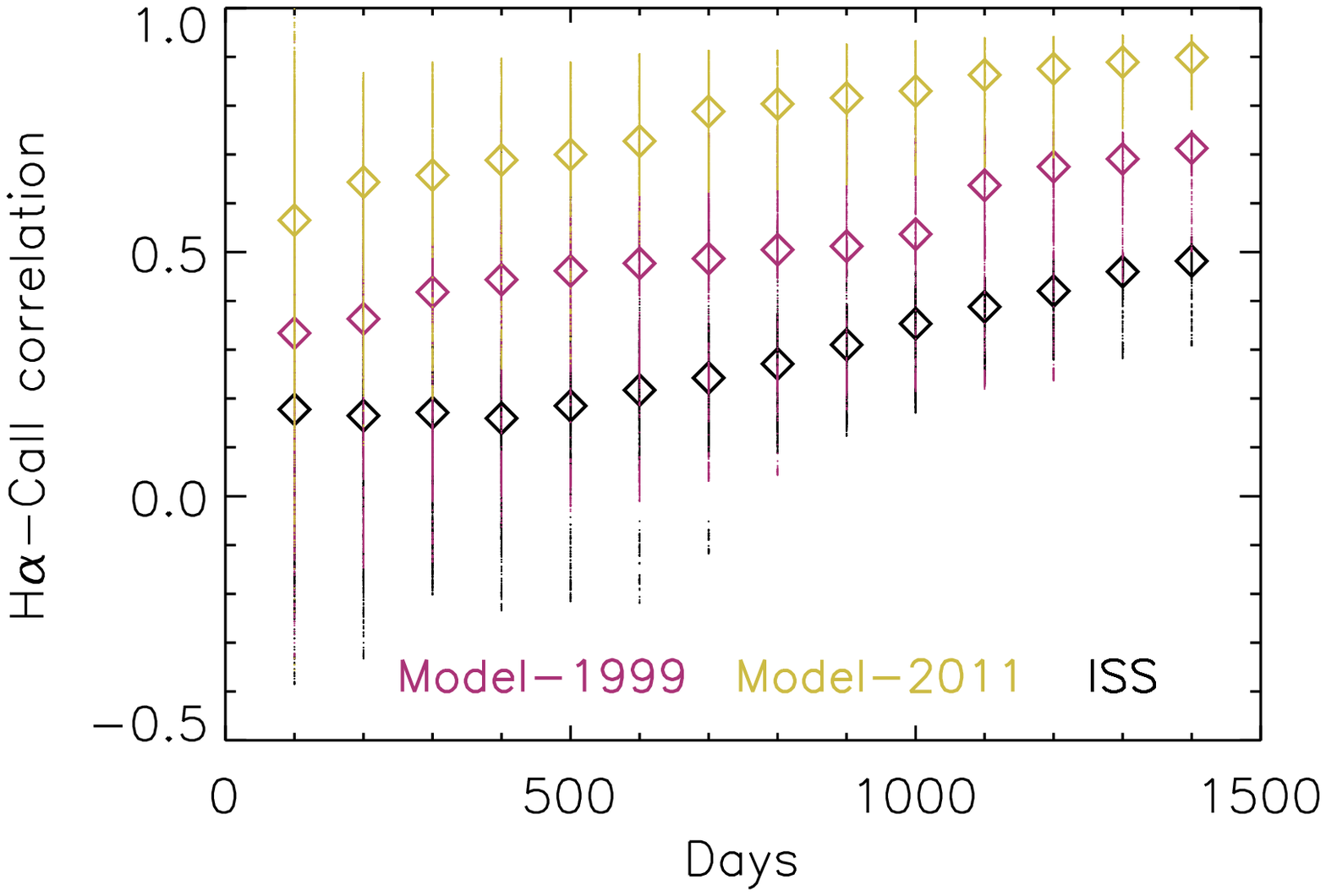}
  \includegraphics[width=5.5cm,trim={1.5cm 0  1.5 0},clip]{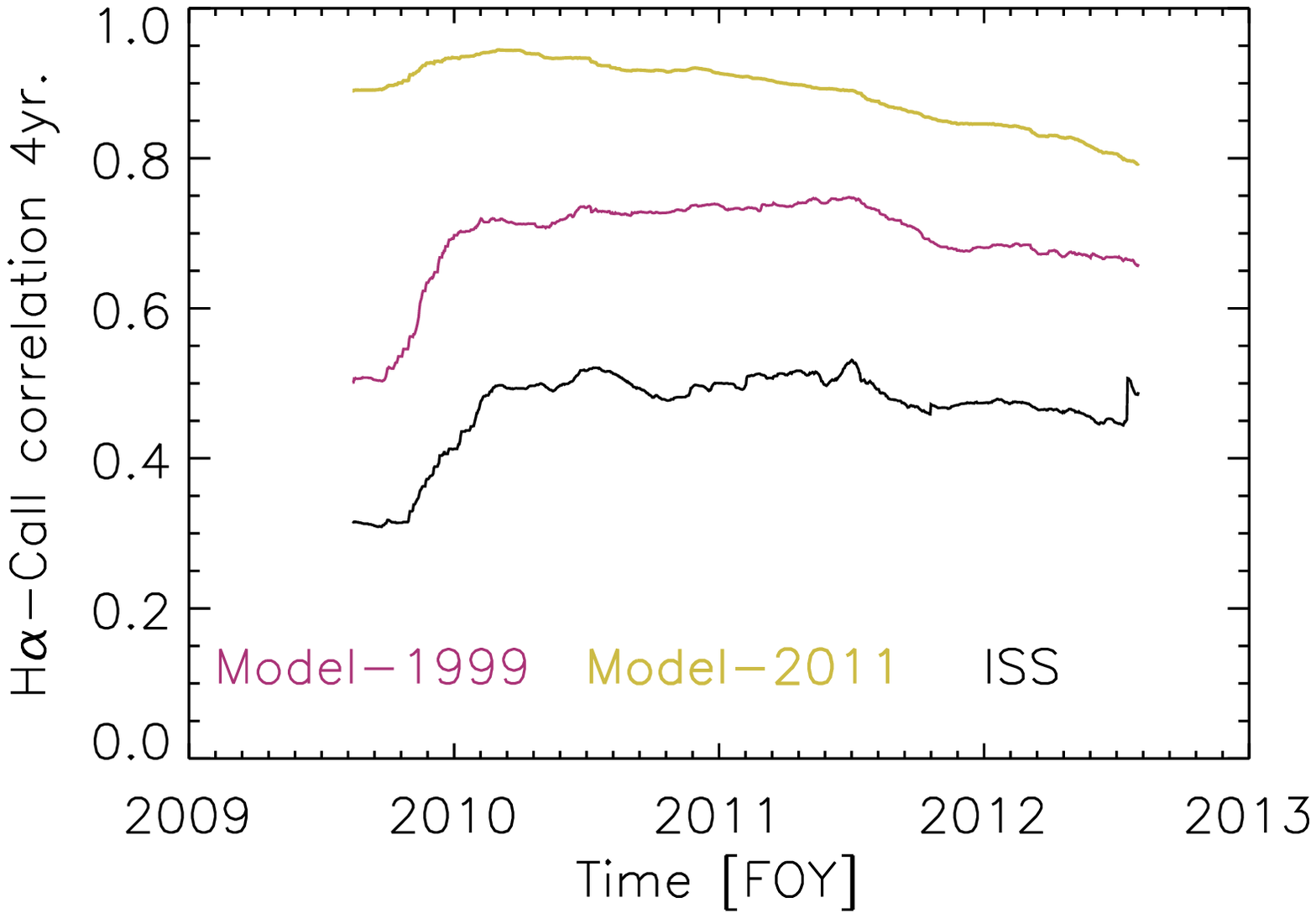}
   \caption{Left: Comparison between the observed H$\alpha$ and the \ion{Ca}{2} K indices, estimated between 2008 and 2014.5. The correlation coefficient is 0.67.  Center: dependence of the H$\alpha$-\ion{Ca}{2} K correlation coefficients on the length of the temporal window used to compute the correlation. Diamonds represent median values. Right: temporal variation of the correlation coefficients computed over a time range of 4 years. }              \label{long_term_corrCa}%
    \end{figure}

To further investigate the relation between the variability of the $H\alpha$ index and other chromospheric indices, we employed the same approach as in \citet{meunier2009} and \citet{maldonado2019}, i.e., we investigated the correlation coefficient between the $H\alpha$ and the \ion{Ca}{2} indices and how this changes with time.
The plot in the left panel of Fig.~\ref{long_term_corrCa} confirms that the H$\alpha$ index is clearly positively correlated with the \ion{Ca}{2} index, the Pearson correlation coefficient being  approximately 0.7 (significant at p$<$0.01), c.f. with approximately 0.8 reported by \citet{meunier2009}. The central plot of the same figure shows the correlation coefficient between the two indices computed on running averages with growing length of temporal window. Diamonds show the median values of the distribution for each temporal window. In agreement with results presented in \citet{meunier2009},  the median value of the correlation increases, while the dispersion of the correlation decreases with the increase of the temporal window duration. In particular, the shorter the temporal window, the more likely it is to find temporal regions in which the H$\alpha$ and \ion{Ca}{2} indices are anti-correlated. The same qualitative trends are found from the analysis of synthetic data, the Model 1999 showing a better agreement with ISS and with results reported in \citet{meunier2009}, who found the highest median correlation value to be approximately 0.8. The plot on the right shows the temporal variation of the correlation coefficient computed over a time span of four years. In qualitative agreement with results obtained by \citet{meunier2009}, the correlation between the two indices is positive and is lowest during periods of minimum magnetic activity; it therefore increases overall with time through this portion of the solar cycle's rising activity. The correlation thus increases during the ascending phase and slightly (and briefly) decreases during periods of high magnetic activity. Even in this case, the models qualitatively reproduce the observed trends, with Model 1999 producing better quantitative agreement. \citet{meunier2009} ascribed the decrease of correlation between the H$\alpha$ and \ion{Ca}{2} indices during periods of minima to the relative increase of the contribution of filaments during periods of low magnetic activity. However, we note that the models employed to reproduce the H$\alpha$ variability do not take into account the contribution of filaments. On the contrary, in the models, the decrease of the correlation is caused by the relative increase during periods of minima of the contribution of quiet and active network, whose line profiles, as shown in Fig.~\ref{Fig_comp_synt_atlas}, only slightly deviate from the one of quiet Sun. %When computed over the whole temporal period, the correlation coefficients between the modelled H$\alpha$ index and the \ion{Mg}{2}, the \ion{Ca}{2} and the TSI are slightly higher than the ones found from the analysis of ISS observations (see Table~\ref{table1}), the best agreement found even in this case for Model 1999.   
Plots presented in Figs.~\ref{long_term_indices} and \ref{long_term_corrCa} both indicate a good qualitative agreement between measurements and models. Quantitatively, differences are found that exceed the nominal uncertainties of measurements. However, a closer inspection reveals that ISS measurements are affected by variations at different temporal scales, imputed to changing observing conditions (e.g. daily and seasonal variations of atmospheric conditions, synodic variations, etc.). These are not present in the models, and that may contribute to a decrease of the correlation coefficients between the H$\alpha$ and \ion{Ca}{2} indices over these decadal timescales.

%%%%%%%%%%%%%%%%%%%%%%%%%%%%%%%%%%%%%%%%%%%%%%%%%%%%
    
%%%%%%%%%%%%%%%%%%%%%%%%%%%%%%%%%%%%%%%%%%%%%%%%%%%%    

%%%%%%%%%%%%%%%%%%%%%%%%%%%%%%%%%%%%%%%%%%%%%%%%%%%%
   \begin{figure}
   \centering
  \includegraphics[width=5.8cm]{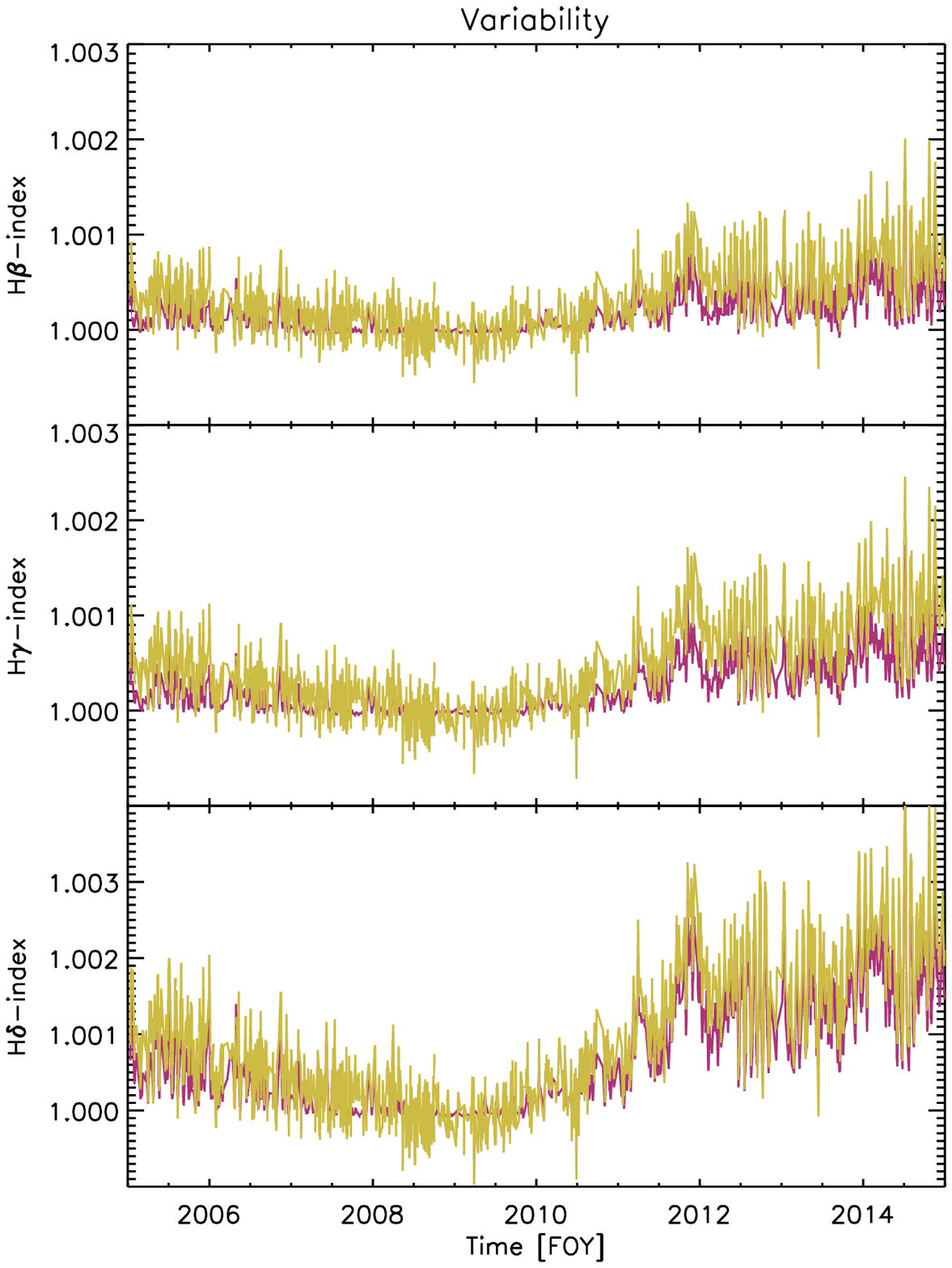}
  \includegraphics[width=5.8cm]{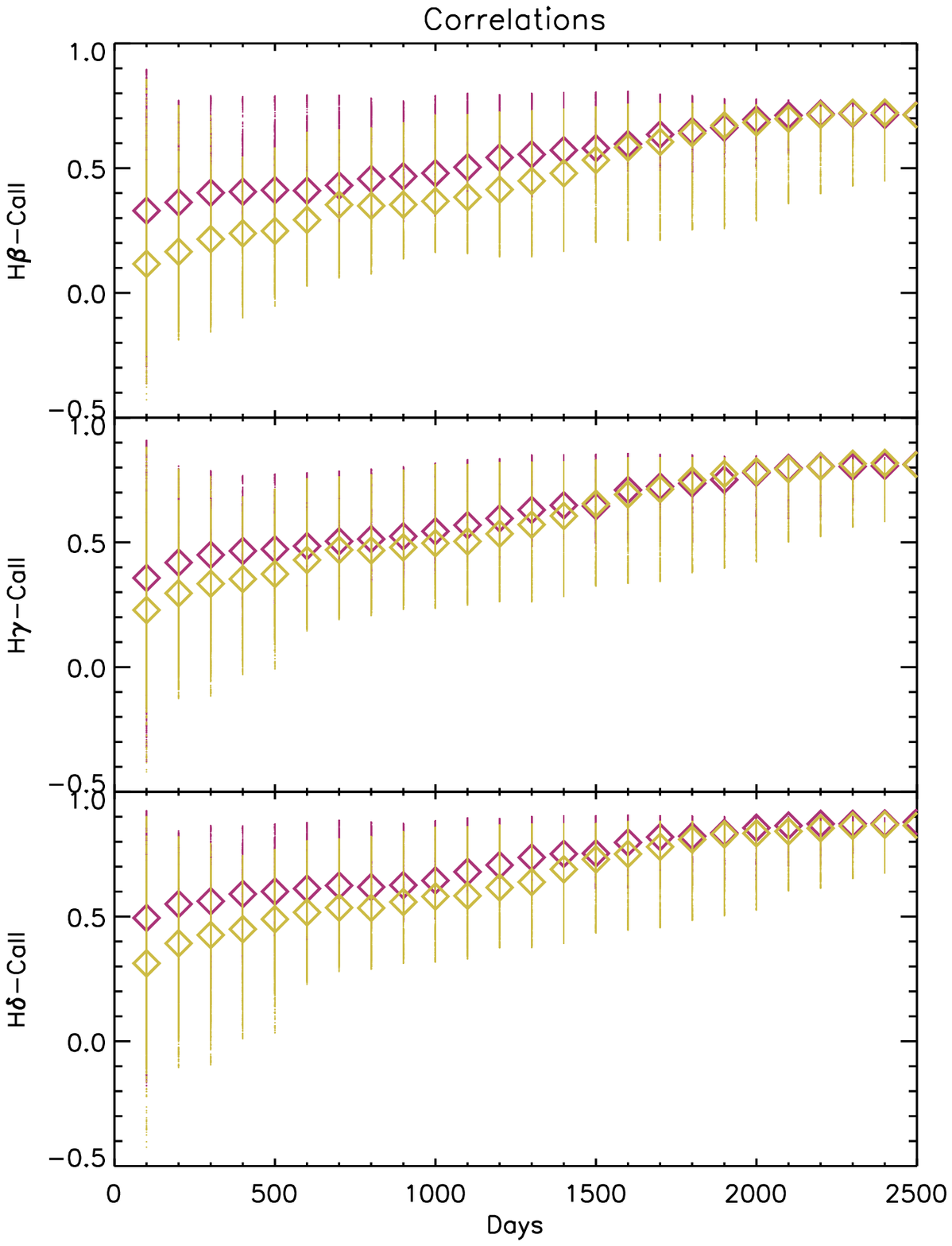}
   \includegraphics[width=5.8cm]{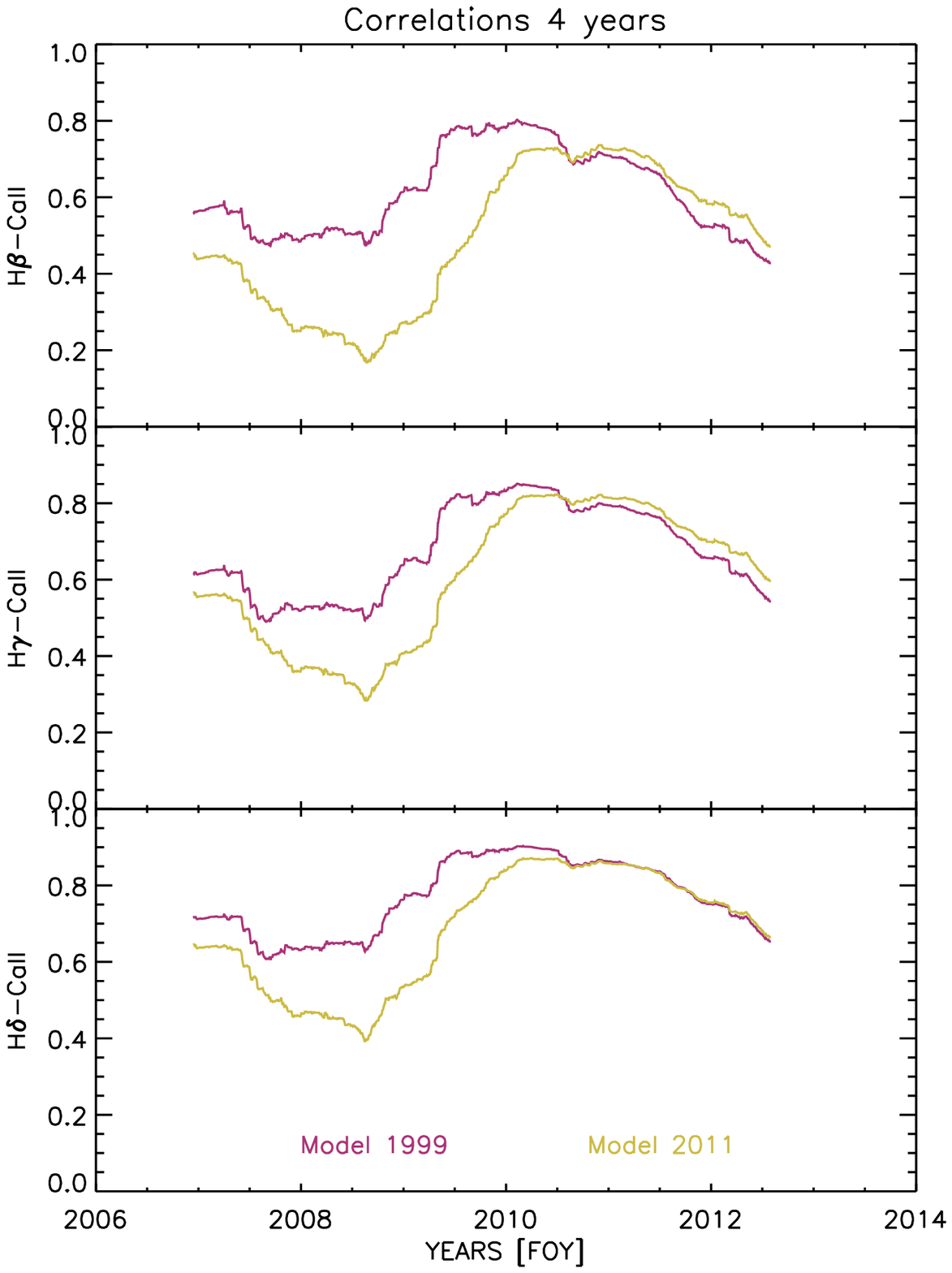}\\

  \caption{Left: Modelled variability of the core-to-wing ratio indices of Balmer lines obtained simulating OMI measurements. Center: dependence of the Balmer-\ion{Ca}{2} K correlation coefficients on the length of the temporal window used to compute the correlation. Right: temporal variation of the correlation coefficients between the Balmer indices and the \ion{Ca}{2} K index computed over a 4-year temporal window.           \label{Fig_Hbalmers}}%
    \end{figure}
    
    %%%%%%%%%%%%%%%%%%%%%%%%%%%%%%%%

%%%%%%%%%%%%%%%%%%%%%%%%%%%%%%%%%%%%%%%%%%%%%%%%%%%%  
\begin{table}[h]

\begin{center}

\begin{tabular}{ |l|c|c|c| } 
 \hline
     & \ion{Mg}{2} & \ion{Ca}{2} & TSI\\
     \hline
H$\alpha$ - ISS & 0.74 & 0.67 & 0.61 \\ 
  H$\alpha$ - Model 1999  & 0.83 & 0.72 & 0.42 \\ 
  H$\alpha$ - Model 2011  & 0.98 & 0.9 & 0.62 \\ 
 H$\beta$ - Model 1999  & 0.83 & 0.70 & 0.29 \\ 
 H$\beta$ - Model 2011  & 0.82 & 0.71 & 0.42 \\ 
  H$\gamma$ - Model 1999  & 0.92 & 0.80 & 0.45 \\ 
 H$\gamma$ - Model 2011  & 0.89 & 0.80 & 0.52 \\ 
  H$\delta$ - Model 1999  & 0.96 & 0.87 & 0.57 \\ 
 H$\delta$ - Model 2011  & 0.93 & 0.86 & 0.59 \\ 
 
 \hline
\end{tabular}
\caption{Solar-cycle timescales:  Pearson correlation coefficients between the observed (ISS: 2009-2014.6) and modelled H$\alpha$ (same epoch, 2009-2014.6) and Balmer core-to-wing ratio indices and various observed activity indices. \label{table3}}
\end{center}
\end{table}
%%%%%%%%%%%%%%%%%%%%%%%%%%%%%%%%%%%%%%%%%%%%%%%%%%%%

\begin{figure}
   \centering
  \includegraphics[width=9cm]{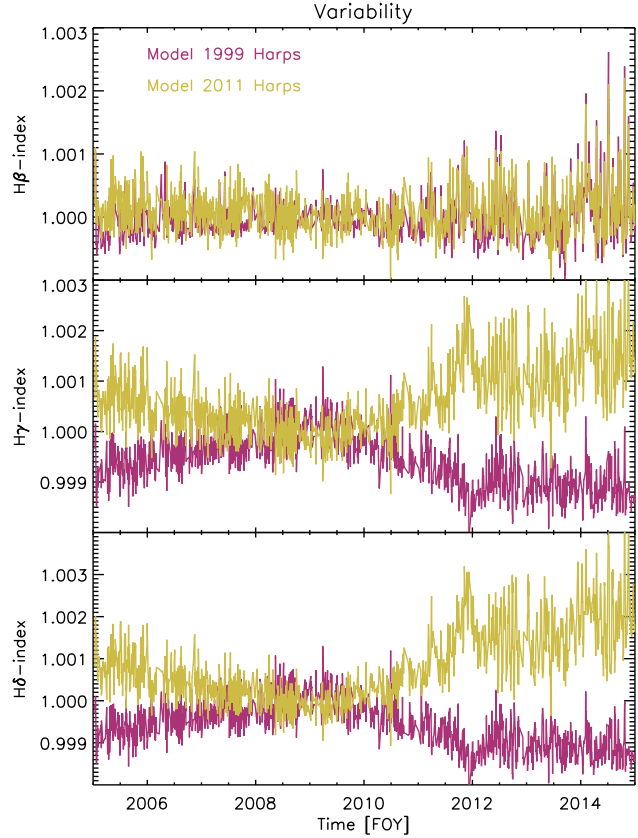}

  \caption{ Modelled variability of the core-to-wing ratio indices of upper Balmer lines obtained simulating HARPS-N measurements.          \label{Fig_Hbalmers_Harps}}%
    \end{figure}

We then investigated the long-term variability of the H$\beta$, H$\gamma$, and H$\delta$ core-to-wing ratio indices derived from the models. To this end, although, for the reasons explained in Sec.~\ref{sec:measurements}, OMI measurements were not suitable for comparison, we used synthetic spectra degraded to the spectral resolution of OMI (approximately 0.5 nm) and followed the definitions of the core-to-wing ratios described in Paper I. The left panel of Fig.~\ref{Fig_Hbalmers} (as well as Table~\ref{table3}) shows that the three indices vary, if modestly, in phase with the magnetic activity cycle. Likewise the H$\alpha$ index, the H$\beta$, H$\gamma$, and H$\delta$ indices are positively correlated with the \ion{Ca}{2} K index, the correlation coefficients being a function of the length of the temporal window (middle panel of Fig.~\ref{Fig_Hbalmers}) and of the particular phase of the cycle (right panel of Fig.~\ref{Fig_Hbalmers}). However, unlike for H$\alpha$, the two models seem to produce rather similar long-term variability and correlations with the magnetic activity indices. These results are not in agreement with the trends observed by \citet{marchenko2014} around the onset of Cycle 24, with the Balmer lines failing to show any changes, in stark contrast to the rapidly changing ensemble of the \ion{Mg}{2}-like transitions. At the egress from Cycle 24, \citet{maldonado2019} found the H$\beta$ and H$\gamma$ indices to be anti-correlated with the magnetic activity indices and the H$\delta$ index showing no correlation. In order to investigate whether this discrepancy could depend on instrumental and observational effects, we computed the indices after degrading the synthetic spectra to the HARPS-N spectral resolution (approximately 0.005 nm) using the indices definition described in \citet{maldonado2019}. Results in Fig.~\ref{Fig_Hbalmers_Harps} show that, in this case, the H$\beta$ presents a slightly positive correlation with the activity for both models. H$\delta$ and H$\gamma$ indices produced with Model 1999 are slightly anti-correlated with activity, while Model 2011 produces indices that are always positively correlated. The anti-correlation in Model 1999 results from the FAL1999 atmosphere models producing network synthetic line profiles slightly dimmer than the quiet model, as shown in Fig.~\ref{models_beta_gamma_delta} (Appendix). Indeed, a reconstruction performed neglecting the network's contribution (which was replaced with the quiet-Sun model) produces an in-phase core-to-wing ratio of H$\gamma$ and H$\delta$ at the HARPS-N resolution (results not shown). This does not mean that the cores of Balmer lines vary in anti-phase with the activity, but that the network components reduce the increase of the line-core intensity, thus producing a net decrease of the core-to-wing ratio with activity.   For instance, for Model 1999 we found that the H$\delta$ core intensity varies approximately 0.5\% from solar minimum to maximum at the OMI spectral resolution and only 0.25\% for HARPS-N, whereas the continua show approximately the same variability of 0.3\% (see also Table~\ref{tab_var_wing_core}). Such dependence on the instrumental characteristics is due to the fact that with a decrease of spectral resolution, the contribution from the far wings (whose intensities are higher in the network than in the quiet Sun) to the core intensity increases up to the point that the core intensity of the network exceeds that of the quiet-Sun model, as in the case of OMI.  The trends shown in Fig.~\ref{Fig_Hbalmers_Harps} are still in partial disagreement with HARPS-N measurements. However, the results found from this analysis indicate that: 1) Instrumental and observational effects critically affect the estimated trends of the Balmer indices; 2) An anti-correlation of the Balmer indices with activity may result from the network component dominating over the plage and sunspot components.   
%%%%%%%%%%%%%%%%%%%%%%%%%%%%%%%%%%%%%%%%%%%%%%%%%%%%
    
    %%%%%%%%%%%%%%%%%%%%%%%%%%%%%%%%  

\section{Discussion}
\label{sec:discussions}

\subsection{Model performance}
\label{sec:modelperf}
Results presented in the previous sections indicate that the models qualitatively reproduce the overall observed variability trends. Judging the quantitative model performance, we use the approaches outlined in Section 3.2: mainly the Pearson correlation coefficients, then MAPE and (observed-modeled) standard deviation statistics (Tables \ref{table_metrics} - \ref{tab:metrics_decadal} in Appendix). These results confirm that the reconstructions obtained from the FAL1999 better reproduce the observations than the reconstructions obtained from the FAL2011 models, particularly regarding the  relatively more modest (over-)~estimates of the contribution of plages.

It is worth noting that the metric values strongly depend on the periods over which they are computed. %, being in general larger during periods of stronger activity and lower during periods of lower activity. 
%, the reconstructions obtained using the FAL1999 models seemingly performing better in reproducing the observations. 
The Pearson correlation between the models and measured H$\alpha$ index ranges, for instance, between r$\simeq$0 in 2009 (near solar minimum) and r$\simeq$0.6 for Model 1999 and r$\simeq$0.5 for Model 2011 in 2014. In all cases, the probability that the measured correlations are generated by a random distribution is less than 10$^{-10}$. Results shown in Table~\ref{table_metrics} also confirm that the reconstructions perform better in reproducing the variability of the Balmer index rather than of the H$\alpha$ index. The standard deviations computed over the detrended data (Table~\ref{table_stddev}) also show that Model 1999 produces better agreement with observations than Model 2011.

Metric values obtained by comparing the models with the ISS measurements over the decadal scale are reported in Table~\ref{tab:metrics_decadal}. As expected, the MAPE suggests Model 2011 better reproduces the observations, while the correlation coefficients are rather similar, confirming that both models reproduce the general trend. As for OSIRIS, we find a very poor correlation of approximately 0.1 for both Model 1999 and Model 2011, most likely due to the increase of the H$\alpha$ index between 2007 and 2009 in OSIRIS measurements. The MAPE is 0.063 and 0.074 for Model 1999 and Model 2011, respectively, thus confirming that the set of FAL1999 better reproduces the observations. 
% In this respect it is worth to note that the fraction of area covered by sunspots is lower than that covered by faculae \citep[e.g.][]{chapman2011, criscuoli2016}, so uncertainties in the facula model are more likely to affect the reconstructions rather than uncertainties in the sunspot model. 
 
 \subsection{On the contribution of networks and plages}
 \label{sec:netplage}
In order to verify that uncertainties in the reconstructions are largely driven by uncertainties in the plage contribution, we performed two tests. Firstly, we computed the reconstruction using the FAL1999 set, but we employed only model H to take into account the two plage components.  We found (results not shown) that this reconstruction produces a MAPE, a standard deviation, and a correlation with measurements of 0.008, 0.9$\times10^{-4}$, and 0.34, respectively, for the H$\alpha$ index. Comparison of these values with those shown in Table~\ref{table_metrics} and Table~\ref{table_stddev} confirms indeed a slightly better agreement with observations compared to the original Model 1999. Similar results are found for the Balmer index. 

In the second test, we weighed the different components of the FAL1999 set using weights assigned in accordance with the multi-regression analysis of H$\alpha$ reported in Table~\ref{tab_multi_reg}. In particular, we weighted by 0 the two network components and by 0.25 and 0.5 the plage and hot-plage components, respectively. This approach also shows the model’s sensitivity to uncertainties in the feature surface-filling factors. Specifically, the masks employed in this study make use of intensity thresholds derived from the FAL2011 atmospheric models, so that the contributions of networks and plages are not estimated in a fully consistent way. For this model the MAPE, standard deviation, and correlation with measurements are 0.006, 0.65$\times10^{-4}$, and 0.45, respectively, which, even in this case, are an improvement with respect to the values found for the original Model 1999. On the other hand, the comparatively small changes of the metric values in spite of the relatively large changes of the filling factors also indicate that uncertainties in the feature identifications did not significantly affect the reconstructed line indices. 

\subsection{On the contribution of filaments and prominences to H$\alpha$ variability}
\label{sec:filaments}
\begin{figure}
   \centering
  \includegraphics[width=6cm,trim={0.45cm 0  1.5 0.9cm},clip]{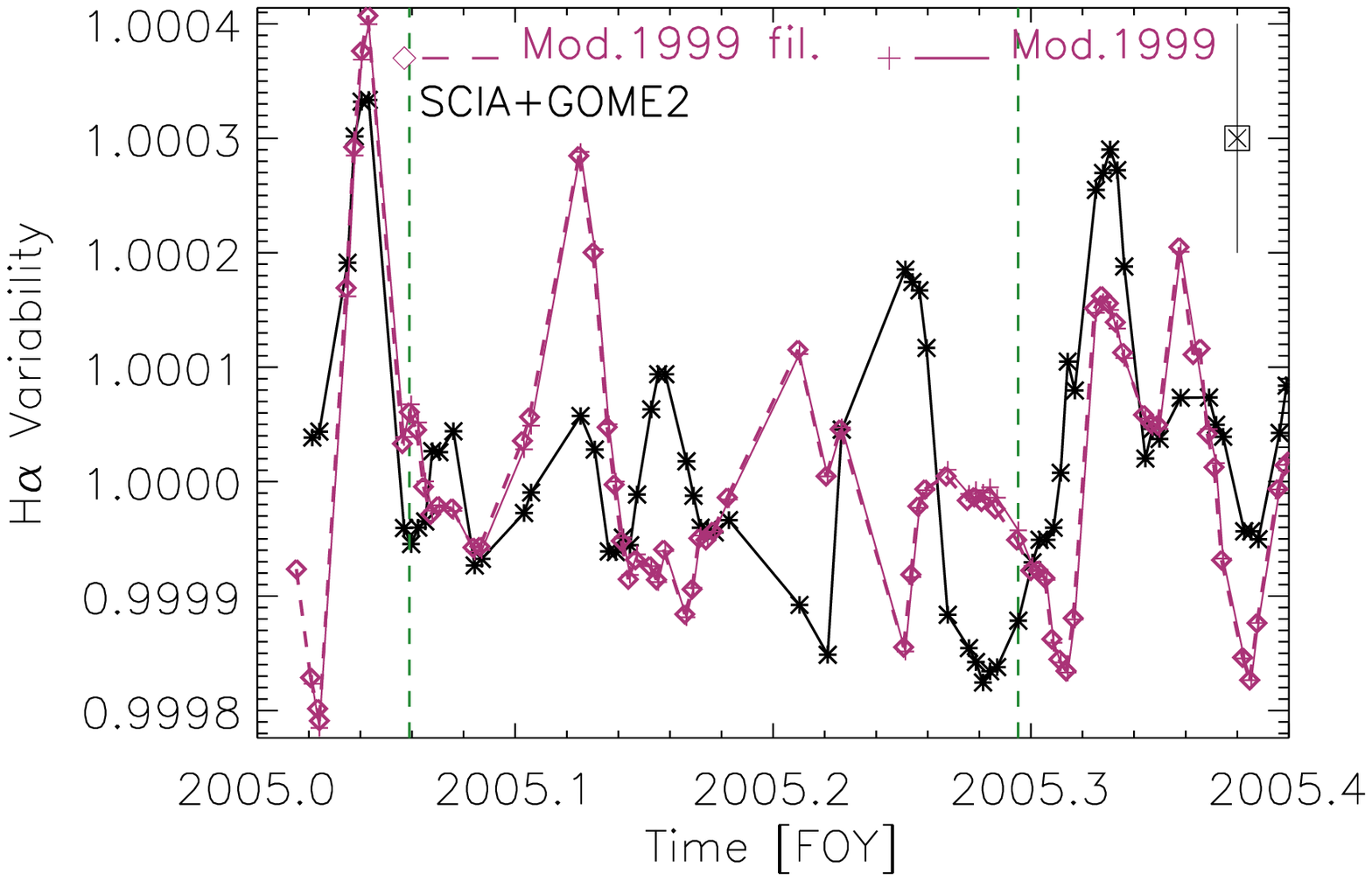}
   \includegraphics[width=5.9cm,trim={0.8cm 0  1.5 0},clip]{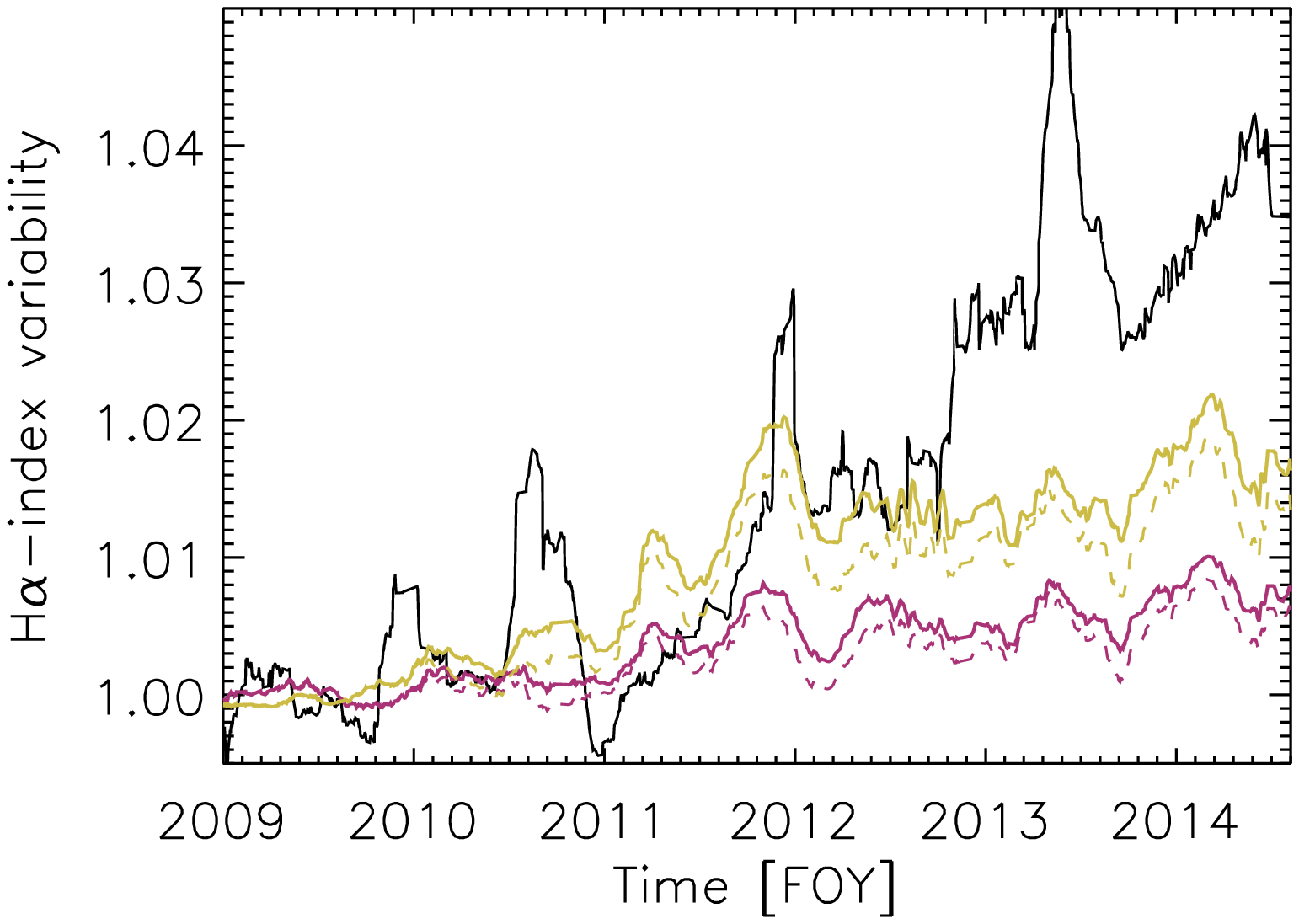}\includegraphics[width=5.9cm,trim={1.2cm 0  1.5 0},clip]{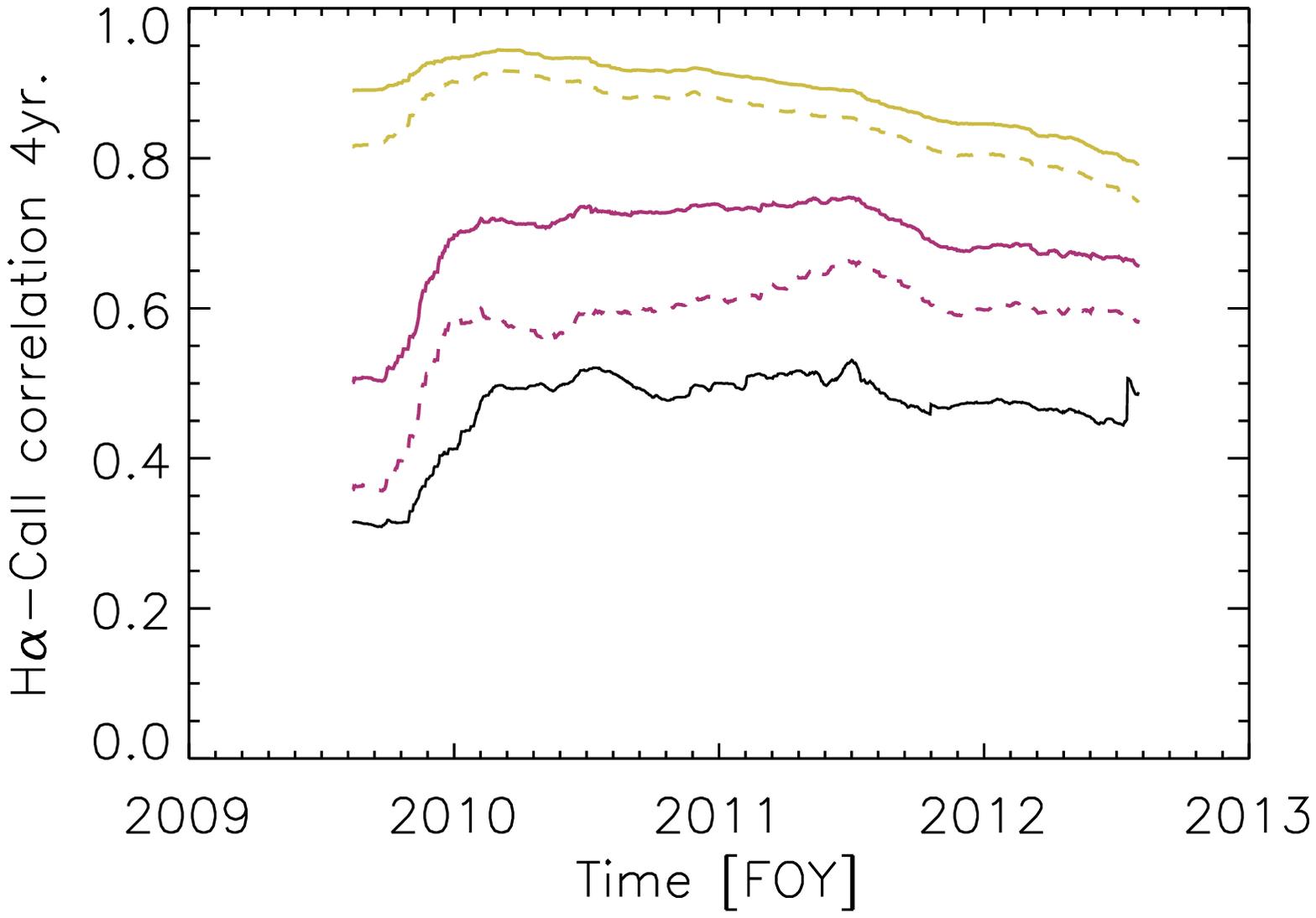}

  \caption{Left: Comparison of the variability of the H$\alpha$ index obtained from measurements (SCIAMACHY and GOME-2) and from Model 1999 with and without the inclusion of filaments (see text). Vertical bar represents the $\pm 1\sigma$ uncertainty of the SCIAMACHY and GOME-2 measurements. Green dashed vertical lines indicate times at which the filaments area peaks. Center: Long-term variation of H$\alpha$ index obtained from the measurements and the models. Right: Comparison of temporal variation of the correlation coefficients between the H$\alpha$ and the \ion{Ca}{2} K indices computed over a 4-year temporal window obtained from ISS measurements and the various models. In central and right panels the color code is as in Fig.~\ref{long_term_corrCa} and dashed lines indicate models taking into account filaments contribution.                 \label{Figvarswithfils}}%
    \end{figure}

In previous sections we showed that the observed trends for the Balmer indices can be explained by a lower sensitivity of these indices to the network contribution. However, in Sec.~\ref{sec:results1}, we showed that the contribution of filaments may occasionally explain the excess of the Balmer indices variability on the rotational scale produced by the models. Moreover, the temporal change of the correlation between the H$\alpha$ and the \ion{Ca}{2} indices presented in Fig.~\ref{long_term_corrCa} is reproduced by the models only qualitatively. Quantitatively, the higher correlation with the \ion{Ca}{2} index found for the models with respect to the observations can be explained either by an overestimate of the plage contribution, by neglecting filaments, or by a combination of both. In order to 
shed some light on this issue, we employed Model 1999 and Model 2011 to construct new models that take into account the contribution of filaments in the following way:

\begin{equation}
     NewModel = \it{ModelXXX}+(I_{fil.}-I_{q})\times f_{fil.}
     \label{eq._withfilssolrot}
\end{equation}
where ModelXXX is either Model 1999 or Model 2011, $I_{fil}$ and $I_{q}$ are the line profiles of filaments and of the Quiet Sun model, respectively, and $f_{fil.}$ is the filling factor of filaments. Here the main assumption is that filaments are 'suspended' over quiet-Sun pixels. This approximation is justified because quiescent filaments are larger and live longer than filaments suspended above active regions; however, this also means that the models most likely underestimate the filament's contribution. Model 1999 is more affected, as the facular continuum intensities are larger in the FAL1999 set than in the FAL2011. The second  main assumption is that a single profile is representative of different filaments and at different positions over the disk, although measurements show that rather different line profiles are found even within the same filament \citep[e.g.][]{maltby1976, chae2006}.  Because, to the best of our knowledge, a database (either synthetic or observed) of filament line profiles is not available, we use and compare models constructed using the shallowest line profile published in \citet{kuckein2016}, whose contrasts value of -0.3  is consistent with typical contrasts reported in  literature \citep[-0.3/-0.1, e.g.][]{chae2006, maltby1976, diercke2022}. Note that \citet{kuckein2016} provided line profiles normalized to the quiet-Sun line profile and that to convert these data to intensities, we used the quiet-Sun model from the FAL2011 set, since this model gives better agreement with the observations. The filament line profile is shown in the bottom, central panel of   Fig.~\ref{Fig_comp_synt_atlas}, along with the FAL2011 Quiet Sun model profile for comparison. The resulting line profiles were then degraded to the spectral resolution of the instruments, and the core-to-wing ratio was computed as explained in Sec.~\ref{sec:model}.  Results are illustrated in Fig.~\ref{Figvarswithfils}.
As expected, on the rotational scale (left panel), the signal is reduced at times when the filament area peaks (marked with vertical green lines in the figure), but the effect is overall small and the standard deviation over the whole 2005-2015 range is unaltered.
Similar results are obtained for Model 2011 and are not shown. A regression analysis performed on these new models confirms that the effect of filaments on the variability on rotational timescales is almost negligible (see also Sec.~\ref{sec:multireg}).  By contrast,  we found that the introduction of filaments produces appreciable effects to the variability on the decadal timescale (central panel of Fig.~\ref{Figvarswithfils}). 
The Pearson correlation coefficients between the H$\alpha$ index and the \ion{Ca}{2} K index over the whole 2009-2014.6 period for Model 1999 decreases from 0.72 to 0.68,  very close to the observed 0.67 value, while for Model 2011 the correlation decreases from 0.9 to 0.88. Finally, the right panel of Fig.~\ref{Figvarswithfils} shows the effect of filaments on the temporal variations of the 4-years correlation coefficients between the \ion{Ca}{2}~K and H$\alpha$ indices. For Model 1999, the introduction of the filaments  produces reasonable agreement with the observations, with the remaining differences most likely resulting from overestimates of the contribution of plages, as described in Sec.~\ref{sec:netplage}, while for Model 2011, the disagreement is still large. We found that for Model 2011 to reproduce the observed 4-years correlation, the contrast of filaments would need to be more than -0.9, which is unrealistic.  
In partial agreement with these findings, \citet{meunier2022} recently estimated that a null or negative correlation with the \ion{Ca}{2} emission, as observed in part of their sample of F-G-K stars,  is found only for filament areas much larger than those typically observed on the Sun, with this result holding at both the rotational and longer temporal scales.

These results suggest that filaments most likely play a role in decreasing the correlation between the H$\alpha$ and the \ion{Ca}{2} K  indices on timescales longer than rotational. Note that we performed further tests (not shown) using the deepest line profile reported in \citet{kuckein2016}, which has a contrast (-0.65) that is approximately more than double what is usually observed.  Even in this case, the correlation between the two indices remains positive.

%%%%%%%%%%%%%%%%%%%%%%%%%%%%%%%%%%%%%%%%%%%%%%%%%5
%%%%%%%%%%%%%%%%%%%%%%%%%%%%%%%%%%%%%%%%%%%%%%%%%%%%%%5

\begin{figure}
   \centering
   \includegraphics[width=6cm,trim={0.48cm 0  1.5 0.8cm},clip]{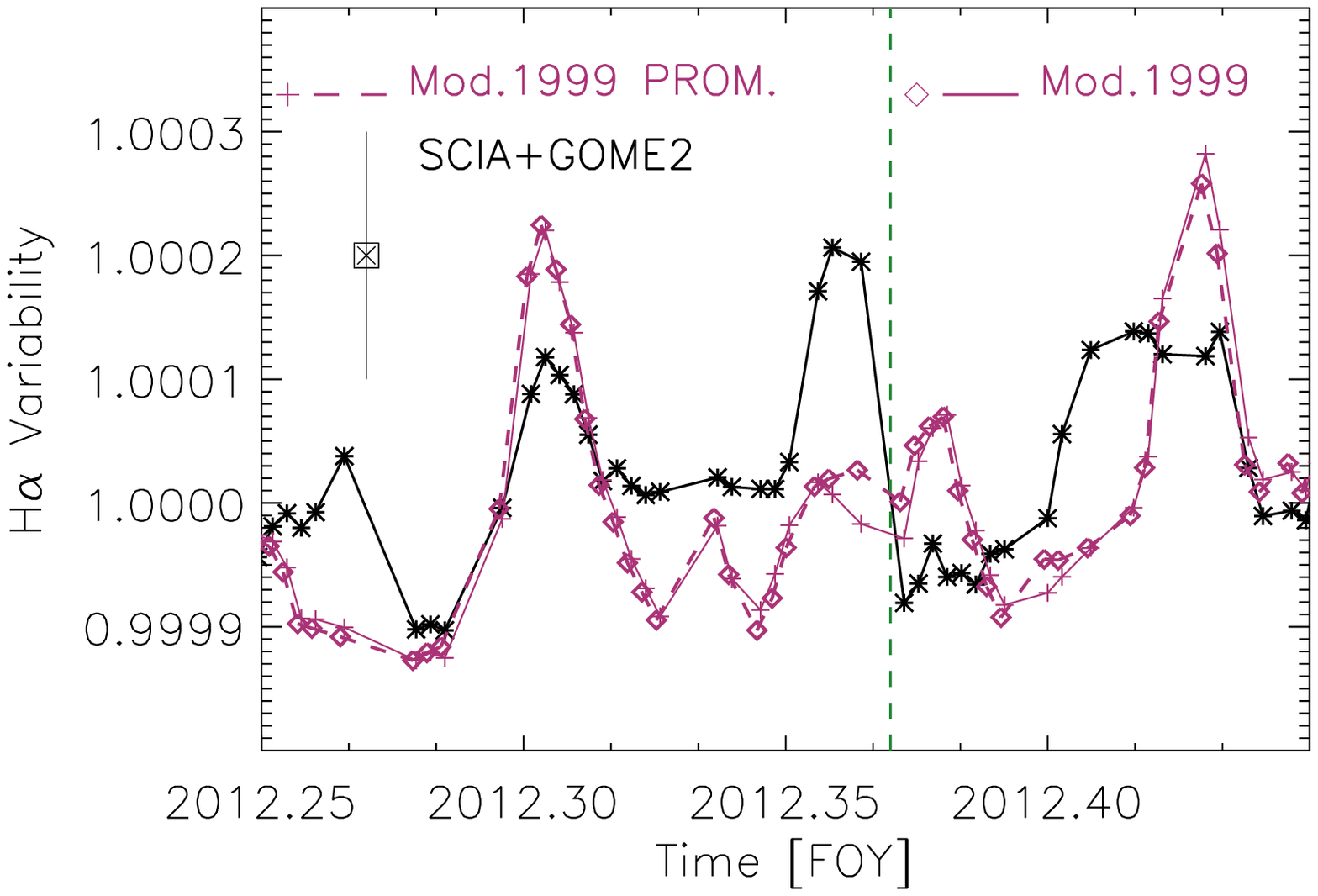}
  \includegraphics[width=5.8cm,trim={1.2cm 0  1.5 0},clip]{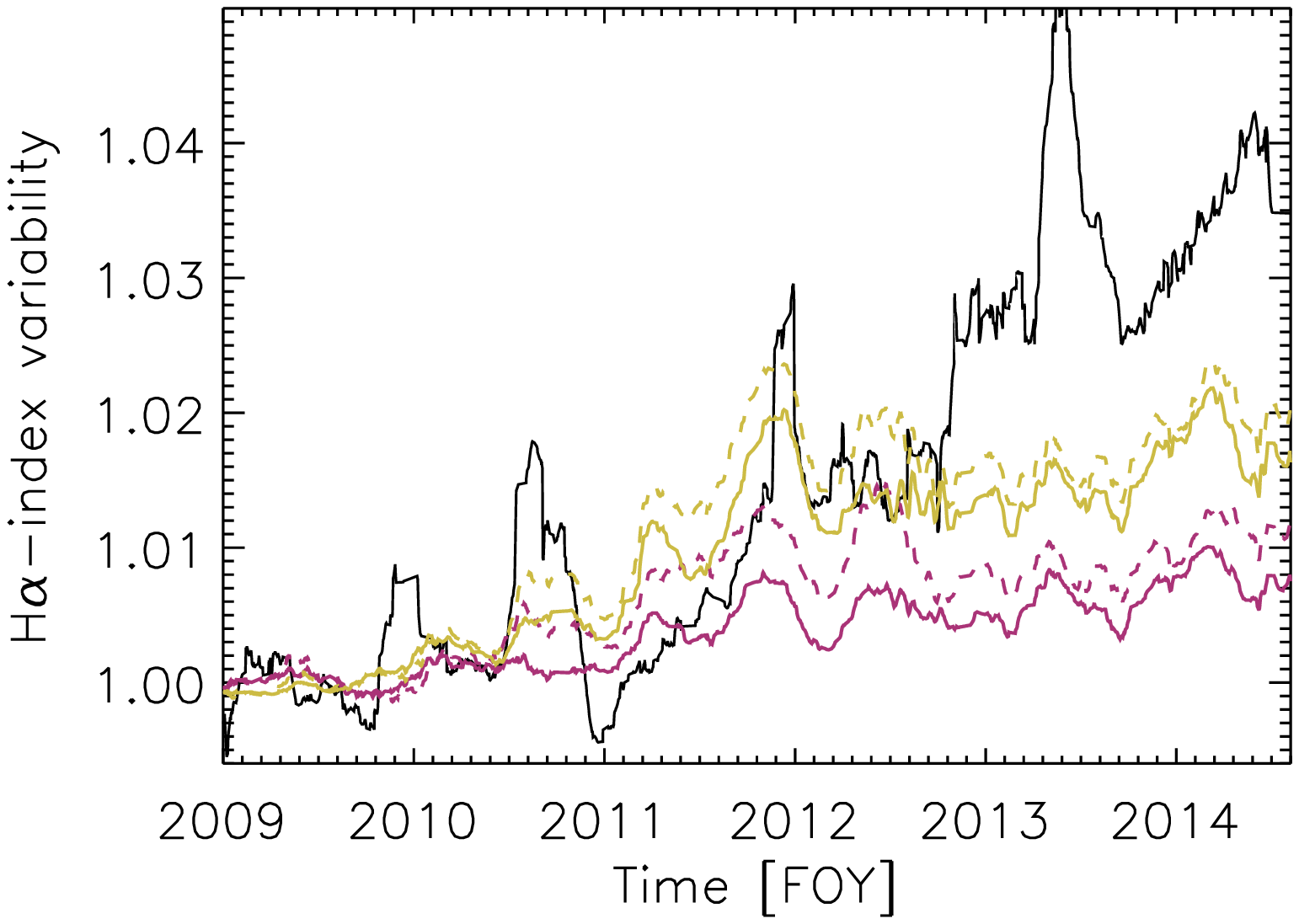}
  \includegraphics[width=5.8cm,trim={1.5cm 0  1.5 0},clip]{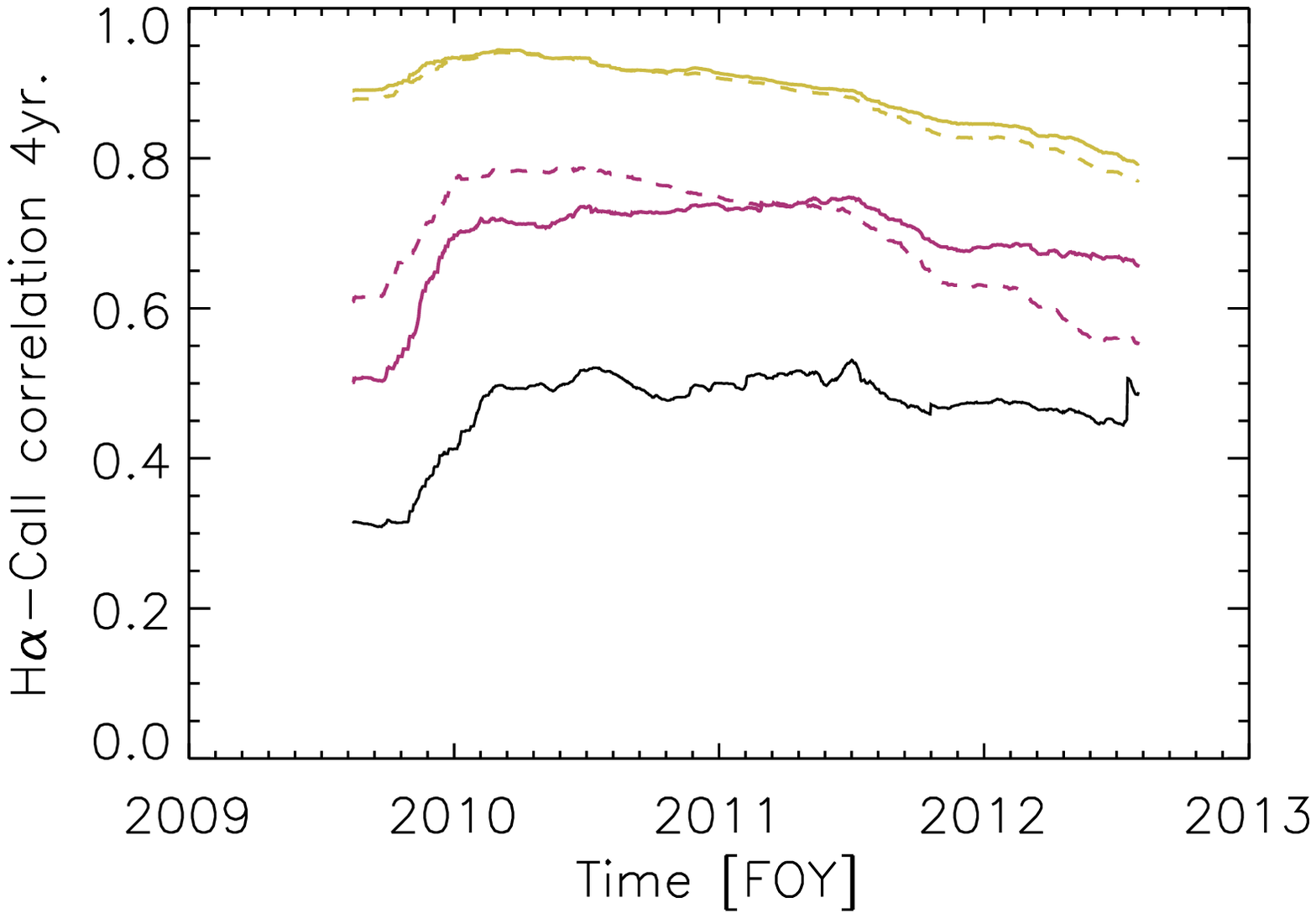}
  \caption{Left: Comparison of the variability of the H$\alpha$ index obtained from measurements (SCIAMACHY and GOME-2) and from Model 1999 with and without the inclusion of prominences (see text).  Vertical bar represents the $\pm 1\sigma$ uncertainty of the SCIAMACHY and GOME-2 measurements. Green dashed vertical line indicates a time at which the contribution of prominences produces noticeable variation of the H$\alpha$ core-to-wing ratio  Center: Long-term variation of H$\alpha$ index obtained from ISS measurements and the models. Right: Comparison of temporal variation of the correlation coefficients between the H$\alpha$ and the \ion{Ca}{2} K indices computed over a 4-years temporal window. In center and right panels the color code is as in Fig.~\ref{long_term_corrCa} and the models with prominence contribution are represented with dashed lines.             \label{Fig_corrs_models_with_prom}}%
    \end{figure}

We also investigated the contribution of solar prominences to the variability of H$\alpha$ by constructing the following model:
\begin{equation}
     NewModel = \it{ModelXXX}+I_{prom}\times A_{prom}
     \label{eq.modwithprom}
\end{equation}

where, like in Eq.~\ref{eq._withfilssolrot}, ModelXXX is either Model 1999 or Model 2011, $I_{prom}$ is the prominence line profile, and $A_{prom}$ is the prominence area. As for filaments, we employed a single line profile for different prominences. Specifically, we employed the line profile shown in Fig.~3 of \citet{heinzel2014}, also plotted in the bottom right panel of Fig.~\ref{Fig_comp_synt_atlas}. For the prominence area, we employed a composite of data derived from Meudon spectroheliograms and the LSO/KSO International H$\alpha$ prominence catalogue. This was necessary to cover the gap between October 29, 2012 and  June 16, 2013 in the Meudon catalogue. Results are shown in Fig.~\ref{Fig_corrs_models_with_prom}. The left panel  shows the variability on the rotational scale obtained from Model 1999. As with filaments, besides some instances at which the area of prominences peaks and causes a small increase of the signal (marked with a green vertical line in the figure), the contribution of prominences to the variability on the rotational scale is overall negligible. Indeed, we found that the standard deviation values reported in Table~\ref{table_stddev} are basically unchanged. On the solar-cycle timescale, prominences induce an increase of variability from approximately 1\% to 1.5\% for Model 1999 and from approximately 2\% to 2.4\% for Model 2011. Finally, the inclusion of prominences seems to leave unaltered the 4-year correlation with the \ion{Ca}{2}~K index for Model 2011, while for Model 1999, the correlation is higher during the minimum/ascending phase and lower at the maximum with respect to the model without prominences, as shown in the right panel of Fig.~\ref{Fig_corrs_models_with_prom}.

%%%%%%%%%%%%%%%%%%%%%%%%%%%%%%%%%%%%%%%%%%%%%%%%%%%%%%

%%%%%%%%%%%%%%%%%%%%%%%%%%%%%%%%%%%%%%%%%%%%%%%%%%%%%%
\subsection{Variability of line wings and line cores on rotational and decadal timescales}
\label{sec:wingcore}
Measurements of the core-to-wing ratio of  lines forming in stellar atmospheres have the advantage of being less sensitive to instrumental effects, as for instance absolute calibration uncertainties. However, such measurements do not allow to distinguish between the wings' and the core's contributions. The widely used line-index definitions use averaged line-wing irradiances. Such 0.1-0.2 nm averages \citep[e.g.][]{maldonado2019} inevitably include scores of weak subordinate lines besides the line-free solar continuum, thus complicating interpretation of the changing core/wing ratios. The two contributions can instead be studied separately using the presented model. For models emulating  radiometric measurements (see Table~\ref{tab_var_wing_core}), we found that the cores and the wings generally show similar variability on both the rotational and the decadal scale (see Fig.~\ref{contribs_with_fil2005}). However, while for Model 1999 the core variability is systematically larger than that of the wings, for Model 2011 the opposite is generally true. Specifically, the standard deviations on the 61-day detrended data for Model 1999 are 5.0$\times10^{-4}$ and 4.7$\times10^{-4}$ for the core and the wings, respectively, while for Model 2011 the standard deviations are 2.9$\times10^{-4}$ and 4.6$\times10^{-4}$, respectively. Similarly, from the solar-cycle minimum to maximum for Model 1999 the core and the wings increase by 0.25$\%$ and 0.17$\%$, respectively, while for Model 2011 the core increases by 0.02$\%$ and the wings decrease by approximately 0.1$\%$. The difference between the two models mostly results from the different wing intensity of the plage models (Fig.~\ref{Fig_comp_synt_atlas}), as discussed in Sec.~\ref{sec:models_contributions}. In the FAL1999 set the plage models produce an excess of intensity with respect to the quiet-Sun model, which compensates for (partially on the rotational scale, fully on the decadal scale) the intensity decrease induced by sunspots. In the FAL2011 set, plage models produce wing intensities that are similar to the ones of the quiet-Sun model, so that the wings' variability is dominated by sunspots.

The contribution of the wings to the variability of the core-to-wing ratio of the upper Balmer lines is qualitatively similar to the one found for H$\alpha$, that is we found that typically the core presents larger variations than the wings, even so slightly.
 %%%%%%%%%%%%%%%%%%%%%%%%%%%%%%%%%%%%%%%%%%%%%%%%%%%%%%%%%%%%%%%%%%%%%%%
  
    \begin{table*}[h]

       \begin{tabular}{cccccccccc}
            \hline
            
               Model  & H$\alpha_{core}$ & H$\alpha_{wing}$  & H$\beta_{core}$ & H$\beta_{wing}$  & H$\gamma_{core}$ & H$\gamma_{wing}$  & H$\delta_{core}$ & H$\delta_{wing}$ \\
           
            \hline
          Model 1999  &  0.25 | 1.3   & 0.17 | 0.3  &   0.3 &0.25 & 0.45 & 0.32 & 0.5 & 0.3  \\
          Model 2011  & 0.02 |2.4   & -0.1 | 0.02  & 0.03 & -0.025 & 0.12 & 0.04 & 0.2 & -0.05   \\
   
            \hline
            \end{tabular}
                 \caption {
                 Minimum to maximum cycle 24 relative variation (in percent) of the core and wings of Balmer lines obtained from the two models. For H$\alpha$ each column report values obtained emulating SCIAMACHY and GOME-2 measurements (left) and ISS observations (right). For the upper Balmer lines values were obtained emulating OMI measurements (as in Paper I).                }
       
    \label{tab_var_wing_core}   
    \end{table*}
    
%%%%%%%%%%%%%%%%%%%
Finally, Table~\ref{tab_var_wing_core} shows also the variability of H$\alpha$  as would be measured by the ISS. In this case, the variability of the core is larger than the variability of the wings for both models due principally to the increased spectral resolution increasing the core variability. The variability of the wing is also different from the variability one would measure with OMI, or more in general using the typical (approximately) 2 nm spectral averages routinely employed for core-to-wing ratios measurements, due principally to the fact that the chosen spectral region in ISS  observations is located in the H$\alpha$ wing, which increases with the magnetic activity even for Model 2011. 

\subsection{Contributions of the model components to the H$\alpha$ core and line wing}
\label{sec:models_contributions}
\begin{figure}[!ht]
   \centering
  \includegraphics[scale=0.63]{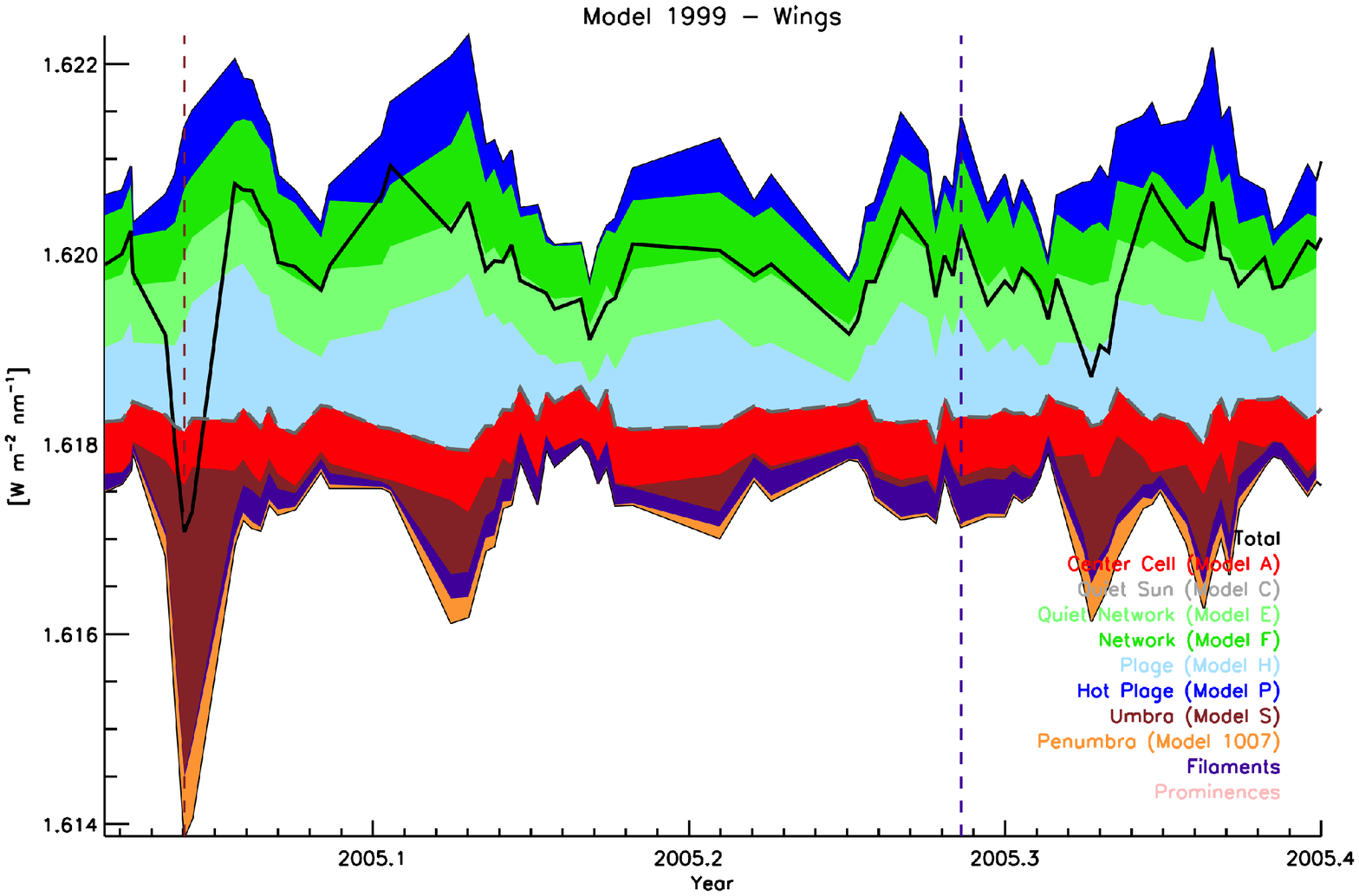}
  \includegraphics[scale=0.63]{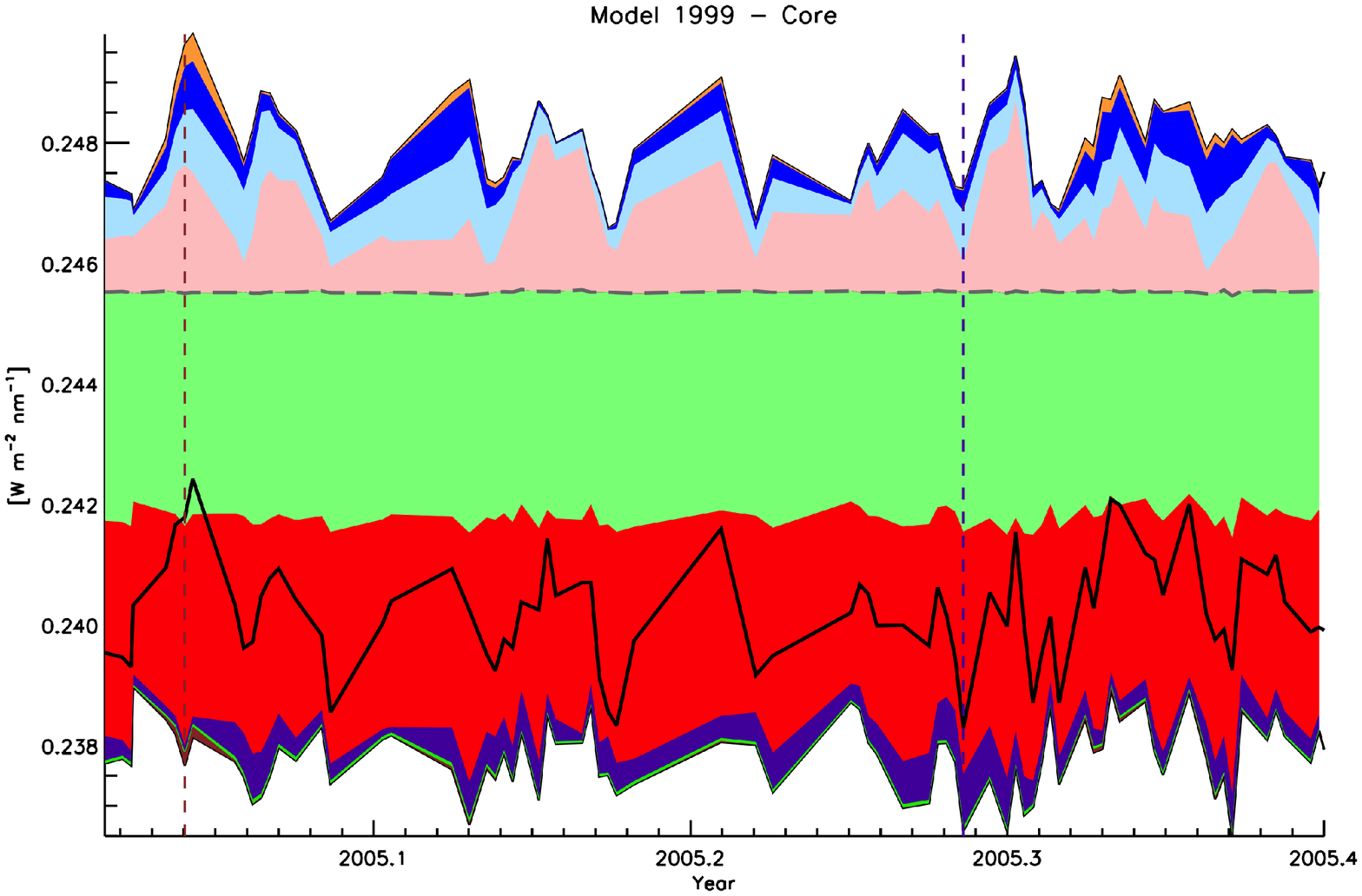}
  \caption{These stacked plots show the contributions of each different component of solar activity type as a function of time computed for Model 1999 modified to take into account the contribution of filaments for both the H$\alpha$ wing and the line core, giving a visual indicator of each component's significance. The contribution from the quiet Sun, which varies due to its areal coverage as other components appear or disappear on the disk, is the dashed grey line, forming a baseline time-varying irradiance. Positive-going components are stacked above this dashed line, ordered from most to least average significance. Negative-going contributions are below the dashed line, ordered similarly. The solid black line gives the spectral irradiance. The dashed brown vertical line shows a time in which variability is dominated by the passage of a large active region (same as in Fig.~\ref{Fig_rotscale_2005}); the  dashed purple vertical line indicates the time at which the contribution of filaments produces the largest decrease of the H$\alpha$ wing (same as in Fig.~\ref{Figvarswithfils}). \label{contribs_with_fil2005}}
    \end{figure}

The discussed models use various components (Table~\ref{tab_core_wings_models}) that may have different impacts on the flux coming from the line core and the line wings, thus inevitably complicating interpretation of variability of the core/wing ratio. Fig.~\ref{contribs_with_fil2005} compares the behavior of the FAL1999 model components and of the filament and prominence models assessed at $\lambda$= 656.28 nm (H$\alpha$ core) and $\lambda$= 654.43 nm (line wing). The single contributions but the prominence one, were computed as the sum of the difference between the line profiles (in irradiance units) of all pixels classified as a particular model component and the line profile of the quiet-Sun model. The prominence contribution is instead an additive component to the irradiance.
In these plots the data were not degraded to the spectral resolution of radiometers.

%%%%%%%%%%%%%%%%%%%%%%%%%%%%%%%%%%%%%%%%%%%%%%%%%%%%
    
    %%%%%%%%%%%%%%%%%%%%%%%%%%%%%%%%

    %%%%%%%%%%%%%%%%%%%%%%%%%%%%%%%%%%%%%%%%%%%%%%%%%%%%
   \begin{figure}[!ht]
   \centering
  \includegraphics[scale=0.63]{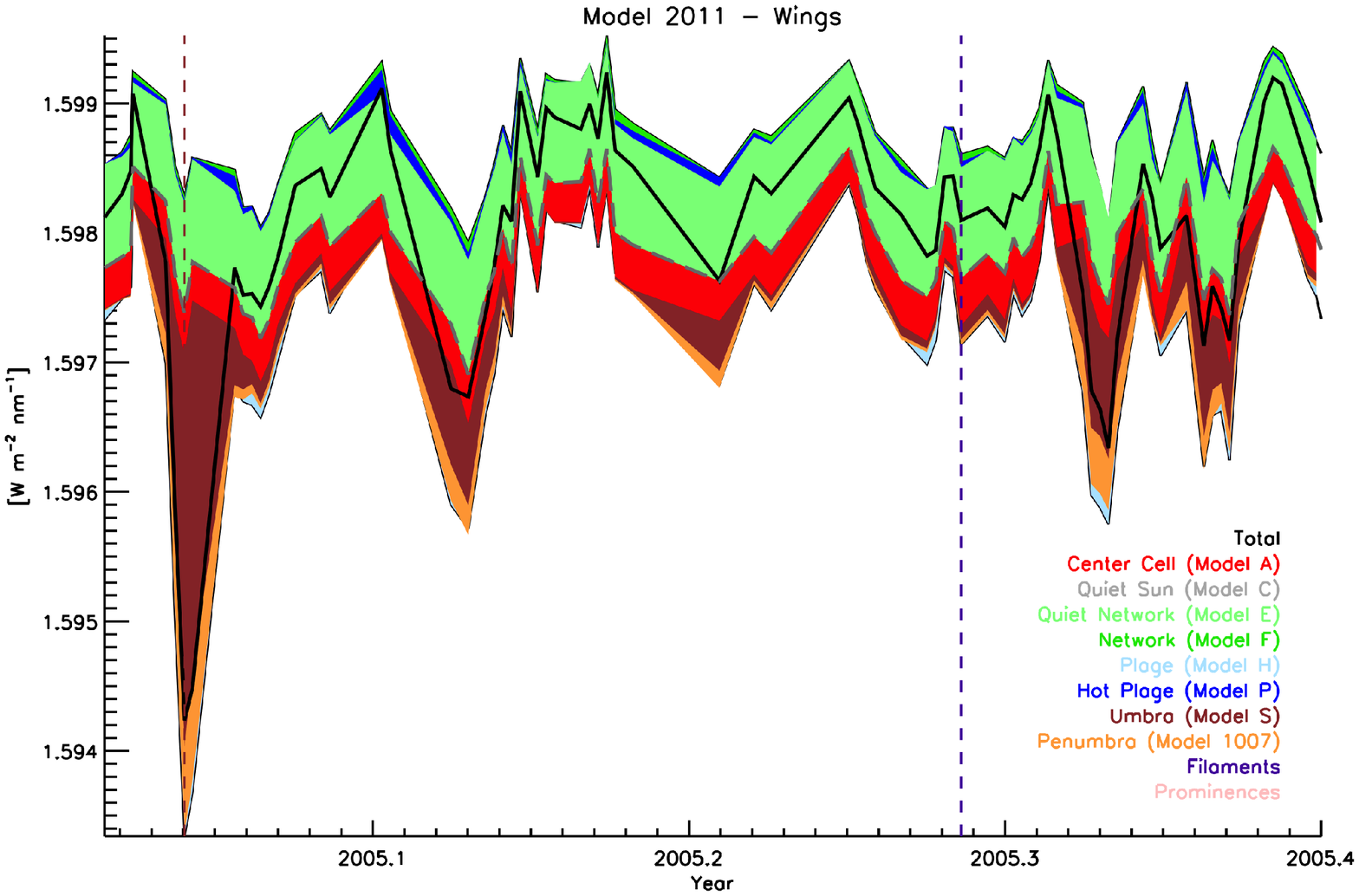}
  \includegraphics[scale=0.63]{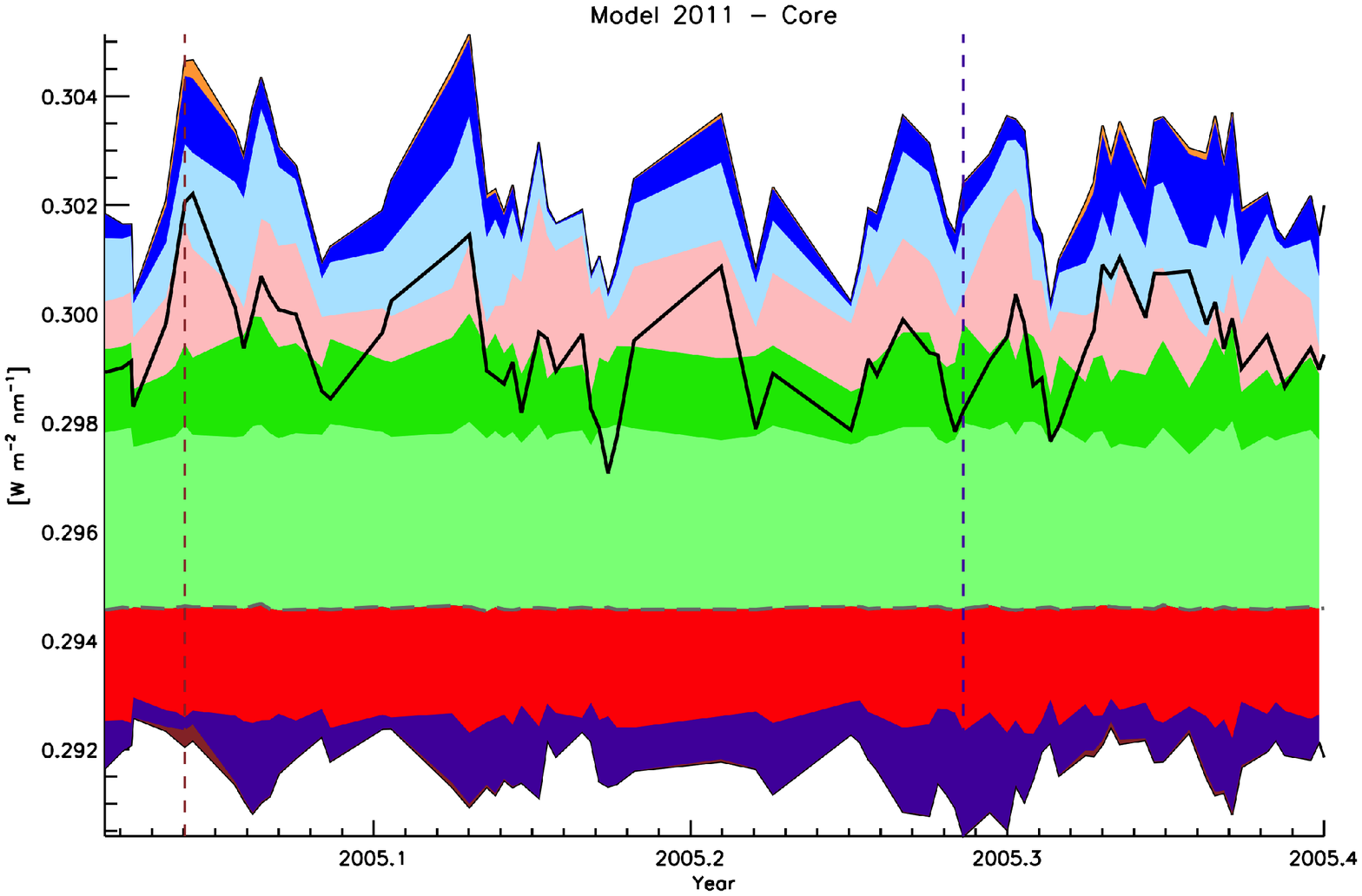}
  \caption{Same as Fig.~\ref{contribs_with_fil2005} but for Model 2011. In comparing the two models, note particularly that the line wings in both models show a lot of variability, both positive and negative, and the line core is dominated by quiet network in both models. The center-cell contributions are comparable in Model 1999. The plage is a minor contributor to the line wings in Model 2011 but a very large contributor in Model 1999. Filaments have comparable contributions to both the line wings and core in Model 1999 but only to the line core in Model 2011. \label{contribs_with_fil2005_mod2011}}
    \end{figure}
        %%%%%%%%%%%%%%%%%%%%%%%%%%%%%%%%

The figure shows that, as expected from inspection of Fig.~\ref{Fig_comp_synt_atlas}, all models but sunspot umbra, penumbra, and center cell contribute positively to the wing intensity. Overall, when sunspots are not present, the major contributors to the wing variability are the plage model followed by the network model. During the passage of  large active regions (e.g. the time marked with the brown dashed vertical line in the plot), the wing intensity decreases, as expected, due to the dominating contribution of the sunspot component, partially balanced, as noted above, by the increase of intensity caused by the two plage contributions. The plot also shows that, besides prominences, only the two plage models and penumbra produce an increase of the core intensity, while all other models provide a negative contribution. During the passage of large active regions, the core intensity increases due to the combined contributions of plages and penumbra, which overcompensate for the decrease of core irradiance caused by the sunspot umbra. After sunspot umbra, the major negative contribution to the variability of the line core is from filaments. However, this negative contribution tends to be balanced and in most cases overcompensated for by the (sum of) plage contributions (e.g. at 2005.12). This result, combined with the small sensitivity of the wings to filaments, explains why filaments do not seem to affect significantly the variability  of the core to wing ratio, even at times in which the filaments contributions is the largest (blue dashed vertical line).
Results presented in Fig.~\ref{contribs_with_fil2005} therefore confirm conclusions drawn in previous sections that the modulation of the core-to-wing ratio is principally driven by the passage of large sunspot groups, due to the combined effect of the reduction of the wing intensity (mostly from the sunspot umbra and penumbra) and the increase of the core intensity (due to penumbra and plage). Note also that the two network models contribute positively to the wing variability and negatively to the core variability, thus producing a net negative contribution to the variability of the core-to-wing ratio. Networks contributions to the variability are however smaller than the contributions of other components, even in the case of the quiet network, which, although large, remains fairly constant over the shown period. 

In Fig.~\ref{contribs_with_fil2005_mod2011} we show the contributions computed for Model 2011. Overall,  conclusions similar to those drawn for Model 1999 can be drawn for Model 2011, with two major differences. The first are the contributions of plage models. In Model 1999 the trends found for the core and wing intensities in the plage models overall correlate, while for Model 2011 they often deviate, the contribution of the plage models being often negative in the wing and the peaks being displaced with respect to the peaks of the cores. This different behaviour results from the steeper temperature gradients of FAL2011 models with respect to FAL1999 models, which produce a larger center-to-limb variation of the wing intensities, and to the lower temperature of the active FAL2011 models with respect to the quiet model at formation heights close to disk center, which results in a negative contribution of plages when these are (predominantly) located at disk center.  Overall, the contribution of plages to the wing variability in Model 2011 is almost negligible, while for the core is higher than for Model 1999, which explains why Model 2011 produces a core-to-wing ratio variability larger then Model 1999. The second major difference  with respect to Model 1999, is the filament contribution in the wings, which is basically null for Model 2011, as expected from the model description in Sec.~\ref{sec:filaments}. Note, however, that even in Model 1999, the contribution of filaments to the wings is very small and does not significantly affect the core-to-wing ratio variability.

Finally, it is important to note that the contribution values are sensitive to the spectral resolution. Specifically, all contribution values decrease with the decrease of the spectral resolution, although the relative contributions of the different models (and therefore the discussion above) remains the same, with the exception of models  A, E, and F, for which the core contributions become slightly positive (at the 10$^{-5}$ - 10$^{-4}$ level) at the radiometer's resolution, and the prominence contribution, which becomes negligible. Because the prominence line profile emission is zero outside of a small spectral region around the H$\alpha$ core, the prominence contribution is particularly sensitive to the spectral resolution and to the spectral range of integration used to define the index. This explains why in Fig.~\ref{Fig_corrs_models_with_prom}, the modeled H$\alpha$ index is largely unaffected by prominences on the rotational scale (obtained at moderate spectral resolution), whereas it is impacted by prominences at longer temporal scales (obtained at high spectral resolution).

\subsection{Chromospheric or photospheric indices?}
\label{sec:chromphot}
Both observations and models suggest that the indices constructed from the core-to-wing ratios of Balmer lines behave more as photospheric rather than chromospheric indices, meaning that they follow more closely the (inverted) TSI and other photospheric indices rather than indices derived from chromospheric lines as the \ion{Mg}{2} or \ion{Ca}{2}. Analyses presented in Sec.~\ref{sec:results} and in Sec.~\ref{sec:models_contributions} suggest this occurs because the core-to-wing ratio is less sensitive to the contribution of network. This, in particular, may contribute to the decrease of correlation between H$\alpha$ and  \ion{Ca}{2} indices reported in the literature. This result is supported by spatially resolved observations. In particular, \citet{cauzzi2009} employed data acquired at the Dunn Solar Telescope with the Interferometric BIdimensional Spectrometer \citep[IBIS,][]{cavallini2006} to show that the core intensity of H$\alpha$ in network regions presents no clear correlation with the underlying magnetic photospheric concentrations and no correlation with the core intensity of the \ion{Ca}{2} 854.2 nm line, which is another broadly employed chromospheric diagnostic. A better chromosphere tracer, the authors noticed, would be the width of the H$\alpha$ core, which shows a clear correlation with both the photospheric field and \ion{Ca}{2} core intensity. Unfortunately, we are not aware of similar studies conducted on other Hydrogen Balmer lines. 

%%%%%%%%%%%%%%%%%%%%%%%%%%%%%%%%%%%%%%%%%%%%%%%%%%%%
   \begin{figure}
   \centering
  \includegraphics[width=9cm,trim={1.5cm 0  1.3 0},clip]{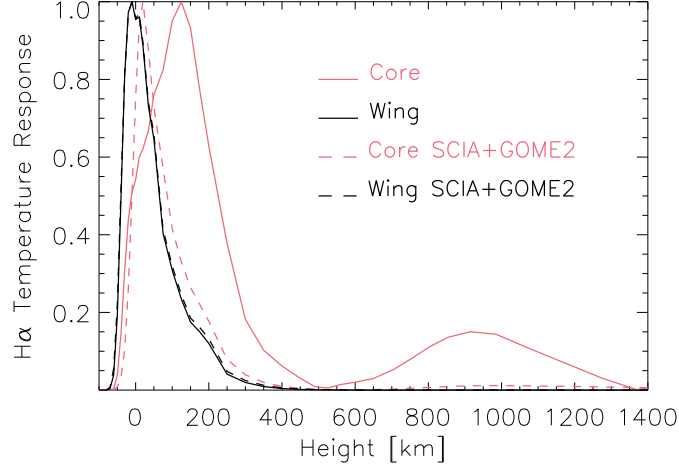}

  \caption{Temperature response function of the H$\alpha$ core and blue wing at the original spectral resolution of the synthesis and degraded to the resolution of radiometric measurements. The core and wing response functions were computed at $\lambda$=656.28 nm  and $\lambda$=654.43 nm, respectively, while for SCIAMACHY and GOME-2 we used the core and wing spectral ranges from Table~\ref{tab_core_wings_models}. 
  \label{Fig_respT_Ha}}%
    \end{figure}
    
    %%%%%%%%%%%%%%%%%%%%%%%%%%%%%%%% 
These results might be surprising considering that it is very well established that  the core of hydrogen Balmer $\alpha$ and $\beta$ lines form in the chromosphere \citep{avrett2008, cauzzi2009,bjorgen2019}. Indeed, the H$\alpha$ core is a widely used chromospheric diagnostic \citep{rutten2007}. However, numerous theoretical studies have pointed out that, as the H$\alpha$ line source function is largely dominated by scattering,  the core intensity of this line is more sensitive to variations of photospheric, rather than chromospheric, physical conditions \citep{socasnavarro2004, cauzzi2009}, up to the point that, unlike other chromospheric lines, the core of H$\alpha$ can be reasonably reproduced even without a chromosphere \citep{przybilla2004,bergemann2016,rutten2012}. The relatively modest spectral resolution of radiometric observations further reduces the chromosperic contribution, as illustrated in Fig.~\ref{Fig_respT_Ha}, which shows the temperature response function of the H$\alpha$ core and blue continuum at the original spectral resolution and degraded to the SCIAMACHY and GOME-2 resolution.
In this respect, it is important to stress that the contrast of network at wavelengths forming in the photosphere is small \citep[e.g.][]{ortiz2002, criscuoli2017} and that consequently the photospheric temperature gradients derived from semi-empirical approaches are rather similar to the ones of the quiet Sun \citep[e.g.][]{cristaldi2017}, thus confirming our finding that the H$\alpha$ core-to-wing ratio index is almost insensitive to the network, which, by contrast, is clearly detectable in \ion{Ca}{2} imagery. 
The fact that the FAL1999 models better reproduce the observed variability of the Balmer core-to-wing ratios, in spite not reproducing the observed properties of the core width (which are instead qualitatively reproduced by the FAL2011 models), therefore suggest that this set of models better reproduces the changes of photospheric temperature gradients induced by magnetic activity. 

Finally, it should be noted that both spatially resolved observations \citep[e.g.][]{rutten2017} and state-of-the-art 3D magnetohydrodynamic simulations of the solar chromosphere \citep{leenaarts2012} indicate that a detailed description of the core of the Balmer H$\alpha$ line must take into account  the 3D, dynamic nature of the solar chromosphere. In particular, the filamentary structures ubiquitous in H$\alpha$ imagery cannot be reproduced in the 1D, static models adopted in this study. It is therefore not surprising that the models better reproduce the index derived from the upper Balmer lines, which have larger contribution from photospheric layers, rather than the H$\alpha$ index. On the other hand, the overall fair agreement of the synthetic indices with the observed ones suggests that the adopted models capture the main physical processes that determine the observed variability.

 \section{Conclusions}
 \label{sec:conclusions}

 In Paper~I, we analyzed the variability of H-$\beta$, -$\gamma$, and -$\delta$ over solar-rotational timescales using high-photometric-precision radiometric  observations from OMI and TROPOMI. In this paper, we extend the analysis to the H$\alpha$ line, a fundamental chromospheric diagnostic, using both space-based and ground-based measurements, obtained with SCIAMACHY/GOME-2 and ISS instruments, respectively. Our results confirm conclusions drawn in Paper~1 that Balmer lines more closely follow the variability of photospheric rather than chromospheric indices on solar-rotational timescales, as they are more sensitive to sunspots than plage and network. Over longer temporal scales, the H$\alpha$ index increases with magnetic activity and is positively correlated with various other activity indices. This result is in agreement with findings by \citet{meunier2009} but disagrees with results presented in \citet{maldonado2019}, who found that the H$\alpha$ index is anticorrelated with the \ion{Ca}{2} index. In agreement with previous studies, we found that the correlation between the H$\alpha$ and the \ion{Ca}{2} indices is a function of the temporal window considered and the level of solar activity.

To validate our conclusions, we compared our measurements with two irradiance reconstruction models, Model 1999 and Model 2011, both based on a semi-empirical approach and on the assumptions that the variability is produced by variations in the areal coverage of quiet, network, faculae, and sunspot components.  On the rotational timescale, Model 1999 produces better agreement with observations than Model 2011. The models seem to overestimate the variability, most likely due to both models overestimating the contribution of plages. On longer temporal scales, a quantitative comparison between the modeled and measured H$\alpha$ indices is hampered by the ISS producing a variability that is most likely unrealistically high. However, in agreement with the ISS observations, both models produce a positive correlation with the \ion{Ca}{2} activity index, and the correlation is a function of the level of activity. % Previous works attributed these trends to the contribution of filaments being relatively higher during periods of activity minima. Because, in our models, we do not take into account the contribution of filaments, we conclude that filaments play a lesser role in affecting the correlation of the Balmer lines indices with chromospheric indices than previously suggested.}

The models allowed us to study the H$\alpha$ core and line wings separately (see Sec.~\ref{sec:wingcore}). Both Model 1999 and Model 2011 produce variability of comparable amplitudes in the core and wings, but Model 1999 produces smaller wing variability due to the positive contributions of plage compensating for the negative contributions of sunspot umbrae and penumbra.

To investigate the contribution of filaments and prominences, we also constructed models that take into account the effects of these two features ( Sec.~\ref{sec:filaments}).   We found that none of the two features significantly affect the H$\alpha$ variability on the rotational timescale, while on the decadal temporal scale filaments produce an appreciable decrease (by approximately 0.2\%)  and prominences an appreciable increase (by approximately 0.5\%). The introduction of filaments also improves the agreement with the observed 4-years correlation with the \ion{Ca}{2} index, while prominences do not seem to produce any significant effect. It is important to note that previous works attributed the observed temporal variation of the H$\alpha$  - \ion{Ca}{2} correlation to the contribution of filaments being relatively higher during periods of activity minima. However, the fact that we found a similar trend  even in models that do not take into account the contribution of filaments (see Sec.~\ref{sec:results2}) suggests that filaments play a lesser role in affecting the correlation of the Balmer lines indices with chromospheric indices than previously suggested.

We investigated the variability of H$\beta$, -$\gamma$, and -$\delta$ over the solar-cycle timescale using the models only, as no reliable disk-integrated observations were available. Our results show that upper Balmer lines indices in counter phase with the activity cycle could be observed with high resolving-power spectrographs, thus suggesting that the variability of these indices depends on the observational characteristics and instrumental properties.

However, our study indicates that the variability of H$\alpha$ is less affected by spectral resolution and the choice of the spectral regions employed to define the core-to-wing ratio. Specifically, we found that spectral resolution does not affect our conclusion that the variability of the $H\alpha$ index is more affected by the passage of large active regions/sunspots and less so by plage regions, as opposed to other chromospheric indices (see Sec.~\ref{sec:models_contributions}). This is due to the relatively large width of the line and to the fact that this line is mostly sensitive to changes in photospheric conditions (see Sec.~\ref{sec:chromphot}), so that its variability is less affected by photospheric contamination, when observed at modest spectral resolutions.    Note, however, that the analysis presented in Sec.~\ref{sec:models_contributions} indicates that the contribution of different features to the H$\alpha$ index variability are affected by spectral resolution. The prominence contribution to the index variability is particularly affected and cannot be appreciated at the spectral resolution of most radiometers, or in general when broad spectral ranges are used to define the core intensity. This is the opposite of what was found for filaments, whose contribution appears to be small even at the fine spectral resolution of the models.

Regarding the discrepancies between our study and \citet{maldonado2019}, we speculate they might be due to the particular temporal window investigated, changing correlation between indices over time or due to data calibration issues.

From the modeling perspective, we find that Model 1999 overall better reproduces the observations. We also find that reconstructed Balmer-line variability is almost insensitive to the particular sunspot atmospheric model employed and that even a Kurucz model without a chromosphere reproduces the observations reasonably well. This result is especially important in the framework of stellar atmospheric modeling, for which spatially resolved observations are not available, and the contribution of different magnetic features is often modelled making use of forward atmospheric models such as those of Kurucz.

\section{Acknowledgments}
{This work utilizes SOLIS data obtained by the NSO Integrated Synoptic Program (NISP), managed by the National Solar Observatory, which is operated by the Association of Universities for Research in Astronomy (AURA), Inc. under a cooperative agreement with the National Science Foundation. DP was partially supported by an NSF grant AGS 1620647. 
The authors are grateful to Dr. Jerry Harder for providing the PSPT masks and to Dr. Adam Kowalski for providing the RH routines utilized for the computation of Balmer lines broadening.  We appreciate the assistance of the OSIRIS instrument team (Dr. Doug Degenstein, Dr. Adam Bourassa, Mr. Chris Roth) in retrieving and processing OSIRIS L1B data. Lomnicky Peak Observatory is operated by the AISAS in Tatranska Lomnica and the Kanzelhoehe Observatory for Solar and Environmental Research is operate by the University of Graz (Austria). This research made use of IDL colour-blind-friendly colour tables \citep[see][]{2017zndo....840393W}. Finally, we express our gratitude to the anonymous referee for their insightful comments, which greatly enhanced the quality of the paper.}

\bibliography{manus}

\appendix
\section{Sensitivity of the H$\alpha$ index to spectral resolution}
\label{sec:resolution}
\begin{figure}[!h]
   \centering
  \includegraphics[width=11cm,trim={0.1cm 0  0 0},clip]{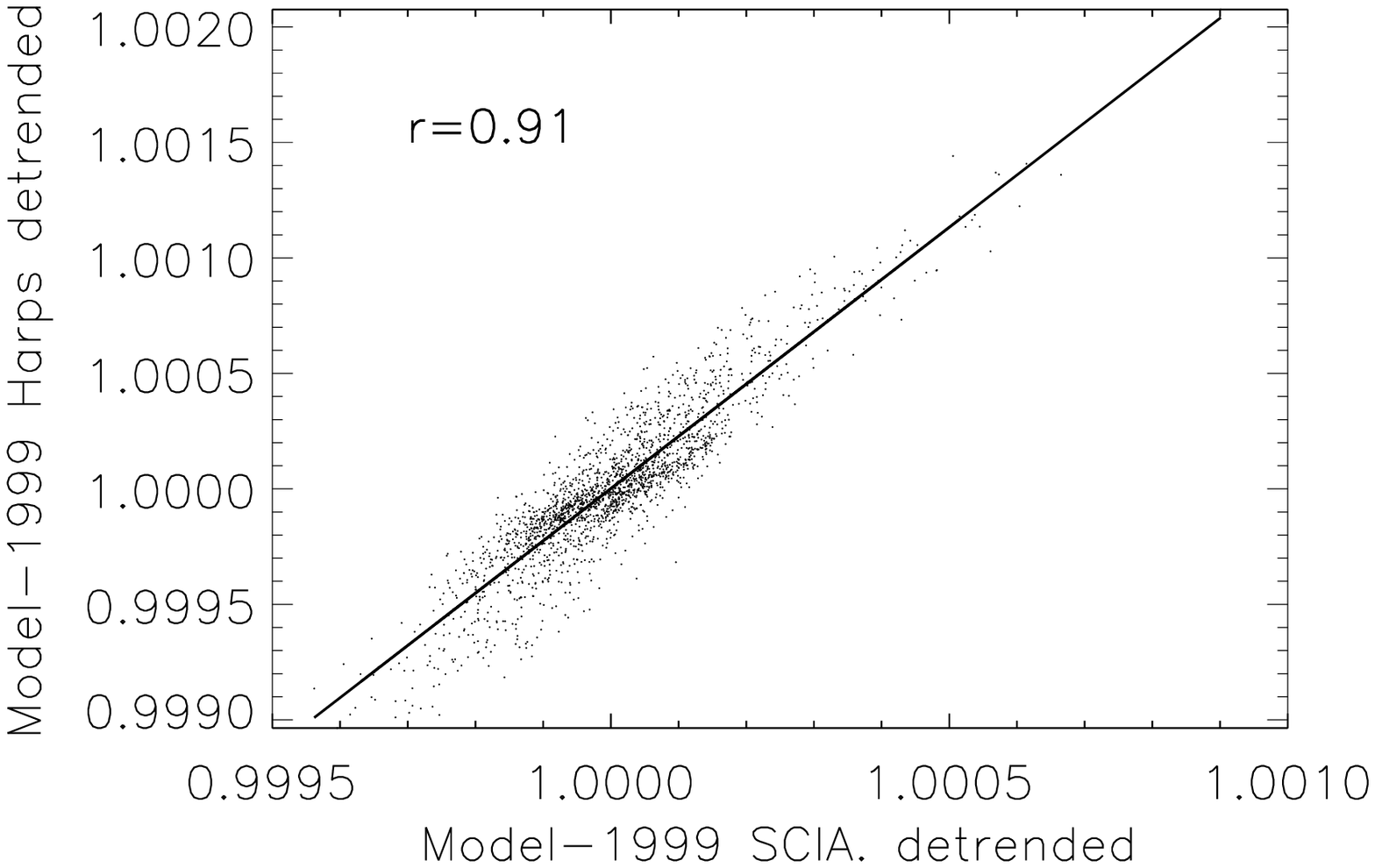}\\
  \includegraphics[width=11cm,trim={0.1cm 0  0 0},clip]{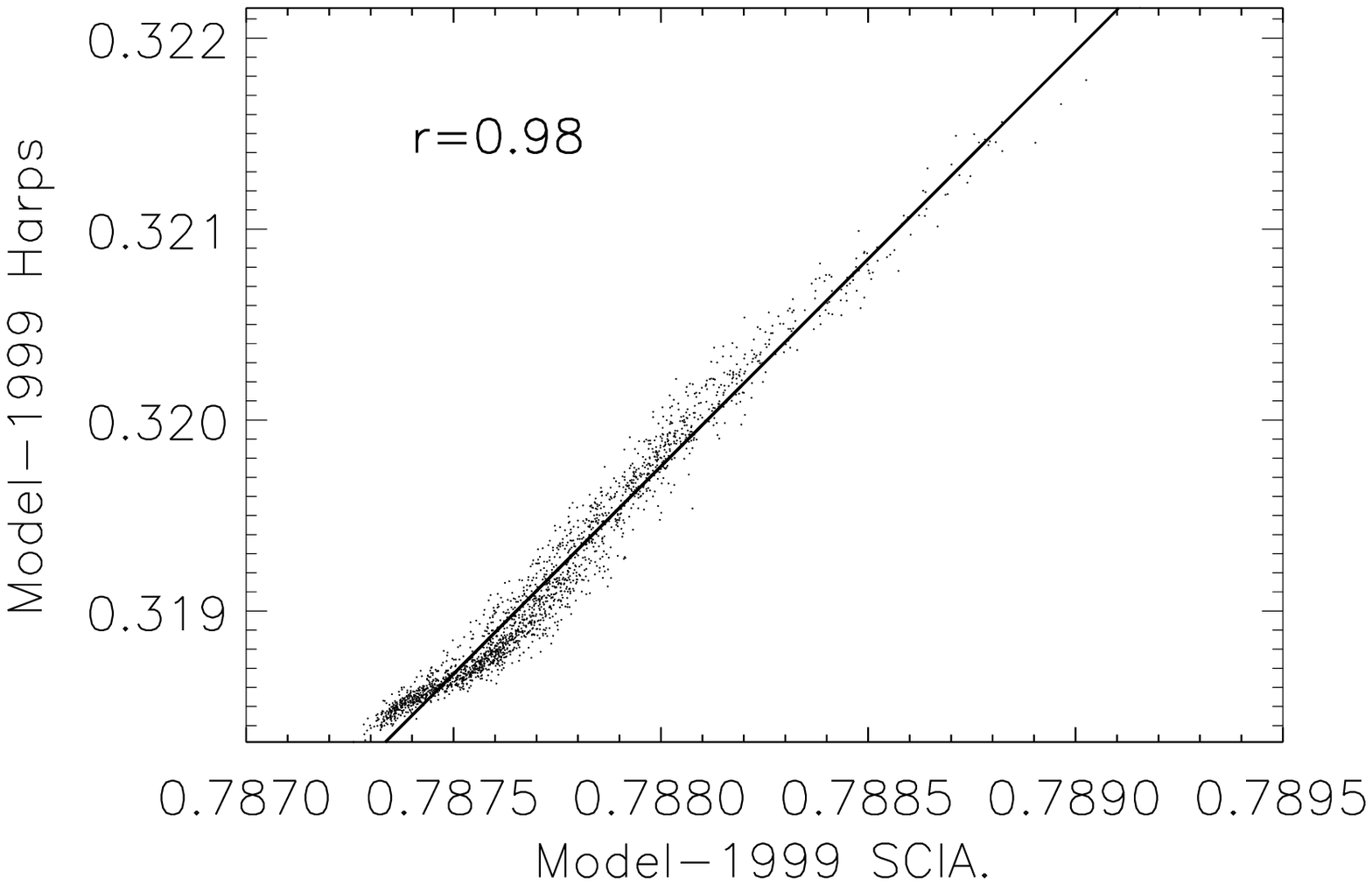}\\

   \caption{Comparison between the H$\alpha$ core-to-wing ratio estimated using Model 1999 simulating HARPS-N and SCIAMACHY measurements. Top: comparison between indices demodulated with a 61-days box (solar-rotational time scale). Bottom: comparison between the original indices (decadal time scale). r is the Pearson correlation coefficient. }             \label{corre_Harps_SCIA}
    \end{figure}

In order to test the sensitivity of estimates of the H$\alpha$ index to instrumental characteristics we modelled the H$\alpha$ core-to-wing ratio by degrading Model 1999 to the HARPS-N and SCIAMACHY spectral resolutions, and computing the index using the definitions described in \citet{maldonado2019} and in Sec.~\ref{sec:measurements}, respectively. The two modelled indices are compared in Fig.~\ref{corre_Harps_SCIA}.

%%%%%%%%%%%%%%%%%%%%%%%%%%%%%%%%%%%%%%%%%%%%%%%%%%%%
 %%%%%%%%%%%%%%%%%%%%%%%%%%%%%%%%%%%%%%%%%%%%%%%%%%%%.%%%%

\section{Synthetic spectra}
%%%%%%%%%%%%%%%%%%%%%%%%%%%%%%%%%%%%%%%%%%%%%%%%%%%%
   \begin{figure}[!h]
   \centering
  \includegraphics[width=8.cm,trim={1.6cm 0  0 0},clip]{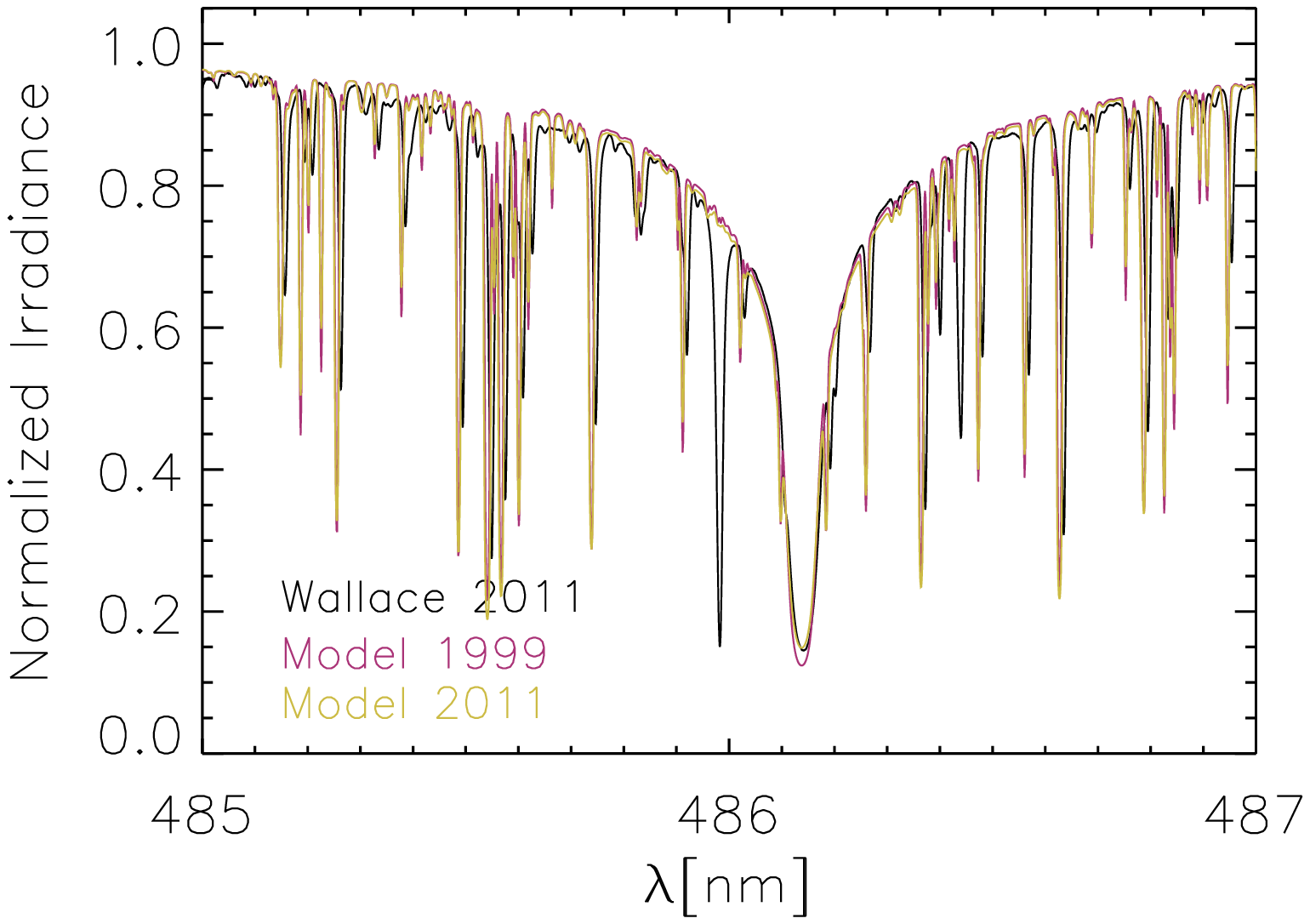}\\
  \includegraphics[width=8.cm,trim={1.6cm 0  0 0},clip]{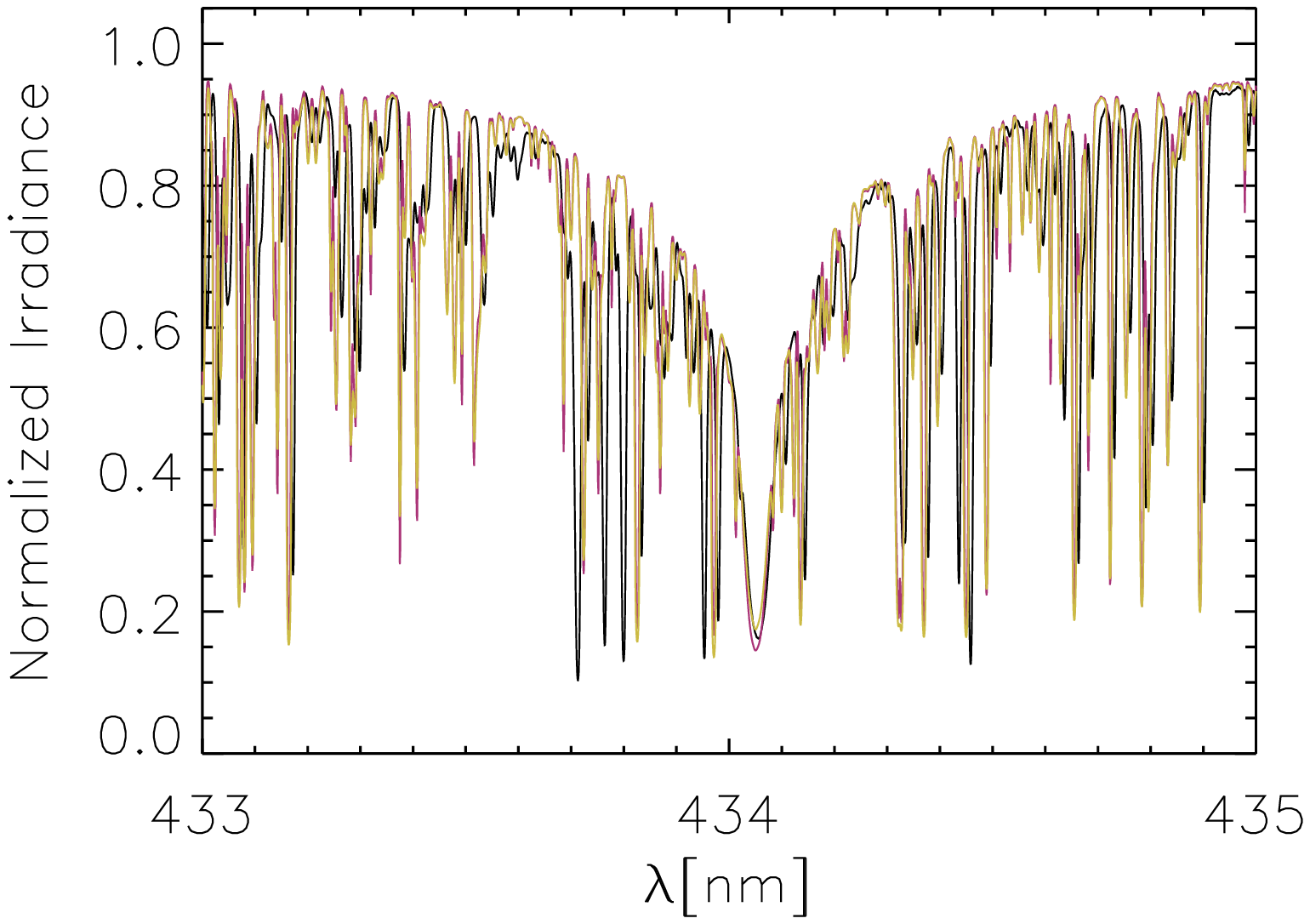}\\
  \includegraphics[width=8.cm,trim={1.6cm 0  0 0},clip]{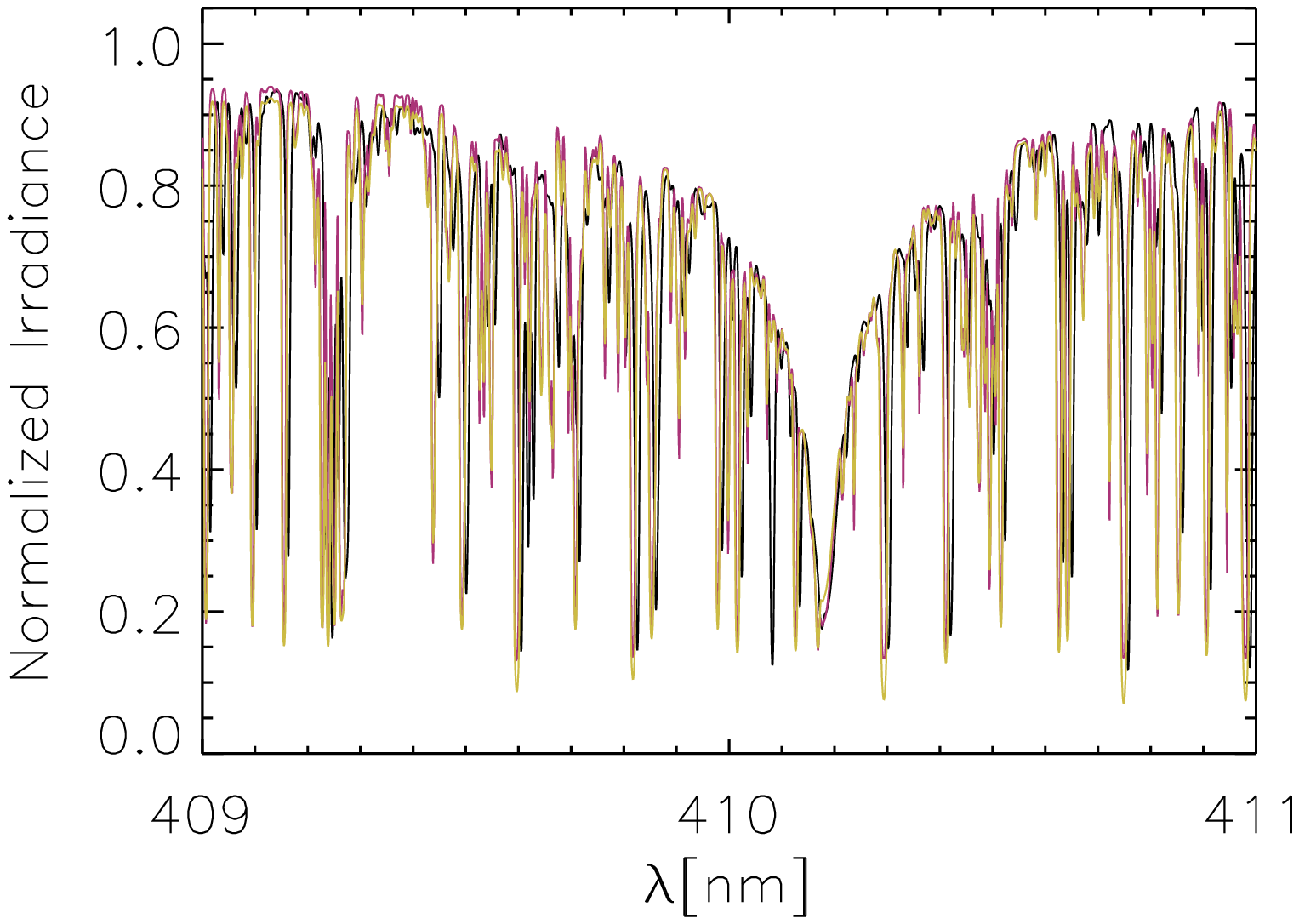}
   \caption{Comparison between synthetic line profiles obtained using quiet Sun models and the observed reference spectrum by \citet{wallace2011}. }             \label{models_vs_wallace}
    \end{figure}
 %%%%%%%%%%%%%%%%%%%%%%%%%%%%%%%%%%%%%%%%%%%%%%%%%%%%.%%%%%%%%%%%%%%%%%%%%%%%%%%%%%%%%%%%%%%%

 %%%%%%%%%%%%%%%%%%%%%%%%%%%%%%%%%%%%%%%%%%%%%%%%%%%%
   \begin{figure}[!h]
   \centering
  \includegraphics[width=8.3cm,trim={1.6cm 0  0 0},clip]{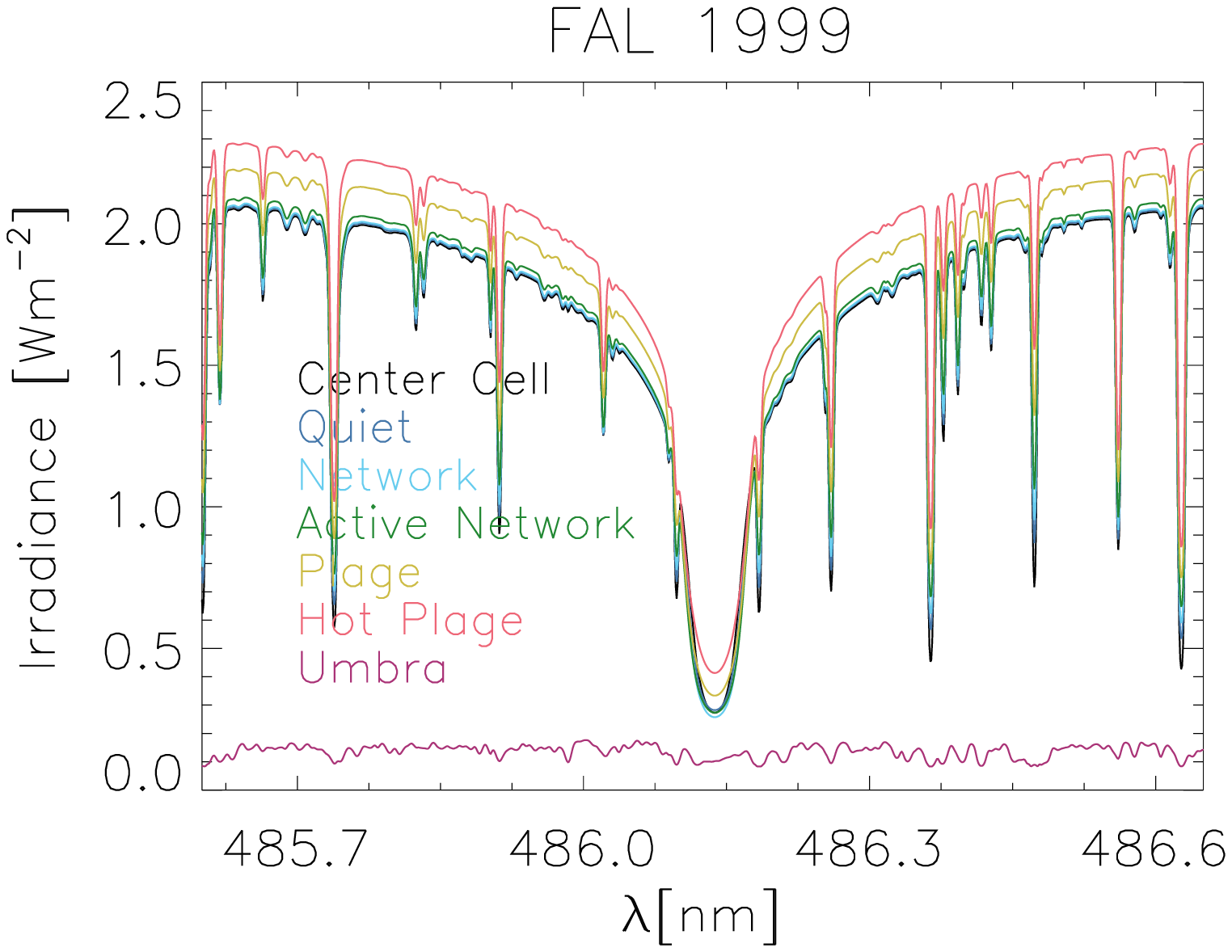}
  \includegraphics[width=8.3cm,trim={1.6cm 0  0 0},clip]{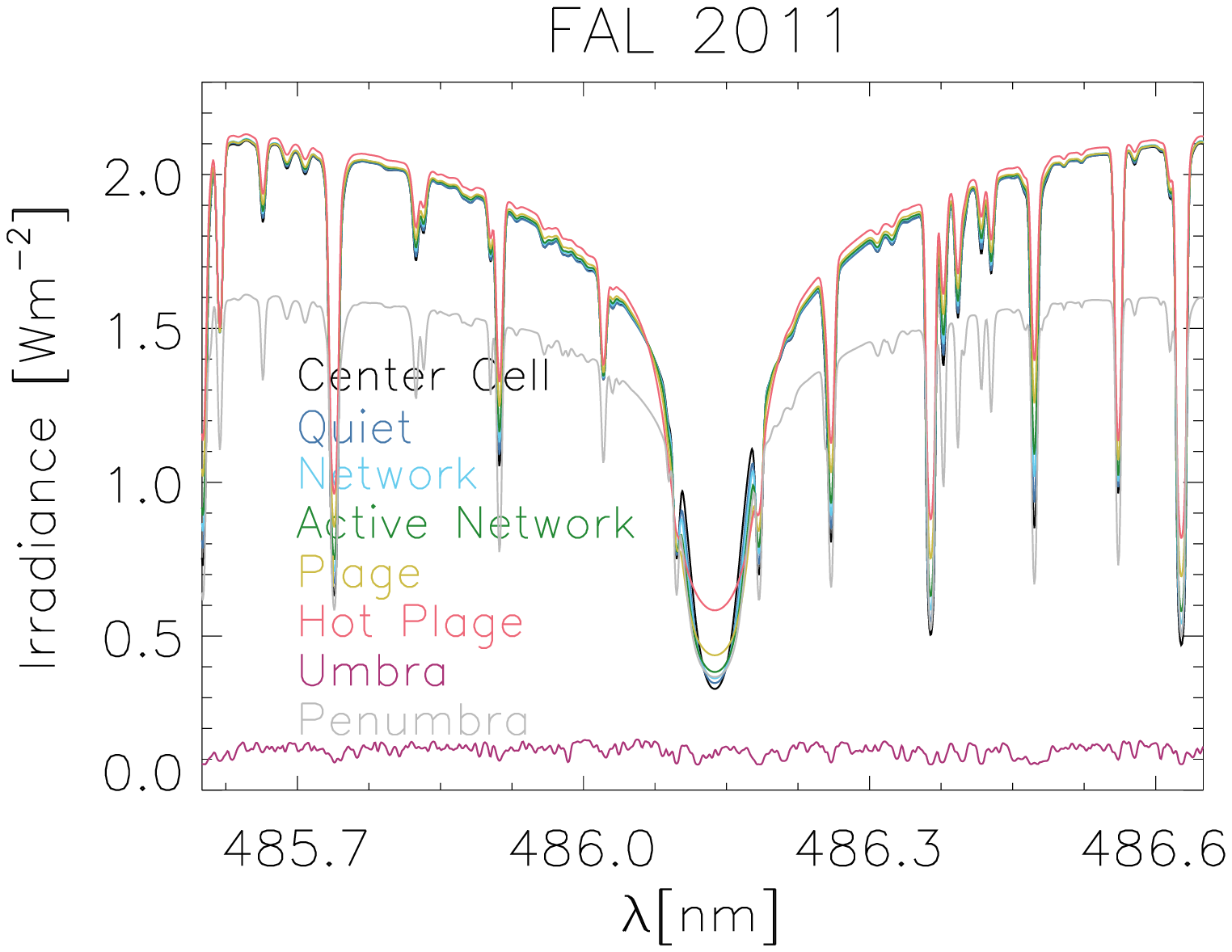}\\
  \includegraphics[width=8.3cm,trim={1.6cm 0  0 0},clip]{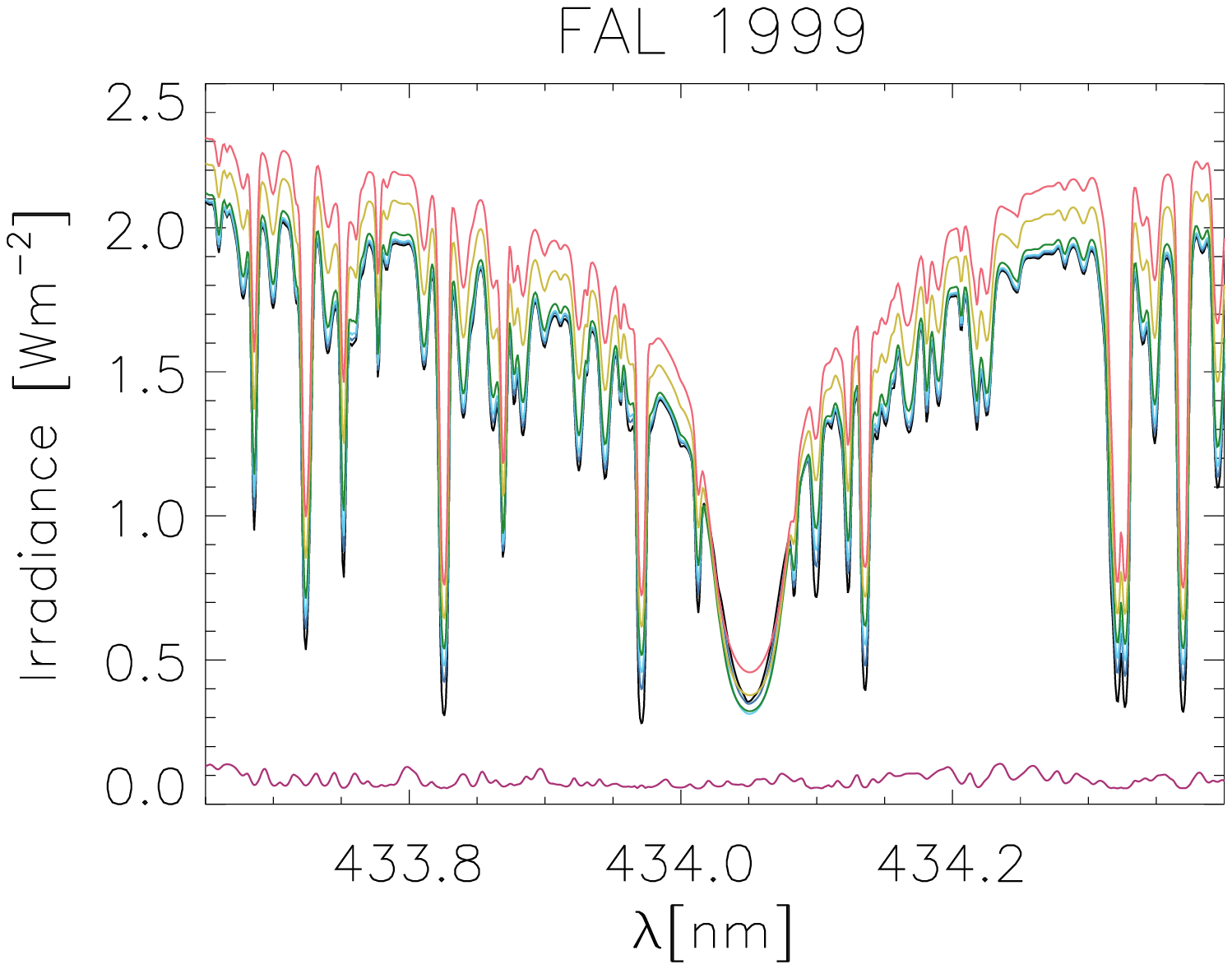}
  \includegraphics[width=8.3cm,trim={1.6cm 0  0 0},clip]{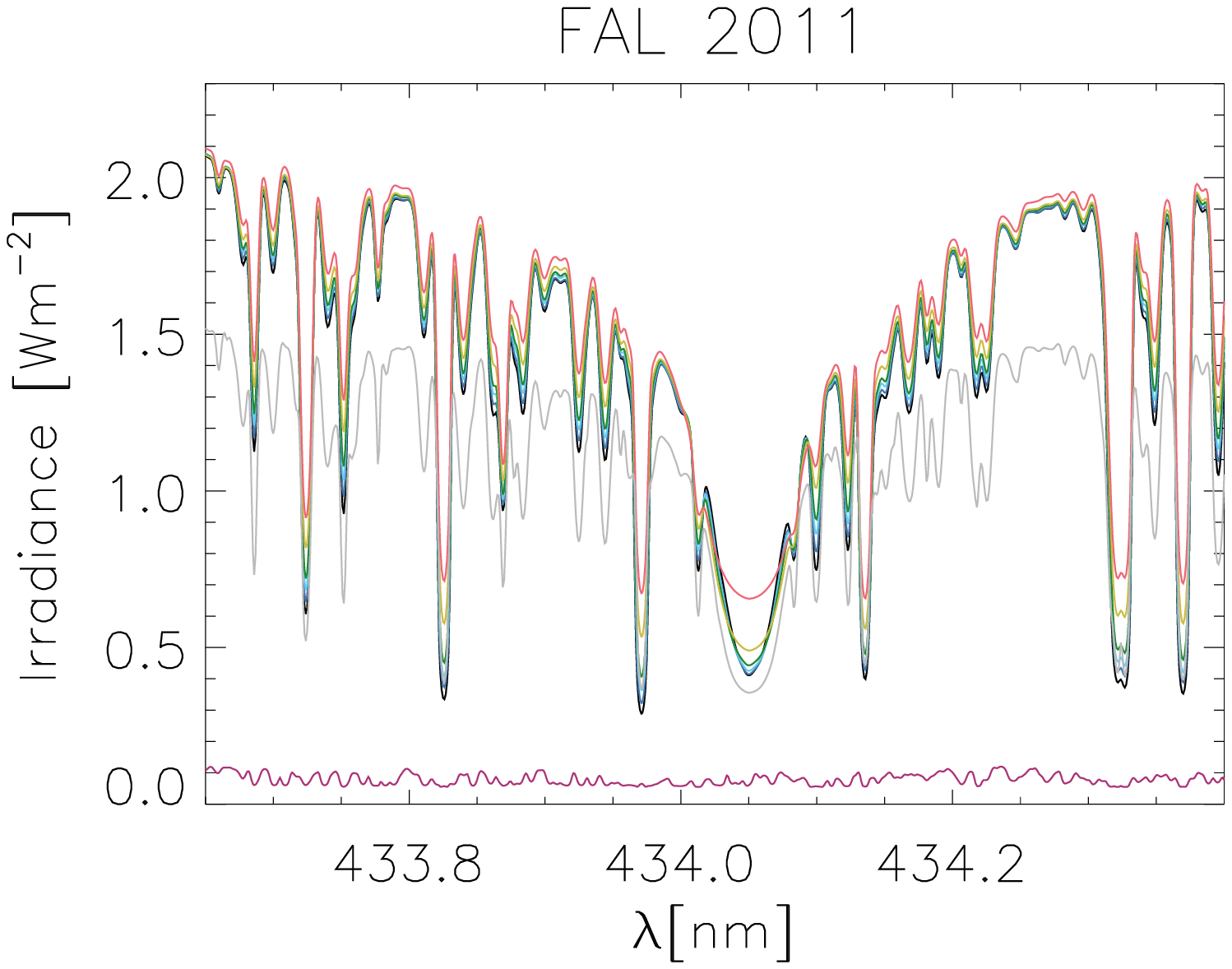}\\
  \includegraphics[width=8.3cm,trim={1.6cm 0  0 0},clip]{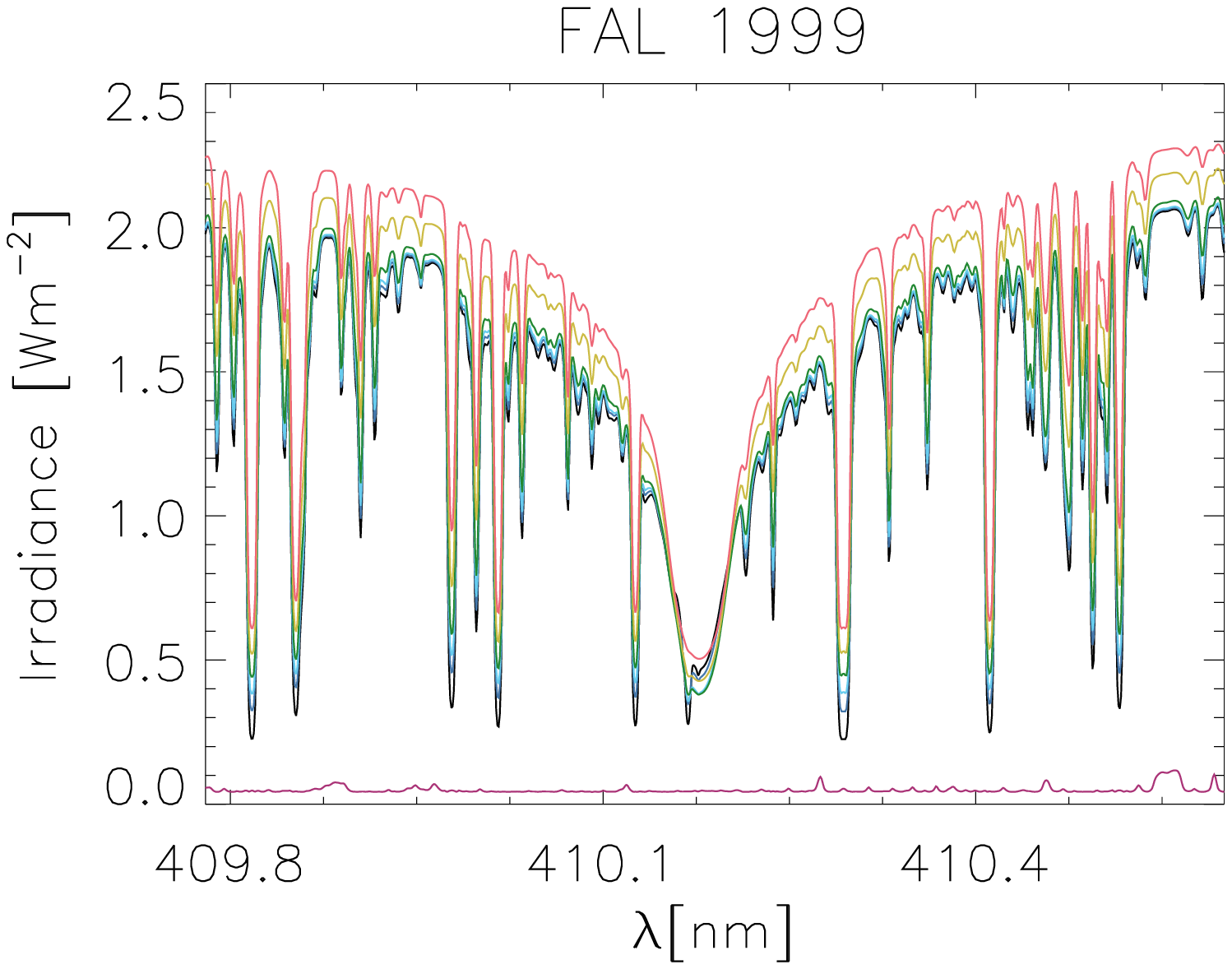}
  \includegraphics[width=8.3cm,trim={1.6cm 0  0 0},clip]{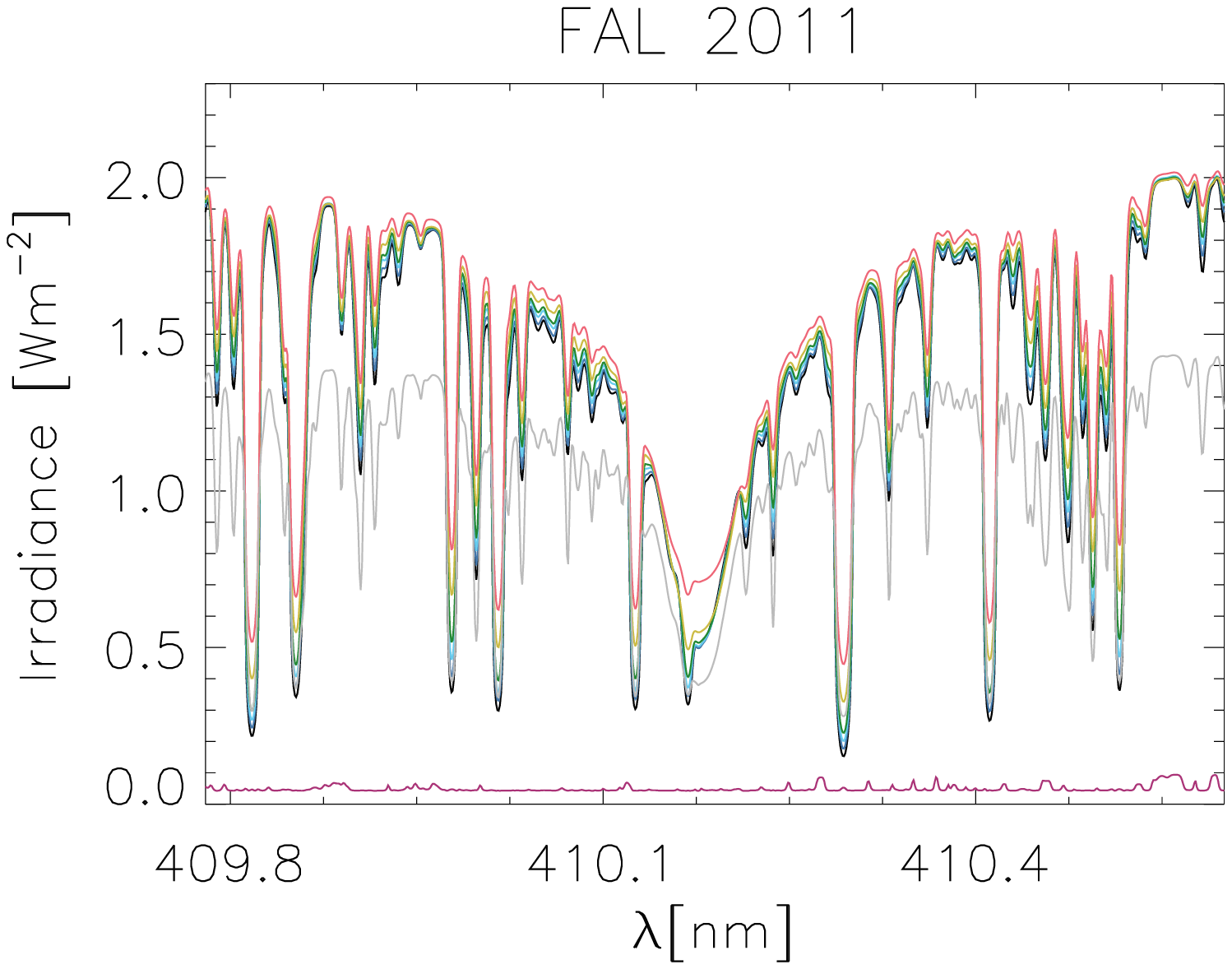}
   \caption{Comparison between synthetic line profiles obtained using the two sets of atmosphere models.} %All profiles were normalized to the average intensity value in a small spectral region of the lines' nearby continua.} 
   \label{models_beta_gamma_delta}
    \end{figure}
 %%%%%%%%%%%%%%%%%%%%%%%%%%%%%%%%%%%%%%%%%%%%%%%%%%%%.%%%%%%%%%%%%%%%%%%%%%%%%%%%%%%%%%%%%%%%
 
 %%%%%%%%%%%%%%%%%%%%%%%%%%%%%%%%%%%

\section{Statistical evaluation of models}
\label{sec:App_metrics}
The following tables report the values of the metrics employed to evaluate the performance of the two sets of models in reproducing the observed variability of Hydrogen Balmer lines. 
 %%%%%%%%%%%%%%%%%%%%%%%%%%%%%%%%%%%%%%%%%%%%%%%%%%%%
\begin{table}[!ht]
\begin{center}
\begin{tabular}{|l|ccc|ccc|}
%    \hline
%    \multicolumn{2}{c}{Multi-column} & testo\\
    \hline
      Model/Observation & \multicolumn{3}{c|}{MAPE} & \multicolumn{3}{c|}{$\rho$}\\
       & 2005-2015 & 2005-2007 & 2011-2015 & 2005-2015 & 2005-2007 & 2011-2015 \\  
     \hline
 Model 1999/H$\alpha$  & 0.009&0.008&0.012 & 0.38&0.38&0.47\\
 %Model 2011/H$\alpha$  & 0.015&0.015&0.016 & 0.2&0.1&0.34\\ 
  Model 2011/H$\alpha$  & 0.008&0.009&0.010 & 0.31&0.25&0.42\\ 
 Model 1999/Balmer & 0.01&0.008&0.014 & 0.49&0.64&0.54 \\ 
 Model 2011/Balmer & 0.017&0.016&0.02 & 0.33&0.39&0.43 \\ 
 \hline
 
\end{tabular}
\caption{MAPE and Pearson correlation coefficients between the models and radiometric measurements detrended with a 61-day average for the period 2005-2015.  \label{table_metrics}}
\end{center}

\end{table}

%%%%%%%%%%%%%%%%%%%%%%%%%%%%%%%%%%%%%%%%%%%%%%%%%%%%%%%%%%%%%%%%%%%%%%%%%%%%%%%
\begin{table}[!ht]
\begin{center}

\begin{tabular}{ |l|l|l|l|l|} 
 \hline
      Index & Observation & Model 1999 & Model 2011 & Model 2011 INTERP.\\
     \hline
 Balmer & 1.1$\times 10^{-4}$ & 1.5$\times 10^{-4}$ & 2.3$\times 10^{-4}$ & \\ 
 H$\alpha$ & 0.94$\times 10^{-4}$ & 1.12$\times 10^{-4}$ & 1.82$\times 10^{-4}$ & 0.95$\times 10^{-4}$\\
 
 \hline
\end{tabular}
\caption{Standard deviations of the indices computed using radiometric measurements and models detrended with a 61-day average for the period 2005-2015.}
\end{center}
\label{table_stddev}
\end{table}
%%%%%%%%%%%%%%%%%%%%%%%%%%%%%%%%%%%%%%%%%%%%%%%%%%%%%%%%%%%%%%%%%%%%%%%%%%%%%%%%%%

\begin{table}[!ht]
\begin{center}
\begin{tabular}{|l|cc|cc|}
%    \hline
%    \multicolumn{2}{c}{Multi-column} & testo\\
    \hline
      Model/Observation & \multicolumn{2}{c|}{MAPE} & \multicolumn{2}{c|}{$\rho$}\\
       & 2008-2014.5 & 2011-2014.5 & 2008-2014.5 & 2011-2014.5 \\  
 \hline
  Model 1999/ISS &19&19.6& 0.67&0.44 \\ 
 Model 2011/ISS &1.79&1.4 & 0.76&0.47 \\ 
 \hline
 
\end{tabular}
\caption{MAPE and Pearson correlation coefficients between the ISS observations and the models.  }
\end{center}
\label{tab:metrics_decadal}
\end{table}
%%%%%%%%%%%%%%%%%%%%%%%%%%%%%%%%%%%%%%%%%%%%%%%%%%%%%%%%%%%%%%%%%

\section{Sensitivity of reconstructed indices to the choice of the sunspot model}
\begin{figure}[!ht]
   \centering
  \includegraphics[width=13cm,trim={1.6cm 0  0 0},clip]{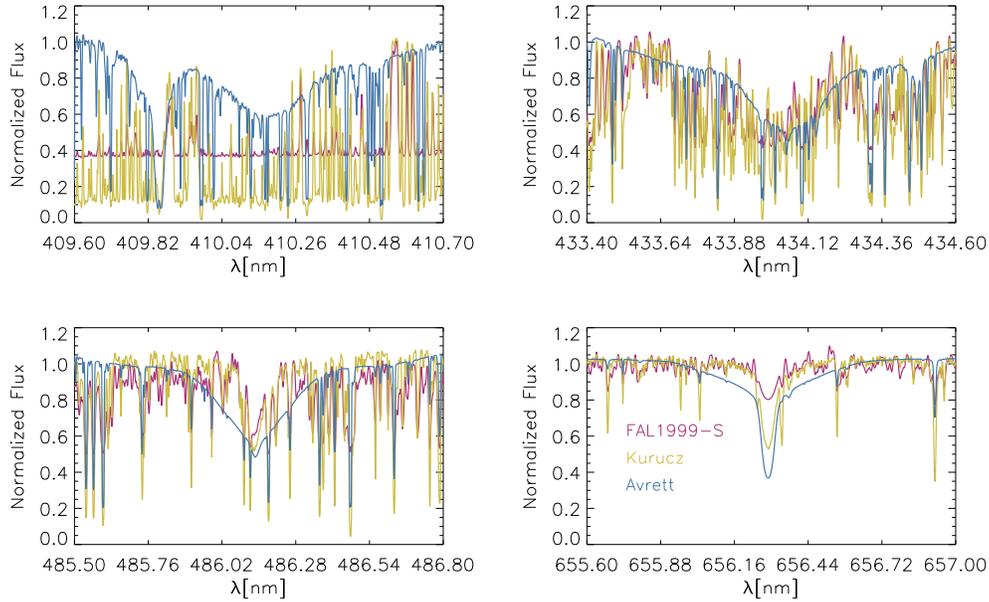}
  
   \caption{Synthetic Balmer lines profiles obtained using different sunspot umbra atmosphere models. All profiles were normalized to the average intensity value in a small spectral region of the lines' nearby continua.}              \label{sunspot_models_flux}
    \end{figure}

\begin{figure}[!ht]
   \centering
  \includegraphics[width=6.cm]{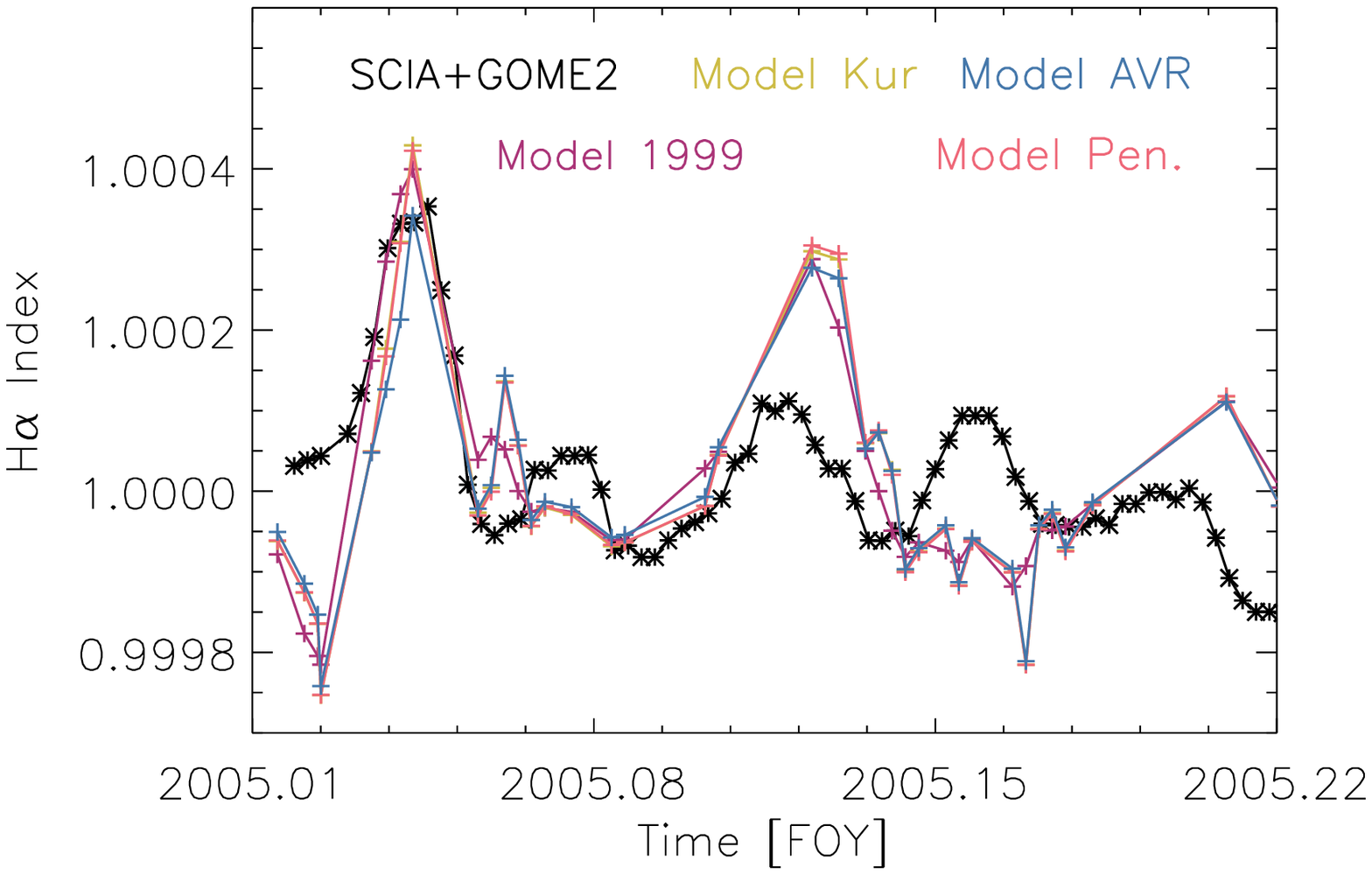} \includegraphics[width=6.cm]{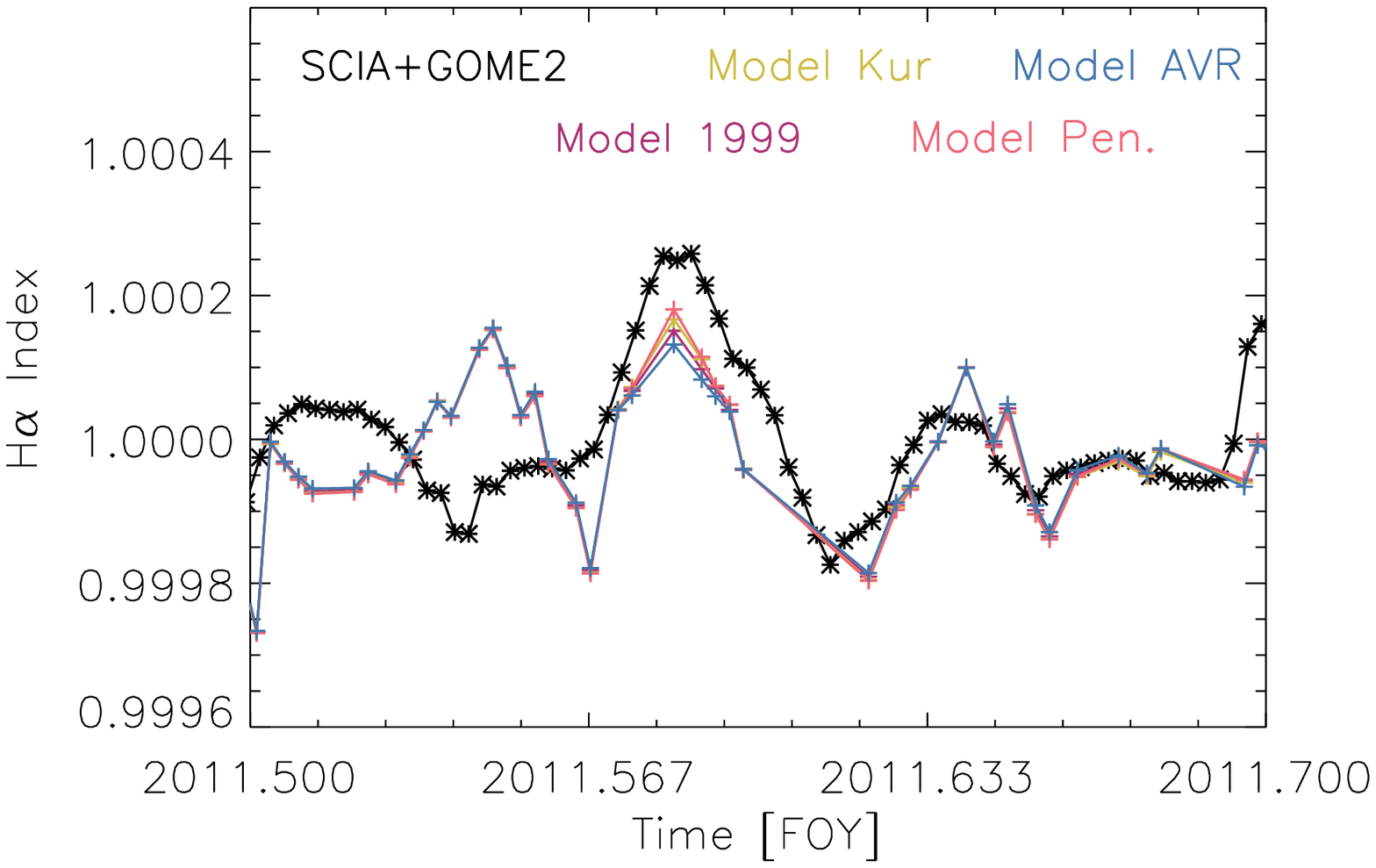}\\
  \includegraphics[width=6.cm]{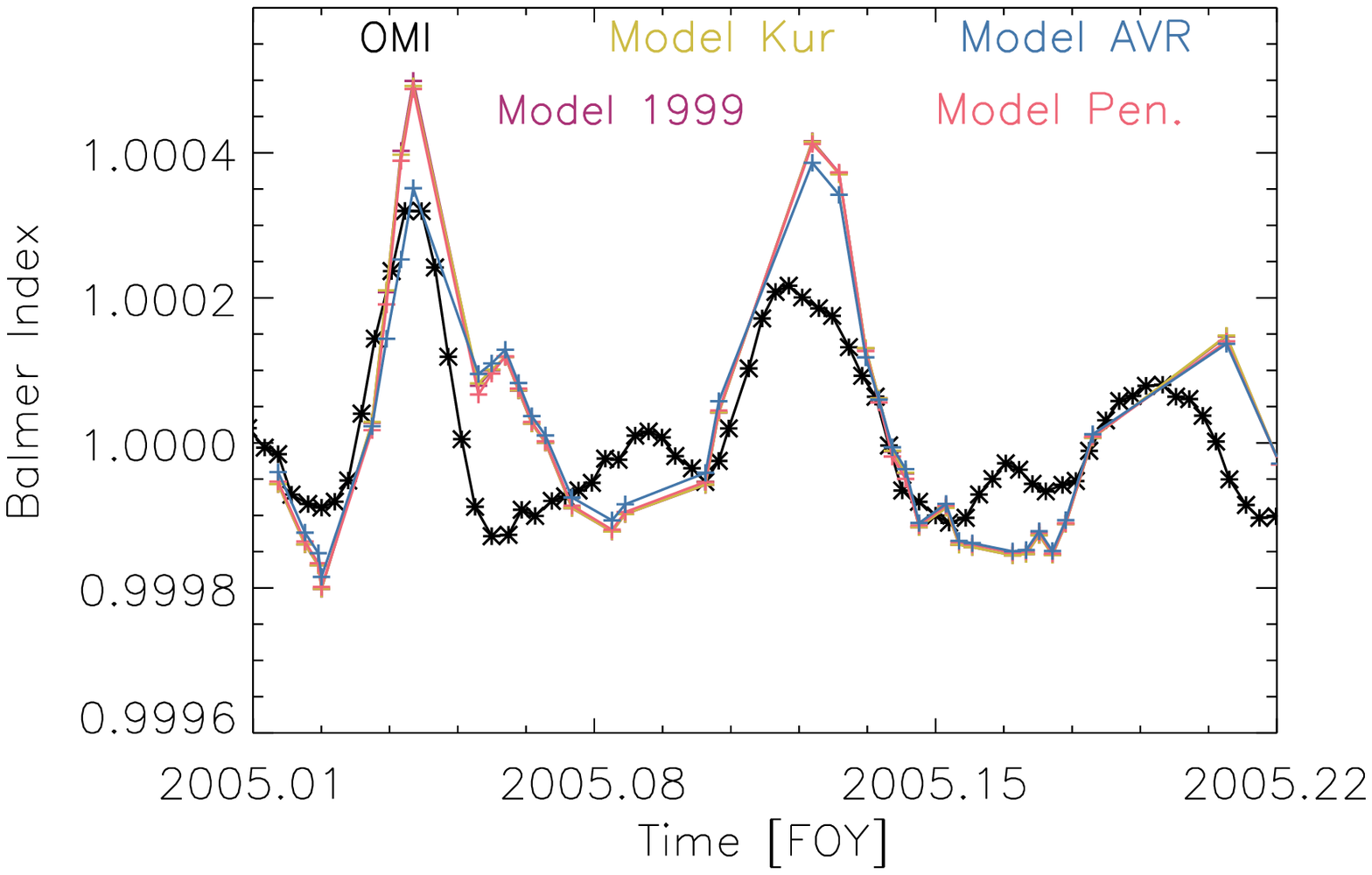}
  \includegraphics[width=6.cm]{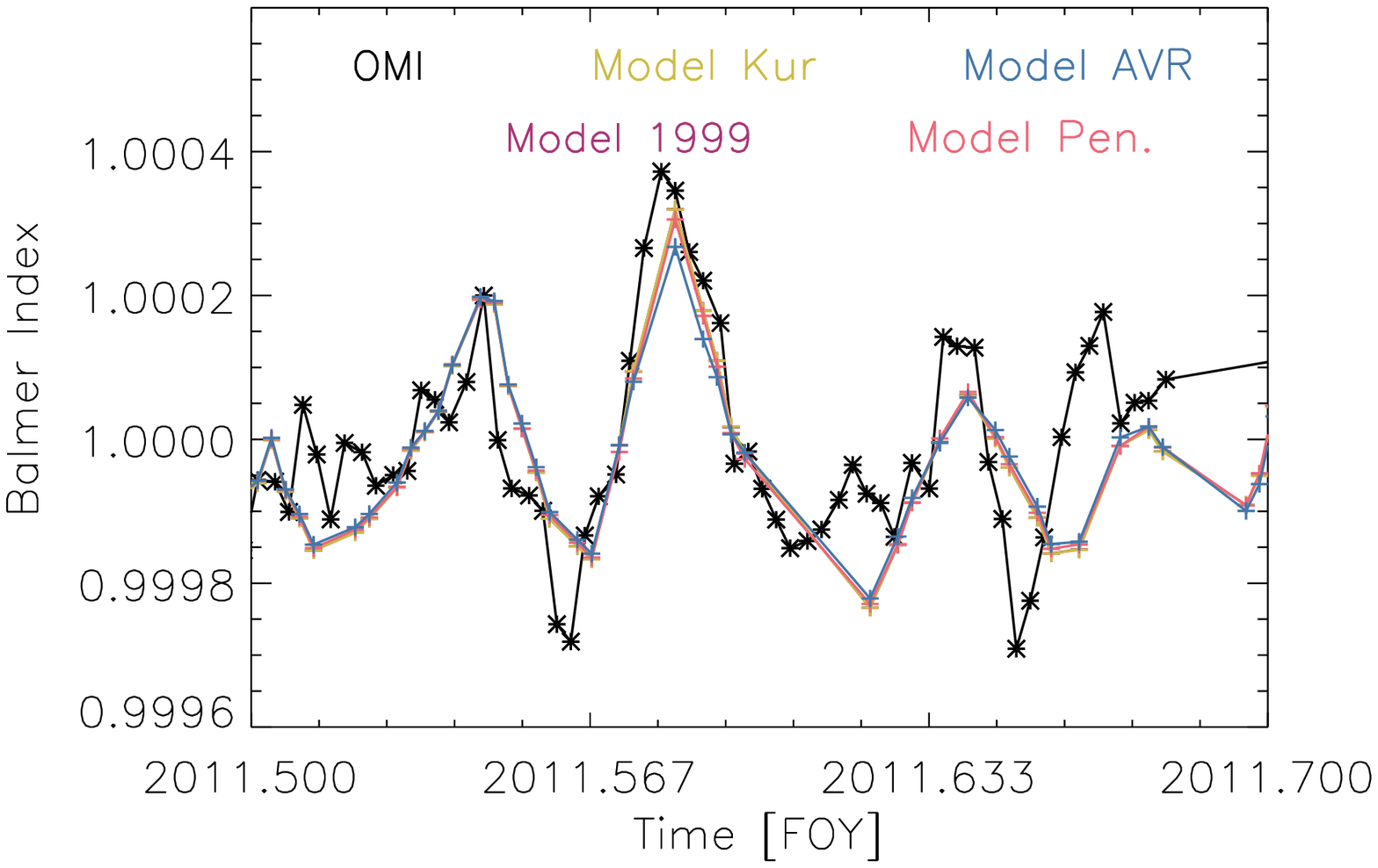}\\
  \caption{Comparison between variability of H$\alpha$ index (top) and Balmer index (bottom) from radiometric measurements and models. All modeled indices were computed making use of a particular sunspot umbra or penumbral model, and the set of FAL1999 atmosphere models to represent other features, as explained in the text.  Model Kur: the sunspot model was substituted with a Kurucz model at 4500K. Model AVR: the sunspot model was substituted with the sunspot model by \citet{avrett2015}. Model Pen: the penumbral model was substituted with the one published in \citet{ding1989}.            \label{Fig_Ha_v2_v3_v4}}%
    \end{figure}

Figure \ref{sunspot_models_flux} shows the synthetic spectra obtained from the different models employed to test the sensitivity of our reconstructions to the adoption of  specific sunspot models.  The variability of reconstructed indices is shown in Fig.~\ref{Fig_Ha_v2_v3_v4}. 
%%%%%%%%%%%%%%%%%%%%%%%%%%%%%%%%%%%%%%%%%%%%%%%%%%%%
 %%%%%%%%%%%%%%%%%%%%%%%%%%%%%%%%%%%%%%%%%%%%%%%%%%%%.%%%%
%%%%%%%%%%%%%%%%%%%%%%%%%%%%%%%%%%%%%%%%%%%%%%%%%%%%
    
    %%%%%%%%%%%%%%%%%%%%%%%%%%%%%%%%

%\section{Contribution of different model components to the H$\alpha$ core and line wing}
%%%%%%%%%%%%%%%%%%%%%%%%%%%%%%%%%%%%%%%%%%%%%%%%%%%%
%   \begin{figure}[!ht]
%   \centering
%  \includegraphics[scale=0.47, angle=90,trim={0.6cm 0  0 3.cm}]{1999V2ContributionsWingandCore2005_v2_withfilv2_SHALLOW.eps}
%  \caption{Contribution of the different components (top 8 eight panels) to the variability of H$\alpha$ wing and core computed for Model 1999 modified to take into account the contribution of filaments (shown in the bottom panel). Green continuous vertical line shows a time in which variability is dominated by the passage of a large active region (same as in Fig.~\ref{Fig_rotscale_2005}); dashed vertical lines indicate times at which the contribution of filaments produce noticeable decrease of the H$\alpha$ core-to-wing ratio (same as in Fig.~\ref{Figvarswithfils}). \label{contribs_with_fil2005}}
%    \end{figure}
    
    %%%%%%%%%%%%%%%%%%%%%%%%%%%%%%%%

    %%%%%%%%%%%%%%%%%%%%%%%%%%%%%%%%%%%%%%%%%%%%%%%%%%%%
%   \begin{figure}[!ht]
%   \centering
%  \includegraphics[scale=0.47, angle=90,trim={0.6cm 0  0 3cm}]{2011V2ContributionsWingandCore2005_v2_withfilv2_SHALLOW.eps}
%  \caption{Same as Fig.~\ref{contribs_with_fil2005} but for Model 2011. \label{contribs_with_fil2005_mod2011}}
%    \end{figure}

\end{document}